\documentclass[12pt, english]{article}

    \usepackage[dvipsnames]{xcolor}
    \usepackage[T1]{fontenc}
    \usepackage[utf8]{inputenc}
    \usepackage[english]{babel}
	\usepackage{amsfonts}
	\usepackage{amssymb}
	    \usepackage{amsmath}
	\usepackage{enumerate}
        \usepackage{enumitem}
	\usepackage{bbm}
	\usepackage{mathtools}
	\usepackage{subfig}
	\usepackage{booktabs}	
	\usepackage{setspace}
	\usepackage{graphicx}
	\usepackage{fullpage}
	
	\usepackage{soul}
	\usepackage{xspace}
	\usepackage{csquotes}
	\usepackage[top=0.75in, bottom=0.75in, left=0.87in, right=0.87in]{geometry}
	\usepackage{placeins}
	\usepackage[footfont=normalsize,font=normalsize]{floatrow}
	\usepackage{lscape}
	\usepackage{longtable}
	\usepackage{mathabx}
    \usepackage{accents}
    \usepackage{float}

    \usepackage{tikz}
    \usetikzlibrary{calc,intersections}

\usepackage{microtype}
    
	\usepackage[longnamesfirst]{natbib}
	\usepackage{bibunits}
	\defaultbibliography{Bibliography.bib}  
	\defaultbibliographystyle{aer}

	\usepackage{hyperref}
	\hypersetup{
    colorlinks=true,
    linkcolor=Red,
    filecolor=magenta,      
    urlcolor=blue,
    citecolor = Blue
}
	\usepackage{cleveref}

    \usepackage{rotating}
	\usepackage[framemethod=tikz]{mdframed}
	\usepackage{todonotes}
	\presetkeys{todonotes}{color=black!5,inline}{} 

    \usepackage{tcolorbox}
    \newtcbox{\feedback}{nobeforeafter,colframe=black,colback=white,boxrule=0.5pt,arc=2pt,
      boxsep=0pt,left=2pt,right=2pt,top=2pt,bottom=2pt,tcbox raise base}
      
    \usepackage{amsthm}

    \newtheorem{asm}{Assumption}[section]
    
    \newtheorem{prop}{Proposition}[section]
    \newtheorem{lem}{Lemma}[section]
    \newtheorem{cor}{Corollary}[section]
    \theoremstyle{definition}
    \newtheorem{rem}{Remark}

\usepackage{array}
\newcolumntype{L}[1]{>{\raggedright\let\newline\\\arraybackslash}m{#1}}
\newcolumntype{C}[1]{>{\centering\let\newline\\\arraybackslash\hspace{0pt}}m{#1}}
\newcolumntype{R}[1]{>{\raggedleft\let\newline\\\arraybackslash\hspace{0pt}}m{#1}}

\usepackage{ragged2e}
\newlength\ubwidth

\newcommand\numberthis{\addtocounter{equation}{1}\tag{\theequation}}

\usepackage{clipboard}

	\newcommand{\bracks}[1]{\left[#1\right]}

	\newcommand{\expesub}[2]{\mathbb{E}_{#1}\bracks{#2}}

	\newcommand{\varsub}[2]{\mathbb{V}\text{ar}_{#1}\bracks{#2}}
	
	\newcommand{\varsubsymbol}[1]{\mathbb{V}\text{ar}_{#1}}
	
	\newcommand{\covsub}[3]{\mathbb{C}\text{ov}_{#1}\bracks{#2, #3}}
	\newcommand{\parens}[1]{\left(#1\right)}

    \newcommand{\prob}[1]{\mathbb{P}\parens{#1}}
    \newcommand{\probsub}[2]{\mathbb{P}_{#1}\parens{#2}}
    
    \newcommand{\expeR}[1]{\mathbb{E}_{R}\bracks{#1}}
    \newcommand{\expeRapprox}[1]{\mathbb{E}_{R}^{approx}\bracks{#1}}
    \newcommand{\probR}[1]{\probsub{R}{#1}}
    \newcommand{\vR}[1]{\mathbb{V}_R\bracks{#1}}
    \newcommand{\vRapprox}[1]{\mathbb{V}_R^{approx}\bracks{#1}}

	\newcommand{\convd}{\stackrel{d}{\longrightarrow}}
	\newcommand{\convp}{\stackrel{p}{\longrightarrow}}

	\DeclareMathOperator*{\argmin}{arg\,min}
	
	\newcommand{\reals}{\mathbb{R}}

	\newcommand{\betahat}{\ensuremath{\hat{\beta}}}

\newcommand{\normnot}[2]{\mathcal{N}\parens{#1,\,#2}}

\newcommand{\tauhat}{\hat{\tau}}
\newcommand{\boldtauhat}{\hat{\boldsymbol{\tau}}}

\newcommand{\ubar}[1]{\underaccent{\bar}{#1}}

\newcommand{\Ytilde}{\tilde{Y}}

\newcommand{\clustersumXX}{\widetilde{XX^\prime_c}}
\newcommand{\clustersumXY}{\widetilde{XY_c}}
\newcommand{\clustersumXeps}{\widetilde{X\epsilon_c}}
\newcommand{\clustersumXepshat}{\widetilde{X\hat{\epsilon}_c}}
\newcommand{\clustersumXXepssq}{\widetilde{X X^\prime \epsilon^2}_c}
\newcommand{\clustersumXXepshatsq}{\widetilde{X X^\prime \hat{\epsilon}^2}_c}

\mathchardef\mhyphen="2D

\usepackage{bbm}
\usepackage{accents}
\usepackage{subfiles}

\usepackage{titlesec}
\titlespacing\section{0pt}{0pt plus 2pt minus 2pt}{0pt plus 2pt minus 2pt}
\titlespacing\subsection{0pt}{0pt plus 2pt minus 2pt}{0pt plus 2pt minus 2pt}
\titlespacing\subsubsection{0pt}{0pt plus 2pt minus 2pt}{0pt plus 2pt minus 2pt}

\usepackage[subtle]{savetrees}
\newclipboard{output-clipboard}

\title{\textbf{Design-Based Uncertainty \\ for Quasi-Experiments}\thanks{We thank Alberto Abadie, Isaiah Andrews, Josh Angrist, Iavor Bojinov, Kirill Borusyak, Kevin Chen, Peng Ding, Avi Feller, Peter Hull, Chuck Manski, Winston Lin, Evan Rose, Pedro Sant'Anna, Yotam Shem-Tov, Neil Shephard, Tymon S\l{}oczy\'{n}ski, Chris Walker, Ruonan Xu, Davide Viviano, and seminar/conference participants at Brown, Ohio State, Cornell, Northwestern, Sciences Po, Toulouse School of Economics, SEA, NASMES, IAAE, and the CEME Young Econometrician's Conference for helpful comments and suggestions. We also thank Haya Alsharif for helpful research assistance. Rambachan gratefully acknowledges support from the NSF Graduate Research Fellowship under Grant DGE1745303.}} 
\author{Ashesh Rambachan\thanks{Massachusetts Institute of Technology. Email: \href{mailto:asheshr@mit.edu}{asheshr@mit.edu}} \and Jonathan Roth\thanks{Brown University. Email: \href{mailto:jonathanroth@brown.edu}{jonathanroth@brown.edu}}}

\begin{document}

{\singlespacing
\maketitle 
\begin{abstract}
Design-based frameworks of uncertainty are frequently used in settings where the treatment is (conditionally) randomly assigned. 
This paper develops a design-based framework suitable for analyzing quasi-experimental settings in the social sciences, in which the treatment assignment can be viewed as the realization of some stochastic process but there is concern about unobserved selection into treatment. 
In our framework, treatments are stochastic, but units may differ in their probabilities of receiving treatment, thereby allowing for rich forms of selection. 
We provide conditions under which the estimands of popular quasi-experimental estimators correspond to interpretable finite-population causal parameters. 
We characterize the biases and distortions to inference that arise when these conditions are violated. 
These results can be used to conduct sensitivity analyses when there are concerns about selection into treatment. 
Taken together, our results establish a rigorous foundation for quasi-experimental analyses that more closely aligns with the way empirical researchers discuss the variation in the data.
\end{abstract}
}

\newpage \clearpage
\section{Introduction}

In the social sciences, researchers often have data on the full population of interest. 
For example, we may observe aggregate data on all 50 U.S. states or administrative data on all individuals in Denmark. Traditional approaches to statistical inference that view the sample as being drawn from a super-population may be unnatural in such settings \citep[][]{ManskiPepper(18)}. One possible alternative in such settings is a model-based approach wherein the units are viewed as fixed, but one develops a statistical model for the outcome. In practice, however, researchers may have difficulty specifying the outcome formation process \citep{AbadieEtAl(22)}. 

The literature on \emph{design-based} inference addresses these difficulties by conditioning on both the units in the finite population and their potential outcomes, and instead viewing the stochastic assignment of treatment as the sole source of randomness in the data. 
This provides an alternative approach to inference in settings where the researcher does not wish to model the statistical process governing the sampling or formation of potential outcomes. 
However, existing work on design-based inference has primarily focused on settings where treatment probabilities are known, as in a randomized experiment \citep[e.g.,][]{neyman_application_1923,imbens_causal_2015, li_general_2017}, or where treatments are determined independently of potential outcomes conditional on covariates \citep[e.g.,][]{abadie_sampling-based_2020, AbadieEtAl(22)}.

In contrast, social scientists often study non-experimental settings in which the assumption of (conditional) random assignment of treatment may be questionable due to concerns about selection into treatment based on unobservable factors. Researchers therefore typically turn to strategies such as difference-in-differences (DID) or instrumental variables (IVs). Researchers often refer to these strategies as ``quasi-experimental'' or ``natural experiments,'' because the treatments are determined in part by factors such as delays in court systems that affect the timing of state-level policy changes \citep[e.g.,][]{jackson_effects_2016}, fluctuations in local weather patterns \citep[e.g.,][]{MadestamEtAl(13), DeryuginaEtAl(19)}, or exposure to natural disasters \citep[e.g.,][]{Hornbeck(12)-DustBowl, HornbeckNaidu(14), Deryugina(17), NakamuraEtAl(21)-Moving} that might reasonably be viewed as stochastic.

In this paper, we develop a design-based approach to inference for such quasi-experimental settings. In line with design-based approaches developed for experiments, we condition on the units in the finite population and their potential outcomes, thus avoiding the need to model the sampling or formation of the potential outcomes. The stochastic nature of the data arises solely from the realization of the quasi-experimental factors, such as court delays or weather shocks, that determine treatment assignment. While we view these factors as stochastic, we importantly do not assume that they generate treatment assignments mimicking a completely randomized experiment. Rather, we view the realization of the quasi-experimental factors as mimicking an unequal-probability experiment wherein each unit $i$ is assigned to treatment with marginal probability $\pi_i$. For example, while it may be reasonable to view court delays as the realization of a stochastic legal process, some states may have a higher probability of realizing such delays than others, leading to heterogeneous $\pi_i$. Of course, if the $\pi_i$ were known, or estimable as functions of observable characteristics, it would be straightforward to adjust for the unequal assignment probabilities. In practice, researchers may not know the $\pi_i$, and they may suspect that they depend on unobservable factors. They therefore proceed using estimators that do not fully adjust for the $\pi_i$.

Our main results concern the properties of common estimators for quasi-experimental settings under this data-generating process. We provide identifying conditions under which common estimators and their associated confidence intervals are valid for finite-population causal estimands. We characterize the biases and coverage distortions that arise when these conditions are violated, and we demonstrate how researchers can conduct sensitivity analyses if they are concerned about possible violations. Altogether, we provide a framework for analyzing quasi-experimental estimators in settings where researchers do not wish to statistically model the sampling or formation of potential outcomes.

As a building block toward understanding popular quasi-experimental estimators, we analyze the difference-in-means (DIM) estimator, which compares the average outcome for the treated and untreated units, under this data-generating process. This allows us to connect our results with the existing design-based literature, which has often focused on the DIM owing to its popularity in randomized experiments. We later generalize our results for the DIM to study least squares regression adjustment, the instrumental variables estimator, and the difference-in-differences estimator.

We derive design-based analogs to the familiar omitted variable bias formula for the DIM.
Its expectation can be decomposed into two terms: a finite-population analog to the average treatment effect on the treated, which we call the expected average treatment effect (EATT), and a bias term that depends on the finite-population covariance between the (unknown) treatment probabilities and the untreated potential outcomes. 
The DIM is unbiased for the EATT if the treatment probabilities are uncorrelated with the untreated potential outcomes in the finite population. 
The DIM is further unbiased for the average treatment effect (ATE) if the treatment probabilities are also uncorrelated with the treated potential outcomes. 

We next establish that the DIM is approximately normally distributed with a particular variance that depends on the finite-population variances of the potential outcomes and treatment effects. 
We provide a finite-population central limit theorem and Berry-Esseen bound, which imply that the DIM is approximately normally distributed when the finite population is large.
We further show that the usual heteroskedasticity-robust variance estimator is consistent for an upper bound on the variance of the DIM. 
These results follow from exploiting connections between our assignment process with unequal probabilities and rejective sampling from a finite population \citep{hajek_asymptotic_1964}.
Taken together, these results imply that when the finite population is large, conventional confidence intervals yield valid but potentially conservative inference for the expectation of the DIM (which corresponds with a causal estimand under the identifying conditions described above).

A novel feature of our setting is that when the individual treatment probabilities $\pi_i$ are heterogeneous across units, conventional standard errors can be strictly conservative even under homogeneous treatment effects. This contrasts with the celebrated result from \citet{neyman_application_1923} for completely randomized experiments, which states that conventional standard errors are strictly conservative if and only if treatment effects are heterogeneous.
As a result, even when the DIM is biased, conventional confidence intervals for the EATT or ATE need not necessarily undercover if the conservativeness of the variance estimator dominates the bias. 
In practice, it is difficult to know which effect will dominate, as neither the conservativeness of the variance estimator nor the bias can be consistently estimated. 

Our results suggest a natural form of sensitivity analysis based on the DIM estimator. Given researcher-specified bounds on the magnitude of selection bias, we show how researchers can construct bounds on and confidence intervals for the EATT or ATE. Researchers can use these bounds to report the ``breakdown'' value of selection bias that would be needed to overturn particular causal conclusions. The (potentially strict) conservativeness of conventional standard errors discussed above implies that such sensitivity analyses yield a (potentially strictly) conservative lower-bound on the robustness of the conclusions to violations of the identifying conditions. 

Our analysis of the DIM estimator immediately applies to the canonical two-period DID estimator \citep[][]{CardKrueger(94), bertrand_how_2004}, one of the most influential quasi-experimental estimators in the social sciences, which can be viewed as a DIM for a first-differenced outcome. Our results imply that the DID estimator is unbiased for the EATT under a design-based analog to the parallel trends assumption, which imposes that the treatment probabilities are uncorrelated with the trends in untreated potential outcomes in the finite population. Our results also enable researchers to conduct sensitivity analyses for violations of this assumption. Similar to the approach in \citet{rambachan_more_2023} from the super-population perspective, we can benchmark reasonable values for the violations of parallel trends using data from pre-treatment periods. 

We illustrate our theoretical results in both a Monte Carlo simulation based on real data and an empirical application. 
In our Monte Carlo simulations, we conduct two-period DID analyses of simulated state-level treatments using aggregated data from Longitudinal Household-Employer Dynamics (LEHD) data from the U.S. Census. 
Since the aggregated data cover over 95\% of all private sector jobs in the United States, the LEHD program writes that ``no sampling error measures are applicable'' \citep[][]{qwi-census}. 
Our simulations therefore analyze uncertainty as arising from the realization of placebo state-level policy changes. 
We allow the state-level treatment probabilities $\pi_i$ to depend on a state's voting results in the 2016 presidential election. 
While the placebo law has no treatment effect for any state, the untreated potential outcomes may vary in a way that is related to state-level voting patterns, leading to violations of the design-based parallel trends assumption.
We illustrate how varying the strength of the relationship between the treatment probabilities $\pi_i$ and state-level voting patterns affects bias and the coverage of conventional confidence intervals for the EATT.
Strengthening the relationship between the $\pi_i$ and state-level voting patterns increases bias but has ambiguous effects on the coverage of conventional confidence intervals, due to its competing effects on bias and the conservativeness of conventional standard errors. Robust confidence intervals that account for the bias have correct coverage for the EATT, but are conservative when the $\pi_i$ differ across units. 

We next revisit empirical work studying the causal effect of Medicaid expansions across U.S. states.
Due to the Affordable Care Act, all U.S. states could expand Medicaid eligibility in 2014, but not all state governments decided to do so. Researchers have used this variation to measure the causal effect of Medicaid expansion on health insurance coverage ($Y_i$) by reporting two-period DID estimates that compare states that expanded Medicaid ($D_i = 1$) against those that did not ($D_i = 0$) \citep[e.g.,][]{wherry_early_2016, MillerWherry(17)}.
We view the 50 U.S. states and their potential outcomes as fixed, and model each state as having an unknown probability of expanding Medicaid based on the realization of stochastic political factors. For example, Ohio famously expanded Medicaid in 2014 only due to a narrow 4-3 ruling by its Supreme Court; but one can imagine that the political process could have played out differently such that Ohio did not expand Medicaid.
Although all states are subject to the vagaries of the political process, some states would require a much rarer realization of the political process in order to adopt Medicaid expansion, leading to potential violations of the design-based parallel trends assumption. 
We conduct sensitivity analyses based on the two-period DID estimator in which we calculate how much the design-based parallel trends assumption must be violated in order to overturn conclusions about the causal effect of Medicaid expansions on health insurance coverage.

We conclude with several extensions that are useful for empirical applications. 
First, we extend our framework to settings with clustered treatments where, for example, we observe individual-level data but treatment is determined in an unknown manner at a more aggregate level (e.g., states or counties).
The cluster-robust variance estimator is valid but potentially conservative, justifying the popular heuristic to cluster standard errors at the level at which treatment is assigned in quasi-experimental settings. 
Second, we provide sufficient conditions under which adjusting for differences in baseline covariates can address the bias of the DIM estimator.
Finally, we study two popular quasi-experimental estimators: instrumental variables (IV) estimators and multi-period difference-in-differences (DID) estimators.
We provide conditions under which their estimands have a causal interpretation and conventional confidence intervals are valid, and we illustrate how researchers can report sensitivity analyses to violations of these assumptions.

Rather than suggesting a new estimator or method for calculating standard errors, our analysis shows that canonical estimators and standard errors can be coherently interpreted from an alternative, design-based perspective.
This perspective aligns with the empirical descriptions provided by researchers, in which statistical uncertainty arises from quasi-experimental factors that partially determine treatments.
Our framework clarifies the identifying conditions under which conventional estimators and standard errors are valid for finite-population causal estimands, and it further provides simple methods for sensitivity analyses based on standard estimators and inferential tools.

\vspace{-1em}
\paragraph{Related work:} We build on the literature on design-based inference, which dates to \citet{neyman_application_1923} and \citet{fisher_design_1935} and has received substantial attention recently. See, for example, \citet{Freedman(2008)-regadj_to_experimental_data, Lin(13), AronowMiddleton(15), li_general_2017,kang_inference_2018, BojinovShephard(19)-ts_experiments, WuDing(21)} in statistics, and \citet{abadie_sampling-based_2020, Xu_2021, Bojinov_Rambachan_Shephard_2020, roth_efficient_2021, AbadieEtAl(22)} in econometrics, among many others. 
Much existing work on design-based inference has focused primarily on settings where treatment probabilities are known to the researcher, as in completely randomized experiments or more complex experimental designs.
By contrast, we analyze a setting in which treatment probabilities are unknown to the researcher and may be related to the potential outcomes.

Our framework is related to the design-based framework in \citet{abadie_sampling-based_2020}, who in Section 3 of their paper consider a setting where treatment assignments are i.n.i.d., and thus can differ across units. \cite{Xu_2021} extends these results to non-linear estimators. 
However, the causal interpretation of the parameters in \citet{abadie_sampling-based_2020} relies on the assumption that treatment probabilities are linear in observable characteristics, whereas we consider estimation and inference for analogs to the ATE or ATT under arbitrary forms of selection. 
We provide a novel analysis of the factors determining the conservativeness of the variance when there is selection into treatment, and the bias and undercoverage that can result from violations of the selection-on-observables assumption.
In the other direction, \cite{abadie_sampling-based_2020} study both binary and continuous treatments, whereas we focus on the binary case only.
Finally, a technical difference between our framework and that in \citet{abadie_sampling-based_2020} is that, as in \citet{neyman_application_1923} and much of the statistics literature that followed, we view the number of treated units $N_1$ as fixed, whereas \citet{abadie_sampling-based_2020} view $N_1$ as stochastic.

\vspace{-1em}
\section{Data-Generating Process}\label{sec:model}

Consider a finite population of $N$ units. 
Each unit is associated with potential outcomes $Y_i(\cdot) := (Y_i(0), Y_i(1))$ corresponding to their outcomes under control and treatment. Individuals also have fixed observable covariates $W_i$. The observed outcome is $Y_i = D_i Y_i(1) + (1 - D_i) Y_i(0)$, where $D_i \in \{0, 1\}$ denotes the treatment of unit $i$. 
The collection of potential outcomes is $Y(\cdot) := \{ Y_i(\cdot) \colon i = 1, \hdots, N \}$ and covariates $W := \{W_i \colon i = 1, \hdots, N\}$ are viewed as fixed (or conditioned on).

Treatment is realized for each unit according to $D_i \sim Bernoulli(p_i)$, where $p_i$ is an unknown, individual-specific treatment probability that may be arbitrarily related to the potential outcomes $Y_i(\cdot)$ and covariates $W_i$. Thus, treatment assignment is determined as if we had an experiment with \emph{unequal} treatment probabilities $p_i$. We analyze the distribution of the treatment vector $D := \left(D_1, \hdots, D_{N}\right)^\prime$ conditional on the number of treated units and the potential outcomes and covariates (see \citet{pashley_conditional_2021} for discussion of why it is desirable to condition on $N_1$). 
\begin{asm}\label{asm: rejective assignment probability}
The treatment vector $D$ satisfies $\prob{D = d \mid \sum_{i=1}^{N} D_i = N_1, W, Y(\cdot)} \propto \prod_i p_i^{d_i} (1 - p_i)^{1-d_i}$ for all $d \in \{ 0,1\}^N$ such that $\sum_{i=1}^{N} d_i = N_1$, and zero otherwise.
\end{asm}
\noindent The special case with $p_i = \bar{p}$ for all $i = 1, \hdots, N$ nests the completely randomized experiment in which any treatment assignment vector with $N_1$ treated units is equally likely. We have in mind that the stochastic treatment assignment $D_i$ corresponds to the realization of some quasi-experimental process, such as court delays or weather. However, some units may be more likely to have a realization of this factor that leads them to adopt treatment than others. This is captured by the individual-specific treatment probability $p_i$. To make this more concrete, we consider the following example.

\vspace{-1em}
\paragraph{Example: Effects of Medicaid expansions across U.S. states.}
As part of the Affordable Care Act, all U.S. states were eligible to expand Medicaid eligibility in 2014, yet not all state governments chose to do so. 
Researchers use this variation in Medicaid expansions across U.S. states to study its effects on state-level health insurance coverage, health care usage, and various health outcomes ($Y_i$) by comparing states that expanded Medicaid ($D_i = 1$) and those that did not ($D_i = 0$) \citep[e.g.,][]{wherry_early_2016, HuEtAl(18), MillerJohnsonWherry(21)}.
Justifying these analyses from a sampling or model-based perspective requires viewing the 50 U.S. states as being drawn from some hypothetical super-population of states or modeling these outcomes as a random process.
By contrast, our framework views the 50 U.S. states ($i = 1, \hdots, 50$) and their potential outcomes $(Y_i(0), Y_i(1))$ as fixed. 
The randomness in the data comes from the realization of state-level expansion decisions $D_i \sim Bernoulli(p_i)$, which we view as the stochastic realization of a state-level political process. For example, Ohio expanded Medicaid in 2014 due to a narrow 4-3 ruling by its Supreme Court, but one could imagine a different realization of the political process in which Ohio chose not to expand in 2014. 
Indeed, similar states such as Wisconsin and Pennsylvania did not expand in 2014. While all states are subject to the whims of their Supreme Court justices and other political processes, we expect the probability of these processes resulting in Medicaid expansion to differ across states. This is reflected in the heterogeneous treatment probabilities $p_i$, which we would expect, for example, to be higher in more liberal states. The $p_i$ are likely to be complicated functions of state characteristics, some of which may be unobserved, and thus we treat them as unknown to the researcher. 
$\blacktriangle$

The treatment assignment process captured in Assumption \ref{asm: rejective assignment probability} is compatible with rich models of \textit{selection bias} (i.e., ``endogeneity'' in econometrics \citep[e.g.,][]{heckman_common_1976, Heckman(78)} or ``non-ignorability'' in statistics \citep[][]{Rubin(78)}), because it allows for the treatment probabilities $p_i$ to be related to the potential outcomes. 
For example, it allows for treatment to be determined by the threshold-crossing model $D_i = 1\{ g(W_i, Y_i(1), Y_i(0)) - \epsilon_i \geq 0 \}$, where $g(\cdot)$ is an arbitrary function of the potential outcomes and covariates, and $\epsilon_i \sim U([0,1]$) is a uniform individual-level shock.
Finally, we emphasize that the interpretation of the treatment probabilities $p_i$ depends on the particular, stochastic determinants of treatment that the researcher has in mind (e.g., court delays or weather); uncertainty is then interpreted relative to that source, holding other determinants of treatment fixed.

\vspace{-1em}
\paragraph{Notation:} 
Let $N_1 := \sum_{i=1}^{N} D_i$ and $N_0 := \sum_{i=1}^{N} (1-D_i)$ denote the number of treated and untreated units, respectively.
We refer to the distribution of $D$ given in Assumption \ref{asm: rejective assignment probability} as the ``randomization distribution'', and we denote probabilities over the randomization distribution by $\probsub{R}{\cdot} := \prob{ \cdot \,|\, \sum_{i=1}^{N} D_i =N_1, W, Y(\cdot) }$. 
We define expectations $\expeR{\cdot}$ and variances $\vR{\cdot}$ analogously. 

While treatment $D_i$ is unconditionally assigned to unit $i$ with probability $p_i$, we conduct our analysis conditional on $N_1 = \sum_i D_i$ (see Assumption \ref{asm: rejective assignment probability}). We denote the marginal probability of treatment for unit $i$ after this conditioning by $\pi_i := \probsub{R}{D_i =1}$. (It turns out that when the finite population is large, the results in \citet[][Theorem 5]{hajek_asymptotic_1964} imply that the $\pi_i$ are approximately equal to the $p_i$ up to a re-scaling; for our results, however, it will typically be easier to work with the $\pi_i$ directly.)

For non-stochastic weights $w_i$ and a non-stochastic attribute $X_i$, we define $\expesub{w}{X_i} := \frac{1}{\sum_{i=1}^{N} w_i} \sum_{i=1}^{N} w_i X_i$ and $\varsub{w}{X_i} := \frac{1}{\sum_{i=1}^{N} w_i} \sum_{i=1}^{N} w_i \left( X_i - \expesub{w}{X_i} \right)^2$ to be the finite-population weighted expectation and variance, respectively.
The finite-population weighted covariance $\covsub{w}{\cdot}{\cdot}$ is defined analogously.
So, for example, $\expesub{1}{Y_i(0)} = \frac{1}{N} \sum_{i=1}^{N} Y_i(0)$ is the equal-weighted average of the untreated potential outcome across the $N$ units in the finite population. 

\vspace{-1em}
\section{Analysis of the Difference in Means Estimator}\label{sec: analysis of DIM}

If the marginal treatment probabilities $\pi_i$ were known to the researcher, it would be straightforward to obtain an unbiased estimate of the average treatment effect using the Horvitz-Thompson estimator, $\frac{1}{N}\sum_i (\frac{D_i}{\pi_i} - \frac{1-D_i}{1-\pi_i}) Y_i$. In practice, however, the treatment probabilities $\pi_i$ are unknown, and may not be consistently estimable if the $\pi_i$ are functions of unobservables. Thus, in practice, researchers will typically estimate a treatment effect using other approaches such as DID or IV that do not explicitly adjust for the differences in treatment probabilities across units.  

As a stepping stone, we study the difference in means (DIM) estimator 
\begin{equation}\label{eqn: defn of tauhat}
\tauhat := \frac{1}{N_1} \sum_{i=1}^{N} D_i Y_{i} - \frac{1}{N_0}\sum_{i=1}^{N} (1-D_i) Y_{i},
\end{equation}
that compares the average outcome for treatment and control units. We derive its expectation and distribution under Assumption \ref{asm: rejective assignment probability}, and show how one can conduct sensitivity analyses that account for bias from non-random assignment. In Section \ref{sec: application to two period DiD}, we show that these results apply immediately to the DID estimator, which can be viewed as a DIM for a first-differenced outcome. For simplicity, we abstract away from observable covariates in this section; see Section \ref{sec: ols adjustment} for an extension to covariate-adjusted estimators. We consider extensions to IV in Section \ref{sec: iv}.  

\subsection{Expectation of the Difference in Means Estimator}\label{sec: expectation of sdim}

We first analyze the expectation of the DIM over the randomization distribution, characterizing its bias for the finite-population average treatment effect and average treatment effect on the treated.

\begin{prop}\label{prop: expectation of sdim}
Under Assumption \ref{asm: rejective assignment probability},
\begin{align*}
\mathbb{E}_{R}[\tauhat] &= \tau_{\text{ATE}} +  \frac{N}{N_0} \covsub{1}{\pi_i}{Y_i(0)} + \frac{N}{N_1} \covsub{1}{\pi_i}{Y_i(1)} \numberthis \label{eqn: expectation of tauhat - ATE} \\
&=\tau_{\text{EATT}} + \frac{N}{N_0} \frac{N}{N_1} \covsub{1}{\pi_i}{Y_i(0)}  \numberthis \label{eqn: expectation of tauhat}
\end{align*}
\noindent where, for $\tau_i = Y_i(1) - Y_i(0)$, $\tau_{ATE} = \frac{1}{N} \sum_{i=1}^{N} \tau_i$ and $\tau_{EATT} = \mathbb{E}\underbrace{\bracks{\frac{1}{N_1} \sum_{i=1}^{N} D_i \tau_i}}_{SATT} = \frac{1}{N_1} \sum_{i=1}^{N} \pi_i \tau_i $.
\end{prop}

Proposition \ref{prop: expectation of sdim} decomposes the expectation of the DIM in two ways. 
First, it equals the finite-population average treatment effect ($\tau_{ATE}$) plus a bias term that depends on the finite-population covariances between the individual treatment probabilities $\pi_i$ and potential outcomes. 
Second, it can also be written in terms of a finite-population average treatment effect on the treated, $\tau_{EATT}$, which we refer to as the \textit{expected} ATT (EATT).
The EATT is the expected value (over the randomization distribution) of what \citet{imbens_nonparametric_2004} and \citet{sekhon_inference_2020} refer to as the ``sample average treatment effect on the treated'' (SATT). Equivalently, it is a convex weighted average of the treatment effects $\tau_i$, with weights proportional to the individual treatment probabilities $\pi_i$. 

Proposition \ref{prop: expectation of sdim} implies that the DIM is unbiased for the EATT if the finite-population covariance between individual treatment probabilities $\pi_i$ and the untreated potential outcomes $Y_i(0)$ is equal to zero, i.e. $\sum_{i=1}^{N} (\pi_i - \frac{N_1}{N}) Y_i(0) = 0$. 
This is satisfied in a completely randomized experiment with $\pi_i \equiv \frac{N_1}{N}$. 
It can also be satisfied if the individual treatment probabilities vary across units but in a way that is not systematically related to the untreated potential outcomes on average in the finite population.
Proposition \ref{prop: expectation of sdim} analogously implies the DIM is unbiased for the finite-population ATE if the finite-population covariance between $\pi_i$ and both potential outcomes is zero.

Since our framework views the potential outcomes as fixed (or conditioned on), we note that both $\tau_{ATE}$ and $\tau_{EATT}$ are functions of the \emph{fixed} potential outcomes for the $N$ units in the population. Such parameters may be easier to interpret than a super-population ATE or ATT in settings where it is difficult to conceptualize sampling from a super-population or the DGP generating the potential outcomes. On the other hand, in many cases researchers may be interested in what the effect of the treatment would be if it were applied in a new, different context, and it may not be entirely obvious how to extrapolate from $\tau_{ATE}$ or $\tau_{EATT}$ to the new setting. As argued in \citet{reichardt_justifying_1999}, however, it is also not entirely clear that imagining the $N$ units as having been drawn from a hypothetical super-population helps with extrapolation to different contexts. We thus view $\tau_{ATE}$ and $\tau_{EATT}$ as coherent, \emph{internally-valid} estimands, while cautioning that they may not be \emph{externally} valid when extrapolated to new settings.

\begin{rem}[Connection to omitted-variable bias formulas]
Proposition \ref{prop: expectation of sdim} can be interpreted as a finite population version of the classic omitted-variable bias formula. 
Define $\varepsilon_i^{Y(0)} = Y_i(0) - \expesub{1-\pi}{Y_i(0)}$ and $\varepsilon^{\tau}_i = \tau_i - \tau_{EATT}$ and rewrite the observed outcome for unit $i$ as $Y_i = \beta_0 + D_i \tau_{EATT} + u_i$, where $\beta_0 = \expesub{1-\pi}{Y_i(0)}$ and $u_i= \varepsilon_i^{Y(0)} + D_i \varepsilon^{\tau}_i$. 
The bias term for $\tau_{EATT}$ given in Proposition \ref{prop: expectation of sdim} is then equal to $\expeR{\frac{\covsub{1}{D_i}{u_i} } {\varsub{1}{D_i}} }$, which coincides with the omitted-variable bias formula for the coefficient on $D_i$ in an OLS regression of $Y_i$ on $D_i$ and a constant. Our results are thus related to those in \cite{Meng(18)}, who analyzes the bias and mean square error of the sample mean under unequal probability sampling. This would correspond to separately analyzing the mean outcome for a single treatment group in our framework.
$\blacksquare$  
\end{rem}

\subsection{Distribution of the Difference in Means Estimator}\label{sec: distribution of DIM}

We next analyze the behavior of $\tauhat$ over the randomization distribution. 
We shown that when the finite population is large, $\tauhat$ is approximately normally distributed with a particular variance and the heteroskedasticity-robust variance estimator is a conservative estimator for this variance. 

Existing results on the distribution of the DIM in randomized experiments \citep{Freedman(2008)-regadj_to_experimental_data, Lin(13), li_general_2017} exploit the fact that random treatment assignment is closely-connected to simple random sampling from a finite-population \citep[][]{Cochran(77)}. Because in our setting treatment probabilities $\pi_i$ differ across units, the DIM estimator no longer corresponds to a sample mean under simple random sampling. 
A key observation for deriving our results, however, is that the DIM is analogous to a Horwitz-Thompson estimator under what is referred to as rejective sampling. We can rewrite the DIM as $\tauhat = \sum_{i=1}^{N} \frac{D_i}{\pi_i} (\pi_i \tilde{Y}_i) - \frac{1}{N_0} \sum_{i=1}^{N} Y_i(0)$, where $\Ytilde_i := \frac{1}{N_1} Y_i(1) +  \frac{1}{N_0} Y_i(0)$. (We can accommodate the case where $\pi_i = 0$ for some $i$, if $\frac{D_i}{\pi_i}$ is defined to be 0 whenever $\pi_i = 0$.) 
The second term, $\frac{1}{N_0} \sum_{i=1}^{N} Y_i(0)$, is non-stochastic, and therefore does not affect the variance or higher-order moments of the distribution of $\hat\tau$. The first term, $\sum_{i=1}^{N} \frac{D_i}{\pi_i} (\pi_i \tilde{Y}_i)$, is a Horvitz-Thompson estimator for $\sum_{i=1}^{N} (\pi_i\Ytilde_i)$ under rejective sampling, which was first studied by \citet{hajek_asymptotic_1964}. Our results on the distribution of $\hat\tau$ below are then obtained by applying results on rejective sampling from \citet{hajek_asymptotic_1964} and others, and then translating these results back into conclusions about the underlying potential outcomes and causal effects (which, in many cases, are non-trivial).

\subsubsection{Comparison of actual and estimated variance}

The exact variance of $\hat\tau$ depends on the second-order treatment probabilities, $\probR{D_i = 1, D_j =1}$, which in general are complicated functions of $(p_1, \hdots, p_N)$. Fortunately,  a simple approximation to the variance is available which becomes accurate when $\sum_{i=1}^{N} \vR{D_i} =\sum_{i=1}^{N} \pi_i (1-\pi_i)$ is large. 

\begin{lem}[Variance of the DIM]\label{lem: variance of tauhat}
Under Assumption \ref{asm: rejective assignment probability},
\begin{equation}
\vR{ \tauhat } (1 + o(1)) = C \left[ \frac{1}{N_1} \varsub{\tilde{\pi}}{Y_i(1)} + \frac{1}{N_0}\varsub{\tilde{\pi}}{Y_i(0)} - \frac{1}{N} \varsub{\tilde{\pi}}{\tau_i} \right] , \label{eqn:asymptotic variance using varw} 
\end{equation}
\noindent where $o(1) \rightarrow 0$ as $\sum_{i=1}^{N} \pi_i (1-\pi_i)  \rightarrow \infty$, $\tilde{\pi}_i := \pi_i (1-\pi_i)$, and $C:=\dfrac{ \frac{1}{N} \sum_{k=1}^{N} \pi_k (1-\pi_k) }{ \frac{N_0}{N} \frac{N_1}{N}  } \leq 1$. \label{prop: formula for asymptotic variance}
\end{lem}

\noindent Lemma \ref{prop: formula for asymptotic variance} shows that the variance of $\tauhat$ depends on the weighted finite-population variances of the potential outcomes and the treatment effects, where unit $i$ is weighted proportionally to the variance of their treatment status, $\vR{D_i} = \pi_i (1-\pi_i)$. 
The leading constant term $C$ is less than or equal to one, with equality when $\pi_i$ is constant across units. In the special case of a completely randomized experiment, the right-hand side of (\ref{eqn:asymptotic variance using varw}) reduces to $\left( \frac{1}{N_1} \varsub{1}{Y_i(1)} + \frac{1}{N_0}\varsub{1}{Y_i(0)} - \frac{1}{N} \varsub{1}{\tau_i} \right) $, matching \citet[][]{neyman_application_1923}'s celebrated formula for completely randomized experiments up to a degrees-of-freedom correction.

We further provide an approximate expression for the expectation of the heteroskedasticity-robust variance estimator $\hat{s}^2$ \citep[][]{White(80)}. 
Define $\hat{s}^2 = \frac{1}{N_1} \hat{s}_1^2 + \frac{1}{N_0} \hat{s}_0^2$, where $\hat{s}_1^2 := \frac{1}{N_1} \sum_i D_i (Y_i - \bar{Y}_1)^2$ and $\hat{s}_0^2 := \frac{1}{N_0} \sum_i (1-D_i) (Y_i - \bar{Y}_0)^2$
for $\bar{Y}_1 := \frac{1}{N_1}\sum_i D_i Y_i$, $\bar{Y}_0 := \frac{1}{N_0}\sum_i (1-D_i) Y_i.$ 

\begin{lem}\label{lem: expectation of Neyman variance}
Under Assumption \ref{asm: rejective assignment probability},
\begin{equation}
\expeR{\hat{s}^2} (1+o(1)) = \frac{1}{N_1} \varsub{\pi}{Y_i(1)} + \frac{1}{N_0} \varsub{1-\pi}{Y_i(0)}, \label{eqn: expectation of shat2}   
\end{equation}
where $o(1)$ is as defined in Lemma \ref{lem: variance of tauhat}.
\end{lem}

By combining the previous two lemmas, our next result shows the heteroskedasticity-robust variance estimator $\hat{s}^2$ is (weakly) conservative for the true variance of $\tauhat$ over the randomization distribution, up to the approximation errors described above. 

\begin{prop}\label{prop: shat conservative}
Let $\vRapprox{\tauhat}$ denote the expression on the right-hand side of (\ref{eqn:asymptotic variance using varw}), and $\expeRapprox{\hat{s}^2}$ the expression on the right-hand side of (\ref{eqn: expectation of shat2}). We have that $\expeRapprox{\hat{s}^2} \geq \vRapprox{\tauhat}$.
Moreover, the inequality holds with equality if and only if 
\begin{equation}
Y_{i}(1) - \expesub{\pi}{Y_i(1)} = \dfrac{(1-\pi_i)/\pi_i}{N_0/N_1} \left( Y_{i}(0) - \expesub{1-\pi}{Y_i(0)} \right) \text{ for all $i$}. \label{eqn: sharpness condition - main text}
\end{equation}
\noindent A closed-form expression for $\expeRapprox{\hat{s}^2} - \vRapprox{\tauhat}$ is given in \eqref{eqn:conservativeness-estimated-variance} in the proof. 
\end{prop}

\noindent In a completely randomized experiment, (\ref{eqn: sharpness condition - main text}) is satisfied if and only if treatment effects are constant, and thus Proposition \ref{prop: shat conservative} nests the well-known result from \citet{neyman_application_1923} that in a completely randomized experiment, the usual variance estimator is weakly conservative and is strictly conservative if and only if there are heterogeneous treatment effects (i.e., $\varsub{1}{\tau_i} >0$). 
Interestingly, Proposition \ref{prop: shat conservative} implies that even when there are constant effects, $\hat{s}^2$ will generally be strictly conservative whenever the marginal treatment probabilities $\pi_i$ differ across units, except in knife-edge cases. 

\begin{cor}\label{cor: shat conservative under constant TEs}
Suppose Assumption \ref{asm: rejective assignment probability} holds and treatment effects are constant, i.e. $Y_i(1) = \tau + Y_i(0)$ for all $i$. Then $\expeRapprox{\hat{s}^2} = \vRapprox{\tauhat}$ only if
\begin{equation}
 \frac{\pi_i}{1-\pi_i} = \frac{N_1}{N_0} \left(1 + \frac{b}{Y_i(0) - \expesub{\pi}{Y_i(0)}} \right)  \label{eqn: sharpness cond without unbiasedness}   
\end{equation}
\noindent for all $i$ such that $Y_i(0) \neq \expesub{\pi}{Y_i(0)}$ and $\pi_i \in (0,1)$, where $b = \expeR{\hat\tau} - \tau$ is the bias of $\hat\tau$. When $\hat\tau$ is unbiased for $\tau$ (i.e., $b=0$), $\expeRapprox{\hat{s}^2} = \vRapprox{\tauhat}$ if and only if $\pi_i = \frac{N_1}{N}$ for all $i$ such that $Y_i(0) \neq \expesub{\pi}{Y_i(0)}$. 
\end{cor}

\noindent Corollary \ref{cor: shat conservative under constant TEs} establishes that when treatment effects are constant and $\hat\tau$ is unbiased, the heteroskedasticity-robust variance estimator is non-conservative if and only if the treatment probabilities $\pi_i$ are equal (as in an experiment) for all units $i$ with $Y_i(0) \neq \expesub{\pi}{Y_i(0)}$. More generally, \eqref{eqn: sharpness cond without unbiasedness} shows that under constant effects (but without unbiasedness of $\hat\tau$), the variance estimator will be strictly conservative unless the odds ratio $\pi_i / (1-\pi_i)$ is exactly proportional to a factor depending on the inverse of $Y_i(0) - \expesub{\pi}{Y_i(0)}$ for all $i$. 

To develop one intuition, note that if $\pi_i$ converges to either zero or one, then $\vR{D_i} = \pi_i (1-\pi_i)$ converges to zero. Thus, when all individual treatment probabilities are close to either zero or one, the variance of $\tauhat$ over the randomization distribution is small. It is less obvious that when treatment effects are constant and $\tauhat$ is unbiased, the variance of $\tauhat$ is in fact maximized when all treatment probabilities are equal (as in a randomized experiment). Notice, however, that the sum of the variances of the treatments, $\sum_{i} \pi_i (1 - \pi_i)$, is maximized when $\pi_i = N_1/N$ for all $i$, by Jensen's inequality.
The proofs of Proposition \ref{prop: shat conservative} and Corollary \ref{cor: shat conservative under constant TEs} establish that this is sufficient for the variance of $\tauhat$ to be maximized under equal treatment probabilities. In Appendix \ref{section: variance conservativeness, conditional vs. unconditional frameworks}, we discuss how the conservativeness of the usual variance estimator is intuitively related to, but distinct from, the well-known fact that a conditional variance must on average be less than an unconditional one by the law of total variance. 

The proof of Proposition \ref{prop: shat conservative} also suggests that the conservativeness of $\hat{s}^2$ will tend to be larger when there is more heterogeneity in $\pi_i$. 
For example, under the setting in Corollary \ref{cor: shat conservative under constant TEs} when $b=0$, $\expeRapprox{\hat{s}^2} - \vRapprox{\tauhat}$ is bounded below by a term proportional to $\varsubsymbol{1}[ (\pi_i - \frac{N_1}{N}) \cdot \allowbreak (Y_i(0) - \expesub{\pi}{Y_i(0)})]$. 
Thus, $\hat{s}^2$ will tend to be quite conservative when the heterogeneity in $\pi_i$ is large, especially if $\pi_i-\frac{N_1}{N}$ is large for units with extreme values of $Y_i(0)$. 
The fact that conventional variance estimates tend to become more conservative when the $\pi_i$ are more heterogeneous has important implications for the coverage of conventional confidence intervals, as we formalize next and explore in Monte Carlo simulations below.

\subsubsection{Asymptotic normality, variance consistency, and confidence intervals}

So far we established that the heteroskedasticity-robust variance estimator is conservative in the sense that its expectation is weakly larger than the true variance of $\hat{\tau}$. 
This suggests standard confidence intervals based on $\hat{s}$ will be conservative for $\expeR{\hat\tau}$ if (i) $\hat\tau$ is approximately normally distributed, and (ii) $\hat{s}^2$ is close to its expectation with high probability. We formalize this argument by considering sequences of finite populations indexed by $m$ of size $N_m$, with $N_{1,m}$ treated units, potential outcomes $\{ Y_{i,m}(\cdot) \,:\, i = 1,...,N_{m} \}$, and assignment probabilities $\pi_{1,m},...,\pi_{N_m,m}$. 
For brevity, we leave the subscript $m$ implicit; all limits are implicitly taken as $m\rightarrow\infty$. 
We provide a central limit theorem (CLT) and variance consistency result under the following mild regularity conditions on the sequence of finite populations. 

\begin{asm}\label{asm: reg conditions for CLT and var consistency}
\hfill
\begin{enumerate}[label=(\alph*)]
\item \label{pi times one minus pi goes to infty} $\sum_{i=1}^N \pi_i (1-\pi_i) \rightarrow \infty$. 
\item \label{lindeberg type condition} Let $\Ytilde_i = \frac{1}{N_1} Y_i(1) + \frac{1}{N_0} Y_i(0)$, and assume $\sigma_{\tilde{\pi}}^2 = \varsub{\tilde{\pi}}{ \Ytilde_i } > 0$ (recall that $\tilde{\pi}_i:=\pi_i(1-\pi_i)$). For all $\epsilon > 0$,
$$\frac{1}{ \sigma_{\tilde{\pi}}^2 } \expesub{\tilde{\pi}}{ \left(\Ytilde_i - \expesub{\tilde{\pi}}{ \Ytilde_i }\right)^2 1\left[ \left|\Ytilde_i - \expesub{\tilde{\pi}}{\Ytilde_i}\right| > \sqrt{ \sum_i \pi_i (1-\pi_i) } \cdot \sigma_{\tilde{\pi}} \epsilon \right]  } \rightarrow 0.$$ 
\item \label{asm for consistent variance estimation} For $m_N(1) := \max_{1 \leq i \leq N} ( Y_i(1) - \expesub{\pi}{Y_i(1)} )^2$ and $m_N(0) := \max_{1 \leq i \leq N} ( Y_i(0) - \expesub{1-\pi}{Y_i(0)} )^2$, $\frac{1}{N_1} \dfrac{m_N(1)}{\varsub{\pi}{Y_i(1)}} \rightarrow 0$ and $\frac{1}{N_0} \dfrac{m_N(0)}{\varsub{1-\pi}{Y_i(0)}} \rightarrow 0$. 
\end{enumerate}
\end{asm}

\noindent Recall  $\pi_i (1-\pi_i)$ is the variance of the Bernoulli random variable $D_i$, so Assumption \ref{asm: reg conditions for CLT and var consistency}\ref{pi times one minus pi goes to infty} implies that the sum of the variances of the $D_i$ grows large. 
It also implies that both $N_1$ and $N_0$ go to infinity, since $\sum_{i=1}^N \pi_i (1-\pi_i) \leq \min\{\sum_i \pi_i, \sum_i (1-\pi_i)\} = \min\{N_1, N_0 \}$. 
Assumption \ref{asm: reg conditions for CLT and var consistency}\ref{lindeberg type condition} is similar to the condition for the Lindeberg central limit theorem, and imposes that the weighted finite-population variance of $\Ytilde_i$ is not dominated by a small number of observations.
Assumption \ref{asm: reg conditions for CLT and var consistency}\ref{asm for consistent variance estimation} bounds the influence that any single observation has on the $\pi$- and ($1-\pi$)-weighted variances of the potential outcomes. Under the conditions introduced above, we have the following finite-population central limit theorem and consistency result for the heteroskedasticity-robust variance estimator. 

\begin{prop}[CLT and Variance Consistency]\label{prop: clt for tauhat and var consistency}
\hfill
\begin{enumerate}
\item Under Assumptions \ref{asm: rejective assignment probability}, \ref{asm: reg conditions for CLT and var consistency}\ref{pi times one minus pi goes to infty} and \ref{asm: reg conditions for CLT and var consistency}\ref{lindeberg type condition},  
$\dfrac{\tauhat - \expeR{\tauhat}}{ \sqrt{\vRapprox{\tauhat}} } \xrightarrow{d} \normnot{0}{1}.$

\item Under Assumptions \ref{asm: rejective assignment probability}, \ref{asm: reg conditions for CLT and var consistency}\ref{pi times one minus pi goes to infty} and \ref{asm: reg conditions for CLT and var consistency}\ref{asm for consistent variance estimation}, 
$
\dfrac{\hat{s}^2}{  \expeRapprox{\hat{s}^2}  } \xrightarrow{p} 1.
$
\end{enumerate}
\end{prop}

These results allow us to formalize the conditions under which conventional confidence intervals of the form $\hat\tau \pm z_{1-\alpha/2} \cdot \hat{s}$ will be valid for $\tau_{EATT}$ (or $\tau_{ATE}$) when the finite population is large, where $z_{1-\alpha/2}$ is the $1-\alpha/2$ quantile of the standard normal distribution.

\begin{prop} \label{prop: local to zero coverage}
Suppose Assumptions \ref{asm: rejective assignment probability} and \ref{asm: reg conditions for CLT and var consistency}\ref{pi times one minus pi goes to infty}-\ref{asm for consistent variance estimation} hold, and that (i) $\frac{b}{ \sqrt{\vRapprox{\hat\tau} } } \rightarrow b^* \in \reals$, where $b = \frac{N}{N_1} \frac{N}{N_0} \covsub{1}{\pi_i}{Y_i(0)}$ is the bias of $\hat\tau$ for the EATT; and (ii) $\sqrt{ \dfrac{\vRapprox{\hat\tau}}{ \expeRapprox{\hat{s}^2} } } \rightarrow r \in (0,1]$.
Then, $\dfrac{\hat\tau - \tau_{EATT}}{ \hat{s} } \xrightarrow{d} \normnot{ b^* \cdot r }{ r^2 },$ 
and $\hat\tau \pm z_{1-\alpha/2} \cdot \hat{s}$ has asymptotic coverage for $\tau_{EATT}$ approaching 
\begin{equation}
\Phi\left( \frac{z_{1-\alpha/2}}{r} -  b^* \right) - \Phi\left( \frac{-z_{1-\alpha/2}}{r} -b^* \right). \label{eqn: asyptotic coverage}
\end{equation}
\noindent The analogous result holds for $\tau_{ATE}$, replacing $b$ with $\frac{N}{N_1} \covsub{1}{\pi_i}{Y_i(1)} + \frac{N}{N_0} \covsub{1}{\pi_i}{Y_i(0)}$.
\end{prop}

\noindent Condition (i) of Proposition \ref{prop: local to zero coverage} imposes that the sequence of finite populations is such that the bias of $\hat\tau$ is of the same order of magnitude as its standard deviation over the randomization distribution (i.e., local to zero). 
Condition (ii) of the proposition imposes that the conservativeness of the typical variance estimator stabilizes asymptotically (recall $\expeRapprox{\hat{s}^2} \geq \vRapprox{\hat{\tau}}$ by Proposition \ref{prop: shat conservative}). 

When $\hat\tau$ is unbiased, so that $b^* = 0$, Proposition \ref{prop: local to zero coverage} shows that confidence intervals based on the normal approximation will have correct but generally conservative coverage. 
Interestingly, it also implies that conventional confidence intervals will maintain correct coverage provided the bias of $\hat\tau$ is sufficiently small relative to the conservativeness of the variance estimator. 
For example, a sufficient condition to ensure at least 95\% coverage is that $|b^*| \leq z_{0.975} \cdot \left(\frac{1}{r}-1\right)$. 
Conventional confidence intervals can therefore accommodate some bias owing to the fact that heterogeneity in treatment probabilities $\pi_i$ or treatment effects $\tau_i$ typically induces conservativeness of the heteroskedasticity-robust variance estimator. In practice which effect dominates will be difficult to gauge, as neither the bias of the estimator nor the conservativeness of the variance are consistently estimable. Nevertheless, this conservativeness has implications for the interpretation of sensitivity analyses that account for the bias, as we discuss in the following section.

In Appendix \ref{section: berry-esseen}, we provide Berry-Esseen type bounds on the approximation quality of the CLT in any finite population of fixed size, applying a result by \citet{berger_rate_1998} for rejective sampling. This result establishes that the distribution of $\hat\tau$ will be approximately normally distributed in sufficiently large finite populations without appealing to a sequence of finite populations of increasing size. 

\subsection{Sensitivity Analyses based on the Difference in Means Estimator}\label{sec: DIM sensitivity analyis}

Our framework lends itself to sensitivity analyses based on the DIM.
While unobserved selection cannot be estimated from the data itself, researchers may place assumptions on the magnitude of selection bias (specifically the finite-population covariance between treatment probabilities $\pi_i$ and potential outcomes). 
Under such assumptions, identified sets for the EATT and the ATE can be obtained and researchers can conduct valid yet conservative inference on the now partially identified, finite-population causal estimands. Concretely, suppose we assume $\covsub{1}{\pi_i}{Y_i(0)}$ lies in the interval $[\underline{b}, \overline{b}]$.
Proposition \ref{prop: expectation of sdim} implies that $\tau_{EATT}$ lies in the interval $[\tau_{EATT}^{lb}, \tau_{EATT}^{ub}]$, where
\begin{equation}\label{eqn: interval for eatt}
    \tau_{EATT}^{lb} = \expeR{\hat{\tau}} - \frac{N}{N_0} \frac{N}{N_1} \overline{b} \mbox{ and } \tau_{EATT}^{ub} = \expeR{\hat{\tau}} - \frac{N}{N_0} \frac{N}{N_1} \underline{b}.
\end{equation}
Natural estimators plug-in the DIM $\hat{\tau}$ for $\expeR{\hat{\tau}}$ in (\ref{eqn: interval for eatt}), yielding unbiased estimates for the bounds $\hat{\tau}_{EATT}^{lb} = \hat{\tau} - \frac{N}{N_0} \frac{N}{N_1} \overline{b}$ and $\hat{\tau}_{EATT}^{ub} = \hat{\tau} -\frac{N}{N_0} \frac{N}{N_1} \underline{b}$. Bounds on the finite-population ATE could be obtained analogously if the researcher also places bounds on $\covsub{1}{\pi_i}{Y_i(1)}$.

By combining our analysis in Section \ref{sec: distribution of DIM} with existing results from the partial identification literature in econometrics, we can obtain valid yet typically conservative confidence intervals for the partially identified EATT. 
In particular, in a super-population setting, \cite{ImbensManski(04)} construct valid confidence intervals for partially identified parameters. Letting $\Delta = \frac{N}{N_0} \frac{N}{N_1} (\overline{b}-\underline{b})$ be the length of the identified set, the Imbens-Manski confidence for the EATT takes the form $[\hat\tau^{lb}_{EATT}- C \hat{s}, \hat\tau^{ub}_{EATT} + C \hat{s}]$, where the constant $C$ is chosen to solve $\Phi\left( \frac{\Delta}{\hat{s}} + C  \right) - \Phi(-C) = 1-\alpha$, for $\Phi(\cdot)$ the standard normal cumulative distribution function. 
In Appendix \ref{section: imbens-manski}, we show that this confidence interval has correct but potentially conservative coverage in our design-based framework.

Altogether, our results imply that researchers can report design-based sensitivity analyses directly based on the DIM and assumptions on the magnitude of selection bias. 
As we illustrate below, a natural statistic to report is the ``breakdown'' value of selection bias needed to overturn their causal conclusions---for example, how large must $\left| \covsub{1}{\pi_i}{Y_i(0)} \right|$ be in order for the Imbens-Manski interval to contain a null effect. \Copy{conservativeness-breakdown}{Since conventional standard errors are conservative for the standard deviation of $\hat\tau$ over the randomization distribution (see Proposition \ref{prop: shat conservative}), such a sensitivity analysis will be conservative about the robustness of causal conclusions. In particular, we show in Appendix \ref{section: imbens-manski} that $\liminf_{N \to \infty} P_R(\hat{b}^* \leq b^*) \geq 1-\alpha$, where $\hat{b}^*$ is the estimated breakdown value using the Imbens-Manski interval and $b^*$ is the true breakdown value for which the identified set includes zero.} 

\Copy{exploit-conservativeness}{An interesting question is whether the conservativeness of the typical variance estimator could be exploited to produce less conservative sensitivity analyses. In general, the conservativess of the variance estimator is not consistently estimable since it is a function of the unknown $\pi_i$ (see Proposition \ref{prop: shat conservative}). However, with some auxiliary assumptions one could potentially obtain a lower bound on the conservativeness of the variance. This strikes us an interesting avenue for future work.} 

\begin{rem}\label{rmk: rosenbaum} 
Such sensitivity analyses are related to, but different from existing finite population sensitivity analyses. 
\citet{Rosenbaum(87), Rosenbaum(02), Rosenbaum(05)} places bounds on the relative odds ratio of treatment between two units (i.e., $\frac{\pi_i ( 1- \pi_j)}{\pi_j (1 - \pi_i)}$ for $i \neq j$) and examines the extent to which the relative odds ratio must vary across units such that we no longer reject a particular sharp null of interest. \citet{AronowLee(2013)} and \citet{MiratrixWagerZubizaretta(18)} bound a finite-population mean under unequal-probability sampling under the assumption that the sampling probabilities are restricted to an interval.
Sensitivity analyses in our framework thus differ in two ways. First, we consider sensitivity of conclusions about a weak null hypothesis about an \emph{average} treatment effect, rather than a sharp null. Second, our approach only requires the researcher to restrict a finite-population covariance, rather than restricting individual-level treatment probabilities. An important implication is that researchers can calibrate such restrictions using the estimated covariance between the treatment and placebo outcomes, as we illustrate in our application to difference-in-differences (Section \ref{subsec: medicaid application}).
\end{rem}

\subsection{Implications for Difference-in-Differences Estimation}\label{sec: application to two period DiD}

Our analysis immediately applies to the classic two-period difference-in-differences estimator \citep[e.g.,][]{CardKrueger(94), bertrand_how_2004}, one of the most influential quasi-experimental estimators. (We show in Appendix \ref{section: multi-period DiD} that our discussion extends directly to non-staggered, difference-in-differences estimators with multiple time periods.) Suppose we observe aggregate outcomes $(Y_{it})$ of U.S. states over two periods $t \in \{1,2\}$.
Some states ($D_i = 1$) are treated beginning in period 2, whereas other states ($D_i=0$) are untreated in both periods. 
The observed outcome for state $i$ in period $t$ is $Y_{it} = D_i Y_{it}(1) + (1-D_i)Y_{it}(0)$. 
In this setting, the DIM estimator for the first-differenced outcome $Y_i := Y_{i2} - Y_{i1}$ is equivalent to the DID estimator between treated and control states, $\hat\tau_{DID} = \frac{1}{N_1} \sum_{i:D_i=1} (Y_{i2} - Y_{i1}) - \frac{1}{N_0} \sum_{i:D_i=0} (Y_{i2} - Y_{i1})$. Under the ``no-anticipation'' assumption that $Y_{i1}(0) = Y_{i1}(1)$, Proposition \ref{prop: expectation of sdim} implies
\begin{equation*}
\expeR{\tauhat_{DID}} = \underbrace{ \frac{1}{N_1} \sum_{i=1}^{N} \pi_i \tau_{i2} }_{\tau_{EATT,2}} + \frac{N}{N_1} \frac{N}{N_0} \covsub{1}{ \pi_i }{ Y_{i2}(0) - Y_{i1}(0) } ,\label{eqn: bias of DID}
\end{equation*}
\noindent where $\tau_{i2} = Y_{i2}(1) - Y_{i2}(0)$ is unit $i$'s treatment effect in period 2. 
The first term is the EATT in period 2. 
The second term is proportional to the finite-population covariance between individual treatment probabilities $\pi_i$ and trends in the untreated potential outcomes. 
Thus, in our framework, the DID estimator is unbiased for $\tau_{EATT,2}$ provided the treatment probabilities $\pi_i$ are uncorrelated in the finite-population with changes in potential outcomes $Y_{i2}(0) - Y_{i1}(0)$. 
This is a finite-population \textit{parallel trends} assumption since it is equivalent to the condition $\expeR{\frac{1}{N_1} \sum_i D_i (Y_{i2}(0) - Y_{i1}(0))} = \expeR{\frac{1}{N_0} \sum_i (1-D_i) (Y_{i2}(0) - Y_{i1}(0))}$.

Furthermore, in this setting, the variance estimator $\hat{s}^2$ is equivalent to the cluster-robust (at the unit level) variance estimator for $\hat{\tau}_{DID}$ from the panel OLS regression $Y_{it} = \alpha_i + \lambda_t + D_i \cdot 1[t=2] \tau_{DID} + \epsilon_{it}.$ Therefore, Proposition \ref{prop: shat conservative} implies that the cluster-robust variance estimator for $\hat{\tau}_{DID}$ is weakly conservative for the variance of the DID estimator over the randomization distribution, and will typically be strictly conservative if treatment probabilities differ across units. 
As a consequence, provided the finite-population parallel trends assumption holds, conventional confidence intervals of the form $\tauhat_{DID} \pm z_{1-\alpha} \cdot \hat{s}$ will be valid (but typically conservative) in our framework. Since empirical researchers are often unsure about the validity of the parallel trends assumption in practice, it will often be useful to conduct sensitivity analyses on conclusions about the EATT under possible violations of the finite-population parallel trends assumption using the approach described in Section \ref{sec: DIM sensitivity analyis} above. We provide an empirical example of this approach in Section \ref{subsec: medicaid application} below.

\vspace{-1em}
\section{Simulations and Application Using Real-World Data}

\subsection{Monte Carlo Simulations}
\label{subsec: qwi monte carlo sims}

We conduct Monte Carlo simulations based on the Quarterly Workforce Indicators (QWI) from the Longitudinal Household-Employer Dynamics (LEHD) Program at the U.S. Census \citep{qwi-census}, which provides aggregate statistics from linked employer-employee microdata covering over 95\% of all private sector jobs in the United States. The LEHD program writes, ``Because the estimates are not derived from a probability-based sample, no sampling error measures are applicable'' \citep{qwi-census}.
Our simulations therefore view uncertainty as arising from the stochastic realization of state-level policy changes.

\vspace{-1em}
\paragraph{Simulation design:} We use aggregate data on the 50 U.S. states and Washington D.C. from the QWI (indexed by $i=1,...,N$) for the first quarter of 2012 and 2016 (indexed by $t=1,2$).
For each state and year, we set the potential outcomes $Y_{it}(1)$ and $Y_{it}(0)$ equal to the state's observed outcome in the QWI ($Y_{it}$). 
Mimicking a two-period DID analysis, we simulate treatment by randomly generating placebo laws across states.
Our simulated treatments have no causal effect for any state, and so $\tau_{EATT,2} = \tau_{ATE,2} = 0$. 
The potential outcomes are held fixed throughout our simulations; the simulation draws differ in that each corresponds with a different realization of the generated placebo laws $D=(D_1,...,D_N)'$. 

Following Assumption \ref{asm: rejective assignment probability}, we draw $D_1, \hdots, D_{N}$ as independent Bernoulli random variables with (unconditional) state-level treatment probabilities $p_i$, discarding any draws where $\sum_i D_i \neq N_1$.
Based on state-level results from the 2016 presidential election \citep{MITElectionData}, the state-level unconditional treatment probabilities $p_i$ are chosen such that, for some $p^1 \in [0,1]$, states that voted for Clinton have $p_i = p^1$, and states that voted for Trump have $p_i = 1 - p^1$. 
When $p^1 =0.5$, all states have the same probability of adopting treatment, as in a completely randomized experiment, whereas when $p^1 > 0.5$, Democratic states are more likely to adopt the treatment. 
We report results as $p^1$ varies over $p^1 \in \{0.50, 0.75, 0.90\}$ and fix the number of treated and untreated states at $N_1=25$ and $N_0=26$, respectively.

For each draw of the assignment vector, we calculate the two-period DID estimator $\tauhat_{DID}$ and a nominal 95\% confidence interval $\tauhat_{DID} \pm z_{0.975} \cdot \hat{s}$, where $\hat{s}$ is the heteroskedasticity-robust standard error for the first-differenced outcome.
We also calculate a nominal \citet{ImbensManski(04)} 95\% confidence interval for the partially identified EATT under the assumption that $|\covsub{1}{\pi_i}{Y_{i2}(0)-Y_{i1}(0)}| \leq \tilde{b}$, as discussed in Section \ref{sec: DIM sensitivity analyis}.
We choose the bound $\tilde{b}$ corresponding to the actual bias of the estimator, $\tilde{b} = |\covsub{1}{\pi_i}{Y_{i2}(0)-Y_{i1}(0)}|$, to evaluate the properties of a robust confidence interval that properly accounts for the bias. We report results for two choices of the outcome $Y_{it}$: the log employment level and the log of state-level average monthly earnings for state $i$ in period $t$. 

\vspace{-1em}
\paragraph{Simulation results:} We first report the bias of the two-period DID estimator. 
While the placebo law has no treatment effect for any state, the change in untreated potential outcomes $Y_{i2}(0) - Y_{i1}(0)$ varies across states in a way that is related to state-level voting patterns in the 2016 presidential election. 
As a result, the design-based parallel trends assumption, $\covsub{1}{\pi_i}{Y_{i2}(0) - Y_{i1}(0)} = 0$, is violated when $p^1 \neq 0.5$, and hence the DID estimator is biased for the EATT over the randomization distribution in these simulations. 
The first row of Table \ref{tab: state level did, main text table} reports the normalized bias of the DID estimator (i.e., $\expesub{R}{\tauhat_{DID}}/\sqrt{\varsub{R}{\tauhat_{DID}}}$) as $p^1$ varies for both of these two outcomes. 
For $p^1 = 0.5$, the bias is zero up to simulation error.
The magnitude of the bias increases as we increase $p^1$, since the average value of $Y_{i2}(0) - Y_{i1}(0)$ differs between Democratic and Republican states for both of our outcomes. 
Appendix Figure \ref{fig: state level did distribution, qwi} plots the distribution of the DID estimator over the randomization distribution.
The distributions are approximately normally distributed, illustrating the finite-population CLT from Section \ref{sec: distribution of DIM}.

\begin{table}[htbp!]
    \centering
    \subfloat[Log employment]{\includegraphics[width=0.45\textwidth]{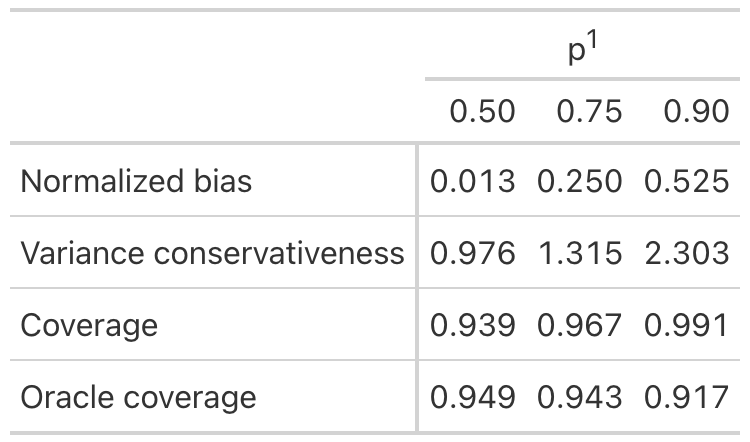}} \hspace{5mm}
    \subfloat[Log earnings]{\includegraphics[width=0.45\textwidth]{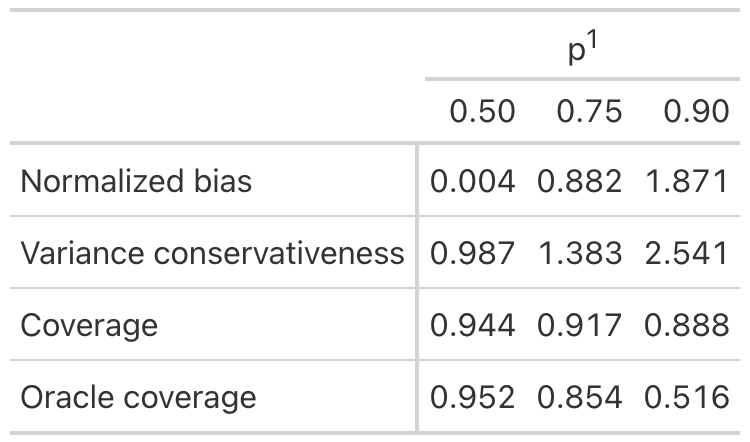}}
    \caption{Normalized bias, variance conservativeness, and coverage in Monte Carlo simulations.}
    \floatfoot{\textit{Notes}: 
    Row 1 reports the normalized bias of the DID estimator ($\expesub{R}{\tauhat_{DID}}/\sqrt{\varsub{R}{\tauhat_{DID}}}$) for the EATT over the randomization distribution.
    Row 2 reports the estimated ratio $\frac{\expesub{R}{\hat{s}^2}}{\varsub{R}{\tauhat_{DID}}}$ across simulations, which measures the conservativeness of the heteroskedasticity-robust variance estimator. 
    Row 3 reports the coverage rate of a nominal 95\% confidence interval of the form $\hat\tau_{DID} \pm z_{0.975} \, \hat{s}$.  Row 4 reports the coverage rate of an ``oracle'' 95\% confidence interval that uses the true variance rather than an estimated one, $\hat\tau_{DID} \pm z_{0.975} \, \sqrt{\vR{\tauhat_{DID}}}$.
    The columns report results as the treatment probability for Democratic states, $p^1$, varies over $\{0.5, 0.75, 0.9\}$.
    The results are computed over 5,000 simulations with $N_1 = 25$.
    }
    \label{tab: state level did, main text table}
\end{table}

\begin{table}[htbp!]
    \centering
    \subfloat[Log employment]{\includegraphics[width=0.45\textwidth]{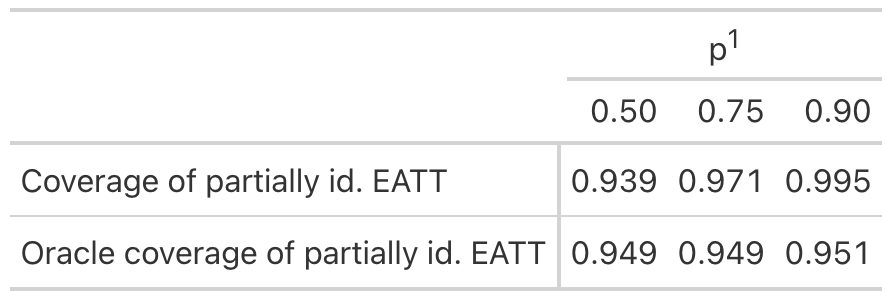}} \hspace{5mm}
    \subfloat[Log earnings]{\includegraphics[width=0.45\textwidth]{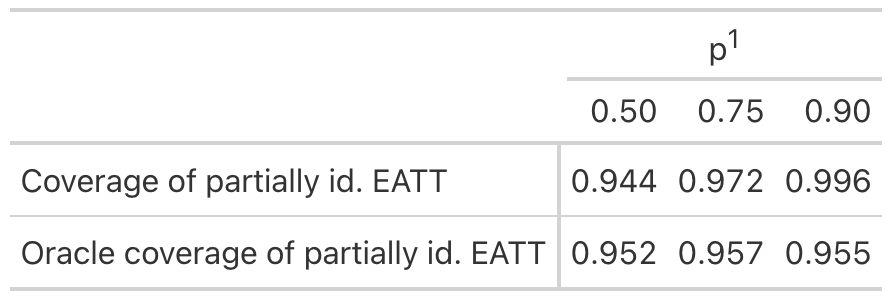}}
    \caption{Coverage for the partially identified EATT in Monte Carlo simulations.}
    \floatfoot{\textit{Notes}: 
    Row 1 reports the coverage rate of a 95\% confidence interval for the partially identified EATT based on the construction in \citet{ImbensManski(04)} (see Section \ref{sec: DIM sensitivity analyis} for details). 
    Row 2 reports the coverage rate of an ``oracle'' 95\% \citet{ImbensManski(04)} confidence interval that uses the true variance rather than an estimated one.
    The imposed upper bound $\tilde{b}$ on $|\covsub{1}{\pi_i}{Y_{i2}(0)-Y_{i1}(0)}|$ is correct in the sense that it is equal to the actual value of $|\covsub{1}{\pi_i}{Y_{i2}(0)-Y_{i1}(0)}|$ in our simulations.
    The columns report results as the treatment probability $p^1$ for Democratic states varies over $\{0.5, 0.75, 0.9\}$.
    When $p^1 = 0.5$, the upper bound $\tilde{b}$ equals zero, and the  \citet[][]{ImbensManski(04)} confidence interval is equivalent to a standard, nominal 95\% confidence interval.
    The results are computed over 5,000 simulations with $N_1 = 25$.
    }
    \label{tab: state level did, id set coverage, main text table}
\end{table}

The conservativeness of the usual heteroskedasticity-robust variance estimator is summarized in the second row of Table \ref{tab: state level did, main text table}, which shows the ratio of the average estimated variance for $\hat\tau$ to the actual variance of the estimator, $\frac{\expeR{\hat{s}^2}}{\varsub{R}{\tauhat}}$. In line with the results in Proposition \ref{prop: shat conservative} and Corollary \ref{cor: shat conservative under constant TEs}, $\hat{s}^2$ becomes conservative when there is variation in the treatment probabilities. 
For simulations with $p^1 = 0.5$, $\hat{s}^2$ is, on average, approximately equal to the true variance of the DID estimator. As $p^1$ increases, however, it becomes more conservative: in the most extreme case when $p^1 = 0.9$, the average estimated variance is approximately 2.5 times as large as the true variance. 
Since there is no treatment effect heterogeneity, this conservativeness is the result of heterogeneity in the $\pi_i$.

The third row of Table \ref{tab: state level did, main text table} reports the coverage of a standard 95\% confidence interval.
When $p^1 = 0.5$, the standard confidence intervals have approximately 95\% coverage for both outcomes. 
As we increase $p^1$, there is a tradeoff between the fact that the estimator is biased (which leads to lower coverage) and the fact that the variance estimator is conservative (which leads to higher coverage), as formalized in Proposition \ref{prop: local to zero coverage}. 
For the log earnings outcome, the bias dominates and coverage decreases in $p^1$---coverage of the EATT is only about 88.8\% when $p^1 = 0.9$. 
By contrast, for the state-level log average employment outcome, the bias is smaller, and so the conservativeness of the variance estimator dominates---the coverage rate is 99.1\% when $p^1 = 0.9$.
For comparison, the last row of Table \ref{tab: state level did, main text table} reports the coverage of an ``oracle'' 95\% confidence interval that uses the true variance of the DID estimator instead of the estimated variance $\hat{s}^2$. When $p^1 = 0.9$ for log-earnings, for example, coverage would be only 51.6\% using the oracle variance, but is 88.8\% using the conventional conservative variance estimator.

Finally, Table \ref{tab: state level did, id set coverage, main text table} highlights the implications of the heteroskedasticity-robust variance estimator's conservativeness for constructing robust confidence intervals for the partially identified EATT, as discussed in Section \ref{sec: DIM sensitivity analyis}. The Imbens-Manski CIs that account for the bias have coverage of at least 93.9\% in all specifications. 
As $p^1$ increases, coverage becomes more conservative---for the state-level log-average employment outcome, the coverage rate is 99.5\% when $p^1 = 0.9$. 
For comparison, we again report the coverage of an ``oracle'' 95\% confidence interval for the identified set that uses the true variance of the DID estimator, which remains approximately 95\% for both outcomes as $p^1$ varies. These results illustrate that robust confidence intervals that account for the bias provide a conservative estimate of how much bias can be accommodated to reach particular conclusions.

Appendix \ref{sec: appendix simulations} presents several extensions. 
We consider simulation designs that vary the number of treated units and finite population sizes. We also consider designs with treatment effect heterogeneity, which we find leads conventional confidence intervals to be even more conservative.

\subsection{Empirical Application: Effects of Medicaid Expansions}\label{subsec: medicaid application}

We return to the example of analyzing the impact of Medicaid expansions introduced in Section \ref{sec:model}.
\citet{wherry_early_2016} study the impact of state-level Medicaid expansions on statewide health insurance coverage using a two-period difference-in-differences estimator that compares the percentage of uninsured individuals ($Y_{it}$) in states that expanded Medicaid in 2014 ($D_i = 1$) against those that did not ($D_i = 0$). 
The authors estimate $\tauhat_{DID} = -7.1$ and report a 95\% CI of $[-11.1, -3.0]$, which implies a standard error of $\hat{s} \approx 2.09$ (see their Table 2). The authors indicate that the standard error is clustered at the state-level. To interpret this standard error from the traditional sampling perspective, we would have to imagine the 50 U.S. states as sampled from an infinite super-population of states. As discussed in Section \ref{sec:model}, it may be more natural to think of the 50 states as fixed, and the state-level treatment assignments as stochastic---e.g. owing to stochastic realizations of state political processes. 
Our framework implies that if the finite-population parallel trends assumption is satisfied, the CI of $[-11.1, -3.0]$ can alternatively be interpreted as a valid, but possibly conservative 95\% confidence for the EATT on the fraction of uninsured individuals. 

\begin{figure}[htbp!]
\includegraphics[width = 3.5in]{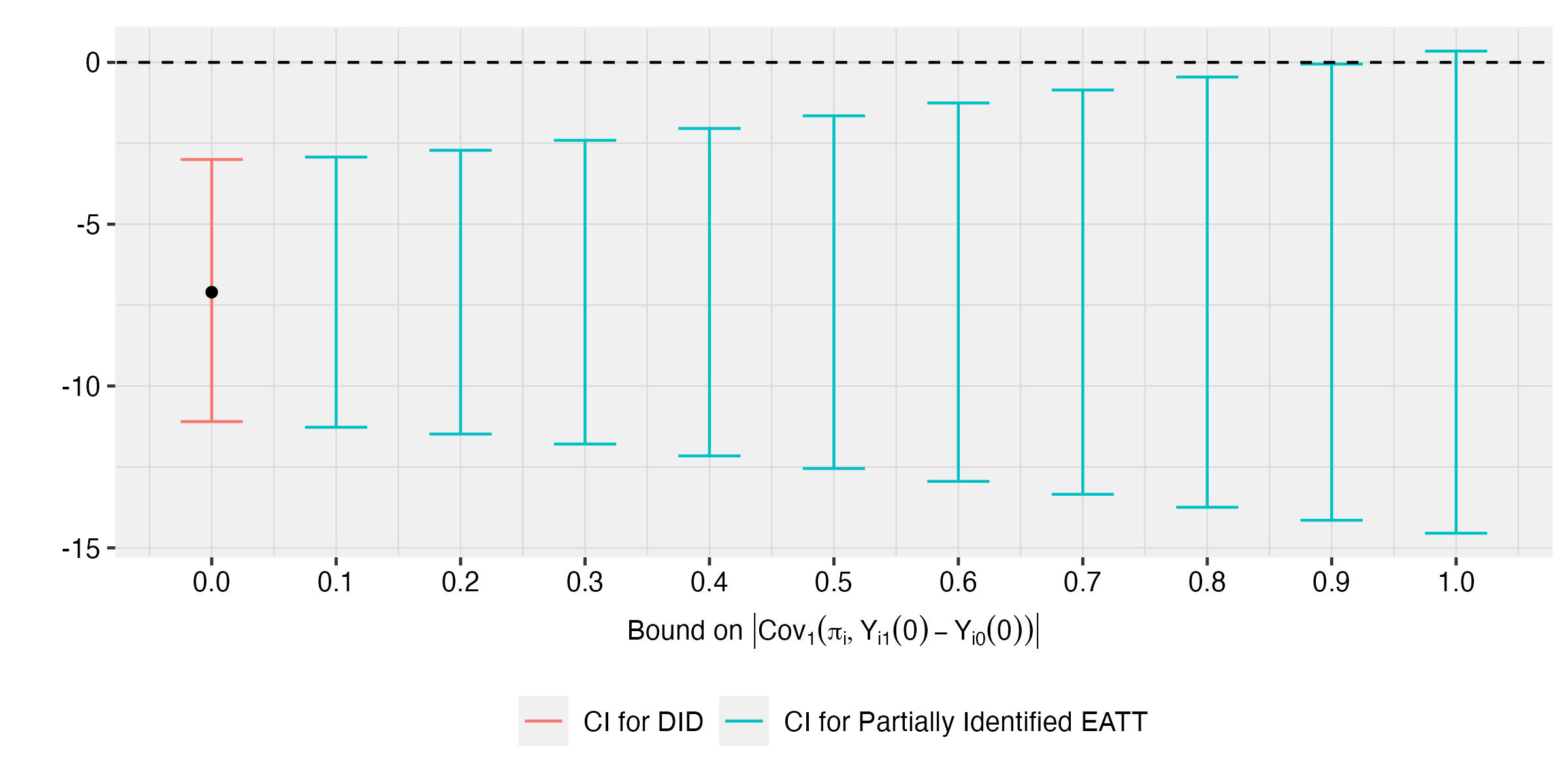}
\caption{Sensitivity analysis for the EATT based on \cite{wherry_early_2016}}
\floatfoot{\textit{Notes}: This figures plots the conventional confidence interval $\hat{\tau}_{DID} + z_{0.975} \, \hat{s}$ for $\tau_{EATT, 2}$ (red) and 95\% confidence intervals for the partially identified EATT under bounds on the magnitude of violations of the design-based parallel trends assumption of the form $\left| \covsub{1}{\pi_{i}}{Y_{i2}(0) - Y_{i1}(0)}\right|  \leq \tilde{b}$ (blue). We report results for $\tilde{b} \in \{0, 0.1, \hdots, 1\}$ and the confidence interval is constructed following \citet{ImbensManski(04)}.
The calculations are based on the estimates reported in Table 2 of \cite{wherry_early_2016}.
}
\label{figure: wherry miller sensitivity analysis}
\end{figure}

We may worry that the finite-population parallel trends assumption is violated---we would expect liberal-leaning states to have higher treatment probabilities than conservative states, and they may have different potential outcomes.
To address such concerns, we conduct a sensitivity analysis on the authors' conclusions about the EATT.
We calculate 95\% Imbens-Manski confidence intervals under the assumption that the covariance between the treatment probabilities $\pi_i$ and trends in potential outcomes $Y_{i2}(0) - Y_{i1}(0)$ is bounded in magnitude by a constant $\tilde{b}$, i.e. assuming $\left| \covsub{1}{\pi_{i}}{Y_{i2}(0) - Y_{i1}(0)}\right|  \leq \tilde{b}$. Figure \ref{figure: wherry miller sensitivity analysis} shows the resulting confidence intervals for different values of $\tilde{b}$. The ``breakdown'' value for concluding there is a significant negative effect is $\hat{b}^* \approx 0.9$, i.e. the robust CI excludes zero for all $\tilde{b} < 0.9$. As discussed in Section \ref{sec: DIM sensitivity analyis}, this is a conservative estimate of the true ``breakdown'' value $b^*$ for which the identified set includes 0. Similar to the analysis in \citet{rambachan_more_2023} from the super-population perspective, we can benchmark the magnitudes of $\tilde{b}$ using data from years prior to treatment. The authors' Appendix Table 6 suggests that the largest in magnitude finite-population covariance between treated probabilities and trends in untreated potential outcomes occurred between 2012-2013, with a point estimate of $-0.37$ (SE $0.48$); the magnitude of this estimate is well below the breakdown value of 0.9, although its 95\% confidence interval includes values larger in magnitude than the breakdown value.

\vspace{-1em}
\section{Extensions}\label{sec: extensions and applications}

In this section, we present several extensions that illustrate practical implications of our framework for empirical research. 
First, we consider the common setting where the researcher has data on individuals but treatment is assigned at a more aggregate level. We show that the cluster-robust variance estimator is valid but potentially conservative, justifying the popular heuristic to cluster at the level at which treatment is determined in quasi-experimental settings.
Second, we provide two sufficient conditions under which adjusting for differences in baseline covariates can address the bias of the DIM estimator.
Finally, we apply our framework to study instrumental variable (IV) estimators, showing conditions under which they have a causal interpretation and how sensitivity analyses can be conducted for violations of these assumptions.
Our analysis also extends directly to non-staggered difference-in-differences estimators with multiple time periods (see Appendix \ref{section: multi-period DiD}).

\subsection{When Should Researchers Adjust Standard Errors for Clustering?} \label{sec: clustering}

We consider the common setting where treatment is determined at a more aggregate level than the unit of observation. Specifically, each unit $i = 1, \hdots, N$ now belongs to one of $C$ clusters, where $c(i)$ denotes the cluster membership of unit $i$. We assume treatment is determined at the cluster level. For example, units $i$ may be individuals living in states $c(i)$, and policy is determined at the state level.

\begin{asm}[Clustered treatment assignment]\label{asm: clustered rejective assignment}
The cluster-level treatment vector, $D := (D_1, \hdots, D_C)^\prime$, satisfies $\prob{D = d \mid \sum_c D_c = C_1, W, Y(\cdot)} \propto \prod_c p_c^{d_c} (1 - p_c)^{1-d_c}$ for all $d \in \{0, 1\}^{C}$ such that $\sum_{c} d_c = C_1$, and zero otherwise.
\end{asm}
\noindent Assumption \ref{asm: clustered rejective assignment} is the cluster-level analog to the assignment mechanism considered throughout the paper (Assumption \ref{asm: rejective assignment probability}). Mirroring our earlier notation, let $C_1 := \sum_{c} D_c$ and $C_0 := \sum_{c} (1 - D_c)$ denote the number of treated and untreated clusters respectively, $\pi_c := \probsub{R}{D_c = 1}$ denote the marginal treatment probability for cluster $c$ under Assumption \ref{asm: clustered rejective assignment}, and $D_i = D_{c(i)}$ denote unit $i$'s treatment assignment. As before, we analyze the behavior of the DIM estimator $\tauhat$ constructed using the outcomes and treatment at the individual level, except we now consider the randomization distribution generated by the clustered treatment assignments. Since the regularity conditions are natural extensions of those in Section \ref{sec: distribution of DIM} to the clustered design, we defer them to Appendix \ref{section: appendix extension to OLS} and summarize the key takeaways here. Proposition \ref{prop: sdim limiting dist, clustering} in the Appendix provides conditions under which $\tauhat$ converges in probability to $\tau_{EATT}^{cluster} + \delta_{cluster}$, where $\tau_{EATT}^{cluster} = \expesub{\pi_{c(i)}}{\tau_i}$ and $\delta_{cluster} = \frac{N}{N - \sum_{i} \pi_{c(i)} } \frac{N}{\sum_i \pi_{c(i)} } \covsub{1}{\pi_{c(i)}}{Y_i(0)}$. The first term, $\tau_{EATT}^{cluster}$, is analogous to the EATT discussed earlier, except it uses the cluster-level treatment probabilities $\pi_{c(i)}$ instead of the individual-level probabilities $\pi_i$. Likewise, the bias $\delta_{cluster}$ is proportional to the finite-population covariance between the cluster-level treatment probabilities $\pi_{c(i)}$ and the potential outcome $Y_i(0)$. 
Proposition \ref{prop: sdim limiting dist, clustering} also shows that $\sqrt{C} (\tauhat - \tau^{cluster}_{EATT} - \delta_{cluster})$ converges to a Gaussian distribution, and the \citet{LiangZeger(86)} cluster-robust variance estimator is consistent for an upper bound on this variance. By contrast, Proposition \ref{prop: EHW invalid for clustered, regression} shows that the heteroskedasticity-robust variance estimator that ignores clustering can be either too large or small, and thus CIs based on this standard error may not have correct coverage even if $\delta_{cluster} = 0$. Taken together, these results imply that if the need for clustering in quasi-experimental settings arises from the stochastic assignment of treatment, then the researcher should cluster at the level at which treatment is assigned.

\begin{rem}
\citet{AbadieEtAl(22)} study a two-step data-generating process in which cluster-level treatment probabilities are initially drawn according to some fixed distribution that is unrelated to potential outcomes. 
Each cluster therefore has the same treatment probability marginalized over the two-step process, and hence the ATE is consistently estimable in their framework. Their results are thus not directly applicable to inference in quasi-experimental settings where treatment probabilities may systematically differ across clusters in ways potentially related to the potential outcomes, and the target parameter may be the EATT rather than ATE. Nevertheless, a similar heuristic applies in both contexts, which is to cluster at the level at which treatment is (independently) determined. Likewise, \cite{SuDing(21)-Cluster} studies clustered assignment mechanisms in which treatments are completely randomized across clusters, and so their calculations are not directly applicable to the quasi-experimental settings we study.
Finally, \citet{Xu_2021} studies clustered standard errors for non-linear estimators from a design-based perspective (although the technical set-up differs somewhat since they do not condition on $C_1$). 
Their results cover inference on a finite-population argmin that is well-defined if units have varying treatment probabilities, although existing results giving a causal interpretation to this parameter require the propensity score to be linear in observable covariates.
\end{rem}

\subsection{When Can Covariate Adjustment Recover Causal Estimands?} \label{sec: ols adjustment}

Suppose each unit $i$ is associated with fixed covariates $W_i \in \mathbb{R}^{k}$, and consider the OLS regression of the observed outcome on a constant, the treatment $D_i$, and the covariates $W_i$. 
This is the ``covariate-adjusted'' DIM studied by \citet{Freedman(2008)-regadj_to_experimental_data} and \citet{Lin(13)}, among others, in the context of completely randomized experiments.
We provide two characterizations of the estimand associated with the OLS coefficient on $D_i$ in our framework.

\begin{prop}\label{prop: interpretation of ols w covariates}
Suppose Assumption \ref{asm: rejective assignment probability} holds.
Let $\beta_{D}$ denote the coefficient on $D_i$ in the best linear projection of $Y_i$ on $(1, D_i, X_i^\prime)^\prime$ over the randomization distribution (see the proof of the proposition for a mathematical definition).
Then, assuming $\expeR{\frac{1}{N} \sum_{i=1}^{N} (1, D_i, W_i^\prime)^\prime (1, D_i, W_i^\prime)}$ is invertible,
\begin{enumerate}[label={(\roman*)}]
\item\label{outcome decomp for ols w. covariate} $\beta_D = \tau_{EATT} + \frac{N}{N_1} \frac{N}{N_0} \covsub{1}{\pi_i}{Y_i(0) - \gamma' W_i}$ for coefficients $\gamma$ defined in the proof.

\item\label{prob decomp for ols w. covariate} $\beta_D = \tau_{OLS} + \expesub{1}{\pi_i (1 - \hat{\pi}_i)}^{-1} \covsub{1}{\pi_i - \hat\pi_i}{ Y_i(0)}$ for $\hat{\pi}_i$ the best linear prediction of $\pi_i$ given a constant and $W_i$, and $\tau_{OLS} = \mathbb{E}_{1}[\pi_i (1 - \hat{\pi}_i)]^{-1} \mathbb{E}_{1}[\pi_i (1-\hat{\pi}_i) \tau_i]$.
\end{enumerate}
\end{prop}

Proposition \ref{prop: interpretation of ols w covariates} gives two decompositions of the OLS estimand $\beta_D$, the first involving an adjusted outcome and the second involving an adjusted treatment probability. Specifically, part \ref{outcome decomp for ols w. covariate} decomposes the covariate-adjusted DIM into the EATT plus a bias term that depends on the finite-population covariance between the treatment probabilities $\pi_i$ and the covariate-adjusted untreated potential outcomes, $Y_i(0) - \gamma^\prime W_i$, where the coefficient $\gamma$ is a weighted average of the projections of each of the potential outcomes onto the covariates. Thus $\beta_D$ corresponds to the EATT if the treatment probabilities $\pi_i$ are orthogonal to the adjusted potential outcomes. Similar to Section \ref{sec: DIM sensitivity analyis}, one could also conduct sensitivity analyses for the bias based on conjectured values for the covariance between $\pi_i$ and the \emph{adjusted} potential outcomes. Part \ref{prob decomp for ols w. covariate} alternatively decomposes the covariate-adjusted DIM into $\tau_{OLS}$, which is a particular weighted average of unit-specific treatment effects, and a bias that depends on the finite-population covariance between $Y_i(0)$ and the residualized treatment probability, $\pi_i - \hat\pi_i$.
The covariate adjusted DIM estimand thus recovers a weighted average of treatment effects whenever the finite-population covariance between the untreated potential outcomes and the residualized treatment probabilities is equal to zero (note that some of the weights could be negative if $\hat\pi_i > 1$ for some $i$). 
If the $\pi_i$ are linear in the covariates, then $\hat\pi_i =\pi_i$ and this bias equals zero.
Part \ref{prob decomp for ols w. covariate} thus nests the known result that when the propensity score is linear, the covariate-adjusted DIM gives a variance-weighted average of treatment effects; see \citet{angrist_estimating_1998} and \citet{abadie_sampling-based_2020} for similar results in a super-population and design-based setting, respectively. Our more general results, however, provide a causal interpretation to the covariate-adjusted DIM if the propensity is not linear in covariates but satisfies the orthogonality conditions described above. Our results also allow us to understand the biases that will result if the propensity score is mis-specified in a way that is related to the potential outcomes.

In Appendix \ref{section: appendix extension to OLS}, we provide regularity conditions under which $\sqrt{N}(\hat\beta_D - \beta_D)$ is asymptotically normally distributed, and show that the typical heteroskedasticity-robust standard errors are consistent for an upper bound on the asymptotic variance. Typical standard errors will thus yield conservative inference on $\beta_D$, and sensitivity analyses for the inference that account for the bias will typically be conservative.

\subsection{Instrumental Variables}\label{sec: iv}
In many settings, the researcher has access to an instrumental variable $Z_i$. In some cases, such as a randomized trial with imperfect compliance, the instrument $Z_i$ is completely randomly assigned. 
However, in other settings the instrument is not explicitly randomized, but the researcher may argue that it is at least partially determined by quasi-experimental factors. 
For example, in studying the effects of childbearing, \cite{angrist_children_1998, Angrist_Lavy_2010} consider having twins at a woman's second birth as an instrument for whether the woman has a third child. The birth of twins $Z_i = 1$ depends on the realization of random biological processes, such as whether a fertilized eggs splits, yet different individuals may have different probabilities of realizing $Z_i = 1$ due to genetic factors, age, or other health risks. 
Our results can be used to interpret and assess the sensitivity of IV estimates when the instrument may not be completely randomly assigned. 

Let $Z_i \in \{0, 1\}$ be a binary instrument, $D_i(z) \in \{0, 1\}$ be the potential treatment status for $z \in \{0, 1\}$, and $Y_i(d)$ be the potential outcome for $d \in \{0, 1\}$. The notation $Y_i(d)$ encodes the exclusion restriction that $Y$ depends on $Z$ only through $d$. We further impose the monotonicity assumption that $D_i(1) \geq D_i(0)$ for all units $i = 1, \hdots, N$.
The observed data is then $(Y_i, D_i, Z_i)$, where $Y_i = Y_i(D_i(Z_i))$ and $D_i = D_i(Z_i)$. We view the instrument as stochastic, holding fixed the potential treatments $D(\cdot) = \{D_i(\cdot) \colon i = 1, \hdots, N\}$ and potential outcomes $Y(\cdot) = \{Y_i(\cdot) \colon i = 1, \hdots, N\}$. We let $N_1^Z$ be the number of units with $Z_i = 1$ and $N_0^Z$ be the number of units with $Z_i = 0$.

\begin{asm}\label{asm: rejective assignment for instrument}
The instrument, $Z := (Z_1, \hdots, Z_N)^\prime$, satisfies $\prob{Z = z \Big| \sum_{i} Z_i = N_1^Z, W, D(\cdot), Y(\cdot)} \propto \prod_i p_i^{z_i} (1 - p_i)^{1-z_i}$ for all $Z \in \{ 0,1\}^N$ such that $\sum_i z_i = N_1^Z$, and zero otherwise.  
\end{asm}

\noindent We write $\probsub{R}{\cdot}$, $\expeR{\cdot}$, $\vR{\cdot}$ as probabilities, expectations, and variances respectively under Assumption \ref{asm: rejective assignment for instrument} and define $\pi_i^{Z} := \probsub{R}{Z_i = 1}$ to be the marginal probability that $Z_i=1$. Similar to Assumption \ref{asm: rejective assignment probability} for the treatment in earlier sections, Assumption \ref{asm: rejective assignment for instrument} models the instrument assignment as a random experiment with \emph{unequal} probabilities. In the example of the twin birth instrument, the stochastic instrument assignment corresponds to the realization of the biological process that determines whether a fertilized egg splits in two. The model, however, allows for different women to have different probabilities of having an egg split in two, owing to different biological risk factors, in ways that may be related to their potential outcomes. By contrast, existing IV frameworks typically assume the instrument to be fully independent of the potential treatments and outcomes (see, e.g., \citet{AngristImbens(94), angrist_identification_1996} for a sampling-based setting, and \citet{kang_inference_2018, hong_inference_2020} for a design-based setting). Our framework will thus allow us to assess the interpretation of the IV estimand in settings where the instrument may not be completely randomly assigned.

We analyze the popular two-stage least-squares (2SLS) estimator, $\betahat_{2SLS} := \tauhat_{RF}/\tauhat_{FS}$, with 
\begin{equation*}
    \tauhat_{RF} = \frac{1}{N_1^Z} \sum_i Z_i Y_i - \frac{1}{N_0^Z} \sum_i (1 - Z_i) Y_i \mbox{ and } \tauhat_{FS} := \frac{1}{N_1^Z} \sum_i Z_i D_i - \frac{1}{N_0^Z} \sum_i (1 - Z_i) D_i
\end{equation*}
\noindent corresponding to the reduced form and first-stage, respectively. Proposition \ref{prop: expectation of sdim} and the monotonicity assumption imply that
\begin{align*}
&\expeR{\tauhat_{RF}} = \frac{1}{N_1^Z} \sum_{i \in \mathcal{C}} \pi_i^Z (Y_i(1) - Y_i(0)) + \frac{N}{N_1^Z} \frac{N}{N_0^Z} \covsub{1}{\pi_i^Z}{Y_i(D_i(0))}\\
&\expeR{\tauhat_{FS}} = \frac{1}{N_1^Z} \sum_{i \in \mathcal{C}} \pi_i^Z + \frac{N}{N_1^Z} \frac{N}{N_0^Z} \covsub{1}{\pi_i^Z}{D_i(0)},
\end{align*}
where $\mathcal{C} := \{i \colon D_i(1) > D_i(0)\}$ is the set of complier units. We define the 2SLS estimand as $\beta_{2SLS} := \frac{\expeR{\tauhat_{RF}}}{\expeR{\tauhat_{FS}}}$. In Appendix \ref{section: appendix extension to multiple outcomes}, we show that under conditions similar to those in Section \ref{sec: distribution of DIM}, $\sqrt{N}(\hat\beta_{2SLS} - \beta_{2SLS})$ converges to a Gaussian distribution, and the usual delta-method standard errors for 2SLS are consistent for an upper bound on this variance. (Note we impose ``strong instrument'' asymptotics where the first-stage is strong relative to sampling variation.) What is the causal interpretation of the estimand $\beta_{2SLS}$? 
If $\pi_i^Z \equiv \frac{N_1^Z}{N}$, so that all units receive $Z_i=1$ with equal probability, then $\beta_{2SLS} = \frac{1}{|\mathcal{C}|} \sum_{i \in \mathcal{C}} (Y_i(1)-Y_i(0))$, which is a design-based local average treatment effect (LATE) \citep{angrist_identification_1996, kang_inference_2018}. 
Our results imply that $\beta_{2SLS}$ maintains a causal interpretation under the weaker orthogonality restriction $\covsub{1}{\pi_i^Z}{ Y_i(D_i(0))} = \covsub{1}{\pi_i^Z}{ D_i(0)} = 0$. 
In this case, $\beta_{2SLS}$ is a weighted average treatment effect among the compliers,
\begin{align*}
\beta_{2SLS} = \frac{1}{\sum_{i\in\mathcal{C}} \pi_i^Z } \sum_{i \in \mathcal{C}} \pi_i^Z \left( Y_i(1) - Y_i(0) \right) \equiv LATE_{\pi^z},
\end{align*}
\noindent where the weights are proportional to $\pi_i^Z$, the probability that $Z_i = 1$ under Assumption \ref{asm: rejective assignment for instrument}. 

Researchers can conduct simple, design-based sensitivity analyses on the two-stage least-squares estimator by placing restrictions on the finite-population covariance between the instrument probabilities and the potential outcomes and treatments to obtain an identified set for the weighted average treatment effect among the compliers. Specifically, assuming $\covsub{1}{\pi_{i}^{Z}}{Y_i(D_i(0))} \in [b_{RF}^{lb}, b_{RF}^{ub}]$ and $\covsub{1}{\pi_i^Z}{D_i(0)} \in [b_{FS}^{lb}, b_{FS}^{ub}]$, our decompositions of the expectations of $\hat{\tau}_{RF}$, $\hat{\tau}_{FS}$ imply the bounds 
\begin{align*}
    & \expeR{\hat{\tau}_{RF}} - \frac{N}{N_1^Z} \frac{N}{N_0^Z} b_{RF}^{ub} \leq \frac{1}{N_1^Z} \sum_{i \in \mathcal{C}} \pi_i^Z \left( Y_i(1) - Y_i(0) \right) \leq \expeR{\hat{\tau}_{RF}} - \frac{N}{N_1^Z} \frac{N}{N_0^Z} b_{RF}^{lb}, \\
    & \expeR{\hat{\tau}_{FS}} - \frac{N}{N_1^Z} \frac{N}{N_0^Z} b_{FS}^{ub} \leq \frac{1}{N_1^Z} \sum_{i \in \mathcal{C}} \pi_i^Z \leq \expeR{\hat{\tau}_{FS}} - \frac{N}{N_1^Z} \frac{N}{N_0^Z} b_{FS}^{lb}.
\end{align*}
Provided the lower bound on $\frac{1}{N_1^Z} \sum_{i \in \mathcal{C}} \pi_i^Z$ is strictly positive, $LATE_{\pi^z}$ must lie in the interval 
$$
    \left[ \frac{\expeR{\hat{\tau}_{RF}} - \frac{N}{N_1^Z} \frac{N}{N_0^Z} b_{RF}^{ub}}{\expeR{\hat{\tau}_{FS}} - \frac{N}{N_1^Z} \frac{N}{N_0^Z} b_{FS}^{lb}}, \frac{\expeR{\hat{\tau}_{RF}} - \frac{N}{N_1^Z} \frac{N}{N_0^Z} b_{RF}^{lb}}{\expeR{\hat{\tau}_{FS}} - \frac{N}{N_1^Z} \frac{N}{N_0^Z} b_{FS}^{ub}} \right].
$$
It is straightforward to estimate these bounds by plugging in $\hat{\tau}_{RF}, \hat{\tau}_{FS}$ in place of the expectations. This will yield consistent estimates of the bounds (under appropriate regularity conditions) in large populations. Likewise, we can further conduct (typically conservative) inference on the bounds based on conventional delta-method standard errors, and we can construct (typically conservative) confidence intervals for $LATE_{\pi^z}$ as in Section \ref{sec: DIM sensitivity analyis}. 

\section{Conclusion}

This paper develops a design-based framework for analyzing quasi-experimental settings in the social sciences in which uncertainty arises from stochastic realizations of treatment assignment, holding fixed the population and their potential outcomes.  
This perspective is natural in settings where the researcher does not wish to model the statistical process governing the sampling or formation of potential outcomes and the researcher describes the variation being used as the result of quasi-experimental factors that influence treatment status. We derive conditions under which conventional estimators and CIs are valid for interpretable causal parameters in this framework and characterize the bias and size-distortions that arise when these conditions are violated. This leads to natural forms of sensitivity analysis. 
Altogether, we show that the design-based perspective can also be coherently applied in quasi-experimental settings where there is concern about selection into treatment. 
While our framework views only treatment assignment as stochastic, an interesting direction for future research could be to study quasi-experimental settings under a finite-population data-generating process that adopts a statistical model for both the outcome and treatment assignments.

\medskip

\noindent \textbf{Disclosure statement}: The authors report there are no competing interests to declare.

\singlespacing
\bibliography{Bibliography.bib}


\singlespacing
\appendix
\newpage
\clearpage

\begin{center}
\vspace{-1.8cm}{\LARGE \textbf{Design-Based Uncertainty for Quasi-Experiments}} \medskip \\ \Large Online Appendix \bigskip \\
\large Ashesh Rambachan \hspace{0.3cm} Jonathan Roth \medskip\\
\today
\bigskip
\end{center}

\section{Proofs for Results in Main Text}\label{section: proofs for main text}

\paragraph{Proof of Proposition \ref{prop: expectation of sdim}}
\begin{proof}
Recall $\expeR{D_i} = \pi_i$ and $\tau_i = Y_i(1) - Y_i(0)$.
Hence, we have that
\begin{align*}
\expeR{\tauhat} &= \expeR{ \frac{1}{N_1} \sum_i D_i Y_i(1) - \frac{1}{N_0} \sum_i (1-D_i) Y_i(0) } \\
&= \frac{1}{N_1} \sum_i \pi_i \underbrace{(Y_i(0) + \tau_i)}_{=Y_i(1)} - \frac{1}{N_0} \sum_i (1-\pi_i) Y_i(0) \\
&= \underbrace{\frac{1}{N_1} \sum_i \pi_i \tau_i }_{=:\tau_{EATT}} + \frac{N}{N_0} \frac{N}{N_1} \underbrace{ \left(  \frac{1}{N}\sum_i \left( \pi_i - \frac{N_1}{N}  \right) Y_i(0) \right) }_{= \covsub{1}{\pi_i}{Y_i(0)}}, \numberthis \label{eqn: expression w eatt - proof}
\end{align*}
\noindent which yields the second expression in the Proposition. To derive the first expression, note that
$$\tau_{EATT} = \frac{1}{N_1}\sum_i (\pi_i - \frac{N_1}{N}) \tau_i + \frac{1}{N} \sum_i \tau_i = \frac{N}{N_1} \covsub{1}{\pi_i}{\tau_i} + \tau_{ATE}.$$ 
Further, since $\tau_i = Y_i(1)-Y_i(0)$, we have that $\covsub{1}{\pi_i}{\tau_i} = \covsub{1}{\pi_i}{Y_i(1)} - \covsub{1}{\pi_i}{Y_i(0)}$, and hence
$$\tau_{EATT} = \tau_{ATE} + \frac{N}{N_1} \covsub{1}{\pi_i}{Y_i(1)} - \frac{N}{N_1} \covsub{1}{\pi_i}{Y_i(0)}.$$ 
Substituting this expression into (\ref{eqn: expression w eatt - proof}) and simplifying then yields 
$$\expeR{\tauhat} = \tau_{ATE} + \frac{N}{N_1} \covsub{1}{\pi_i}{Y_i(1)} + \frac{N}{N_0} \covsub{1}{\pi_i}{Y_i(0)},$$ 
as needed.
\end{proof}

\paragraph{Proof of Lemma \ref{lem: variance of tauhat}}
\begin{proof}
Since $\tauhat$ can be represented as a Horvitz-Thompson estimator under rejective sampling, Theorem 6.1 in \citet{hajek_asymptotic_1964} implies
\begin{equation}\vR{\tauhat} [1+o(1)] = \left[ \sum_{k=1}^{N} \pi_k (1-\pi_k) \right] \varsub{\tilde{\pi}}{\tilde{Y}_i} = \left[ \sum_{k=1}^{N} \pi_k (1-\pi_k) \right] \varsub{\tilde{\pi}}{\frac{1}{N_1} Y_i(1) + \frac{1}{N_0} Y_i(0)}.\label{eqn:var tauhat using ytilde}\end{equation}
Standard decomposition arguments for completely randomized experiments (e.g. \citet{imbens_causal_2015}), modified to replace unweighted variances with weighted variances, yield
$$\varsub{\tilde{\pi}}{\frac{1}{N_1} Y_i(1) + \frac{1}{N_0} Y_i(0)} = \frac{N}{N_1 N_0} \left(  \frac{1}{N_1} \varsub{\tilde{\pi}}{Y_i(1)} + \frac{1}{N_0}\varsub{\tilde{\pi}}{Y_i(0)} - \frac{1}{N} \varsub{\tilde{\pi}}{\tau_i} \right), $$ 
which together with the previous display yields the desired result.
\end{proof}

\paragraph{Proof of Lemma \ref{lem: expectation of Neyman variance}}
\begin{proof}
We will show that $\expeR{\hat{s}_1^2}(1+o(1)) = \varsub{\pi}{Y_i(1)}$. The equality $\expeR{\hat{s}_0^2}(1+o(1)) = \varsub{1-\pi}{Y_i(0)}$ can be obtained analogously, from which the result is immediate. Observe that 
\begin{align*}
\expeR{\hat{s}_1^2} &=\expeR{\frac{1}{N_1} \sum_i D_i Y_i^2 - \bar{Y}_1^2 } = \expeR{\frac{1}{N_1} \sum_i D_i Y_i^2 - (\bar{Y}_1 - \expesub{\pi}{Y_i(1)} + \expesub{\pi}{Y_i(1)})^2} \\
&= \expeR{\frac{1}{N_1} \sum_i D_i Y_i^2} - \expesub{\pi}{Y_i(1)}^2 - 2 \expesub{\pi}{Y_i(1)} \expeR{\bar{Y}_1 - \expesub{\pi}{Y_i(1)}} - \expeR{(\bar{Y}_1 - \expesub{\pi}{Y_i(1)})^2} \\
&= \varsub{\pi}{Y_i(1)} - \vR{\bar{Y}_1},
\end{align*}
\noindent where the last equality is obtained using the fact that $\expeR{D_i} = \pi_i$, and hence $\expeR{\frac{1}{N_1} \sum_i D_i Y_i^2} = \expesub{\pi}{Y_i(1)^2}$ and $\expeR{\bar{Y}_1 - \expesub{\pi}{Y_i(1)}} = 0 $. Applying Theorem 6.1 in \citet{hajek_asymptotic_1964} as in the proof to Lemma \ref{lem: variance of tauhat}, we see that
$$\vR{\bar{Y}_1}(1+o(1)) = \left[\sum_k \pi_k (1-\pi_k) \right] \varsub{\tilde{\pi}}{Y_i(1)/N_1} .$$
\noindent Next, observe that
\begin{align*}
\left[\sum_k \pi_k (1-\pi_k) \right] \varsub{\tilde{\pi}}{Y_i(1)/N_1}&= \frac{1}{N_1^2} \sum_i \pi_i (1-\pi_i) (Y_i(1) - \expesub{\tilde{\pi}}{Y_i(1)})^2 \\
&\leq \frac{1}{N_1^2} \sum_i \pi_i (1-\pi_i) (Y_i(1) - \expesub{\pi}{Y_i(1)})^2 \\
&\leq \frac{1}{N_1^2} \sum_i \pi_i (Y_i(1) - \expesub{\pi}{Y_i(1)})^2 = \frac{1}{N_1} \varsub{\pi}{Y_i(1)} \\
&\leq \left[\sum_k \pi_k (1-\pi_k) \right]^{-1} \varsub{\pi}{Y_i(1)} = o(1) \varsub{\pi}{Y_i(1)}
\end{align*} 
\noindent where the first inequality uses the fact that $\expesub{\tilde{\pi}}{Y_i(1)} = \argmin_u \sum_i \pi_i (1-\pi_i) (Y_i(1) - u)^2$, the second inequality uses the fact that $\pi_i (1-\pi_i) \leq \pi_i$, and the third inequality uses the fact that $N_1 = \sum_i \pi_i \geq \sum_i \pi_i (1-\pi_i)$. Combining the previous three displays, we see that $\expeR{\hat{s}_1^2} = (1+o(1)) \varsub{\pi}{Y_i(1)},$ as we wished to show.
\end{proof}

\paragraph{Proof of Proposition \ref{prop: shat conservative}}
\begin{proof}
From (\ref{eqn:var tauhat using ytilde}), we have that 
$$\vRapprox{\tauhat} = \sum_{i=1}^N \pi_i (1-\pi_i) \left( \frac{1}{N_1} Y_i(1) + \frac{1}{N_0} Y_i(0) - \expesub{\tilde{\pi}}{\frac{1}{N_1} Y_i(1) + \frac{1}{N_0} Y_i(0)} \right)^2.$$
\noindent Since for any $X_i$ and constant $c$, we have that $\expesub{\tilde{\pi}}{(X_i-c)^2} = \expesub{\tilde{\pi}}{(X_i-\expesub{\tilde{\pi}}{ X_i })^2} + (\expesub{\tilde{\pi}}{ X_i } - c)^2$, it follows that
\begin{align*}
\vRapprox{\tauhat} &= \sum_{i=1}^N \pi_i (1-\pi_i) \left( \frac{1}{N_1} Y_i(1) + \frac{1}{N_0} Y_i(0) - \left(\expesub{\pi}{\frac{1}{N_1} Y_i(1)} + \expesub{1-\pi}{\frac{1}{N_0} Y_i(0)}\right) \right)^2  \\ &- \left(\sum_i \pi_i (1-\pi_i) \right) \cdot \left(\expesub{\pi}{\frac{1}{N_1} Y_i(1)} + \expesub{1-\pi}{\frac{1}{N_0} Y_i(0)} - \expesub{\tilde{\pi}}{\frac{1}{N_1} Y_i(1) + \frac{1}{N_0} Y_i(0)} \right)^2 . \label{eqn: intermediate upper bound on varw}\end{align*}
\noindent Let $\dot{Y}_{i}(1) = Y_i(1) - \expesub{\pi}{Y_i(1)}$ and $\dot{Y}_{i}(0) = Y_i(0) - \expesub{1-\pi}{Y_i(0)}.$ Then the expression on the first line in the previous display can be written as
\begin{align*}
&\sum_{i=1}^N \pi_i (1-\pi_i) \left( \frac{1}{N_1} \dot{Y}_{i}(1) + \frac{1}{N_0} \dot{Y}_{i}(0)   \right)^2 \\
&= \left[ \frac{1}{N_1^2} \sum_{i=1}^N \pi_i \dot{Y}_i(1)^2 + \frac{1}{N_0^2} \sum_{i=1}^N (1-\pi_i) \dot{Y}_i(0)^2  - \right. \\ &\left. \frac{1}{N_1^2} \sum_{i=1}^N \pi_i^2 \dot{Y}_i(1)^2 - \frac{1}{N_0^2} \sum_{i=1}^N (1-\pi_i)^2 \dot{Y}_i(0)^2 + \frac{2}{N_1 N_0} \sum_{i=1}^{N} \pi_i (1-\pi_i) \dot{Y}_i(1) \dot{Y}_i(0)  \right] \\
&=  \underbrace{\frac{1}{N_1} \varsub{\pi}{Y_i(1)} +  \frac{1}{N_0}\varsub{1-\pi}{Y_i(0)}}_{ =\expeRapprox{\hat{s}^2} } - \frac{1}{N^2} \sum_{i=1}^N \left( \dfrac{\pi_i}{N_1 /N} \dot{Y}_i(1) - \dfrac{1-\pi_i}{N_0 /N} \dot{Y}_i(0) \right)^2.
\end{align*}

\noindent Combining the previous two displays, we see that
\begin{equation}
\begin{split}
 \expeRapprox{\hat{s}^2} - \vRapprox{\hat{\tau}} =& \left(\sum_i \pi_i (1-\pi_i) \right) \left(\expesub{\pi}{\frac{1}{N_1} Y_i(1)} + \expesub{1-\pi}{\frac{1}{N_0} Y_i(0)} - \expesub{\tilde{\pi}}{\frac{1}{N_1} Y_i(1) + \frac{1}{N_0} Y_i(0)} \right)^2 + \\ 
&\frac{1}{N^2} \sum_{i=1}^N \left( \dfrac{\pi_i}{N_1 /N} \dot{Y}_i(1) - \dfrac{1-\pi_i}{N_0 /N} \dot{Y}_i(0) \right)^2  \geq 0.
\end{split}
\label{eqn:conservativeness-estimated-variance}
\end{equation}

\noindent and the inequality holds with equality if and only if both
\begin{equation}
\expesub{\tilde{\pi}}{\frac{1}{N_1} Y_i(1) + \frac{1}{N_0} Y_i(0)} = \frac{1}{N_1}\expesub{\pi}{Y_i(1)} + \frac{1}{N_0} \expesub{1-\pi}{Y_i(0)}   \label{eqn: sharpness cond1}
\end{equation}
and
\begin{equation}
\frac{\pi_i}{N_1/N} Y_i(1) - \frac{1-\pi_i}{N_0/N} Y_i(0)  
 = \frac{\pi_i}{N_1/N} \expesub{\pi}{Y_i(1)} - \frac{1-\pi_i}{N_0/N} \expesub{1-\pi}{Y_i(0)} \text{ for all } i. \label{eqn: sharpness cond2}  
\end{equation}

\noindent We have thus shown that $\expeRapprox{\hat{s}^2} \geq \vRapprox{\hat\tau}$, with equality if and only if \eqref{eqn: sharpness cond1} and \eqref{eqn: sharpness cond2} both hold. Note that \eqref{eqn: sharpness condition - main text} is just a re-arrangement of the terms in \eqref{eqn: sharpness cond2}. To complete the proof, it thus suffices to show that \eqref{eqn: sharpness cond2} actually implies \eqref{eqn: sharpness cond1}. To do this, we multiply both sides of \eqref{eqn: sharpness cond2} by $(1-\pi_i)/N$ and sum across $i$ to obtain that
$$s \cdot \expesub{\tilde\pi}{ \frac{1}{N_1} Y_i(1) + \frac{1}{N_0} Y_i(0) } - \expesub{1-\pi}{Y_i(0)} = \frac{s}{N_1} \expesub{\pi}{Y_i(1)} - \frac{1}{N_0} \sum_i (1-\pi_i)^2 \expesub{1-\pi}{Y_i(0)} ,$$ 
\noindent where $s = \sum_i \pi_i (1-\pi_i)$. Re-arranging terms, we obtain that
\begin{equation}
 \expesub{\tilde\pi}{ \frac{1}{N_1} Y_i(1) + \frac{1}{N_0} Y_i(0) } = \frac{1}{N_1} \expesub{\pi}{Y_i(1)} + \frac{1}{N_0} \frac{1}{s} \left(N_0 - \sum_i (1-\pi_i)^2  \right)  \expesub{1-\pi}{Y_i(0)}. \label{eqn: only need one cond - intermediate}   
\end{equation}
\noindent Note, however, that
$$N_0 - \sum_i (1-\pi_i)^2 = N_0 - \sum_i(1-\pi_i) + \sum_i \pi_i (1-\pi_i) = s ,$$
\noindent and thus \eqref{eqn: only need one cond - intermediate} implies that
$$\expesub{\tilde\pi}{ \frac{1}{N_1} Y_i(1) + \frac{1}{N_0} Y_i(0) } = \frac{1}{N_1} \expesub{\pi}{Y_i(1)} + \frac{1}{N_0} \expesub{1-\pi}{Y_i(0)},$$
\noindent as needed.
\end{proof}

\paragraph{Proof of Corollary \ref{cor: shat conservative under constant TEs}}
\begin{proof}
\noindent From Proposition \ref{prop: shat conservative}, $\expeRapprox{\hat{s}^2} = \vRapprox{\tauhat}$ if and only if \eqref{eqn: sharpness condition - main text} holds. Rearranging terms in \eqref{eqn: sharpness condition - main text}, we see that $\expeRapprox{\hat{s}^2} = \vRapprox{\tauhat}$ if and only if

$$\frac{\pi_i}{N_1} (Y_i(1) - \expesub{\pi}{Y_i(1)}) - \frac{1-\pi_i}{N_0} (Y_i(0) - \expesub{1-\pi}{Y_i(0)}) = 0 \text{ for all } i.$$

\noindent Since $Y_i(1) = Y_i(0) + \tau$, it follows that $Y_i(1) - \expesub{\pi}{Y_i(1)} = Y_i(0) - \expesub{\pi}{Y_i(0)}$. Hence, the previous display can be written as 
\begin{equation}
  \frac{\pi_i}{N_1} (Y_i(0) - \expesub{\pi}{Y_i(0)}) - \frac{1-\pi_i}{N_0} (Y_i(0) - \expesub{1-\pi}{Y_i(0)}) = 0 \text{ for all } i. \label{eqn: sharpness cond - rewritten}  
\end{equation}

To establish the first part of the result, note that rearranging terms in \eqref{eqn: sharpness cond - rewritten} implies that
\begin{align*}
 \frac{\pi_i}{1-\pi_i} = \frac{N_1}{N_0} \left(1 + \frac{ \expesub{\pi}{Y_i(0)} - \expesub{1-\pi}{Y_i(0)}}{Y_i(0) - \expesub{\pi}{Y_i(0)} } \right)   
\end{align*} 
\noindent for all $i$ such that $Y_i(0) - \expesub{\pi}{Y_i(0)} \neq 0$ and $\pi_i \in (0,1)$. From the second equation in display \eqref{eqn: expression w eatt - proof}, we see that when $\tau_i =  \tau$ for all $i$, $\expeR{\tauhat} = \tau + \expesub{\pi}{Y_i(0)} - \expesub{1-\pi}{Y_i(0)}$, and hence $b = \expesub{\pi}{Y_i(0)} - \expesub{1-\pi}{Y_i(0)}$. Substituting this expression for $b$ into the previous display yields \eqref{eqn: sharpness cond without unbiasedness} given in the corollary. 

To establish the second part of the result, observe that since $b = \expesub{\pi}{Y_i(0)} - \expesub{1-\pi}{Y_i(0)}$, when $b=0$ we have that $\expesub{\pi}{Y_i(0)} = \expesub{1-\pi}{Y_i(0)}$. Hence, when $b=0$, \eqref{eqn: sharpness cond - rewritten} can be written as 
$$\left( \frac{\pi_i}{N_1} - \frac{1-\pi_i}{N_0} \right) \left(Y_i(0) - \expesub{\pi}{Y_i(0)} \right) = 0  \text{ for all } i,$$

\noindent which holds if and only if $\pi_i = \frac{N_1}{N}$ for all $i$ such that $Y_i(0) \neq \expesub{\pi}{Y_i(0)}$. 
\end{proof}

\paragraph{Proof of Proposition \ref{prop: clt for tauhat and var consistency}}
\begin{proof}

First, viewing $\hat\tau$ as a Horwitz-Thompson estimator under rejective sampling as in Section \ref{sec: distribution of DIM}, the central limit theorem follows immediately from Theorem 1 in \citet{berger_rate_1998}.
\citet{hajek_asymptotic_1964} states a similar result where the Horvitz-Thompson estimator uses an approximation to the marginal probabilities $\pi_i = \expeR{D_i}$ in terms of the underlying probabilities $p_i$.

Second, to show convergence of $\hat{s}^2/\expeRapprox{\hat{s}^2}$, it suffices to show that $\dfrac{\hat{s}_1^2}{ \varsub{\pi}{Y_i(1)} } \rightarrow_p 1$ and $\dfrac{\hat{s}_0^2}{ \varsub{1-\pi}{Y_i(0)} } \rightarrow_p 1$. We provide a proof for the former; the latter proof is analogous. For notational convenience, let $v_{1} = \varsub{\pi}{Y_i(1)} .$ From the definition of $\hat{s}_1^2$, we can write
$$\frac{\hat{s}_1^2}{v_{1}} = \frac{1}{v_{1}} \left( \left(\frac{1}{N_1} \sum_i D_i (Y_i(1) - \expesub{\pi}{Y_i(1)})^2  \right) - (\bar{Y}_1 - \expesub{\pi}{Y_i(1)})^2 \right).$$

Now, $\frac{1}{N_1} \sum_i D_i (Y_i(1) - \expesub{\pi}{Y_i(1)})^2$ can be viewed as a Horvitz-Thompson estimator of $\frac{1}{N_1}\sum_i \pi_i (Y_i(1) - \expesub{\pi}{Y_i(1)})^2 = v_{1}$, and thus by Theorem 6.1 in \citet{hajek_asymptotic_1964}, its variance is equal to $$(1+o(1)) \left(\frac{1}{N_1^2} \sum_i \pi_i (1-\pi_i) \right) \cdot \varsub{\tilde{\pi}}{(Y_i(1) - \expesub{\pi}{Y_i(1)})^2}.$$

Note further that 
\begin{align*}
\left(\frac{1}{N_1^2} \sum_i \pi_i (1-\pi_i) \right) \cdot \varsub{\tilde{\pi}}{(Y_i(1) - \expesub{\pi}{Y_i(1)})^2} &\leq \frac{1}{N_1^2} \sum_i \pi_i (1-\pi_i) (Y_i(1) - \expesub{\pi}{ Y_i(1) })^4\\
&\leq \frac{1}{N_1^2} m_N(1) \sum_i \pi_i (Y_i(1) - \expesub{\pi}{ (Y_i(1)})^2  \\
&= \frac{1}{N_1} m_N(1) \varsub{\pi}{ Y_i(1) }.
\end{align*}
\noindent Applying Chebyshev's inequality, we have
$$ \frac{1}{N_1} \sum_i( D_i (Y_i(1) - \expesub{\pi}{Y_i(1)})^2  - v_{1} = O_p\left(\sqrt{\frac{1}{N_1} m_N(1) \varsub{\pi}{ Y_i(1) } } \right).$$

\noindent 

Next, viewing $\bar{Y}_1$ as a Horvitz-Thomson estimator, we see that its variance is $(1+o(1)) \left( \frac{1}{N_1^2} \sum_i \pi_i (1-\pi_i) \right) \cdot \varsub{\tilde{\pi}}{Y_i(1)}$,  which by similar logic to that above is bounded above by $(1+o(1))\frac{1}{N_1} \varsub{\pi}{Y_i(1)}$. Thus, by Chebyshev's inequality, $$\bar{Y}_1 - \expesub{\pi}{Y_i(1)} = O_p\left(\sqrt{\frac{1}{N_1} \varsub{\pi}{Y_i(1)}}\right).$$

Combining the results above, it follows that
\begin{align*}
\frac{\hat{s}_1^2}{v_{1}} &= \frac{1}{v_{1}} \left( v_{1} + O_p\left( \sqrt{ \dfrac{m_N(1) v_1}{ N_1} }\right) + O_p\left( \frac{1}{N_1} v_1 \right) \right) = 1 + O_p\left( \sqrt{ \dfrac{m_N(1)}{ v_1 N_1} }\right) + O_p\left( \frac{1}{N_1} \right).
\end{align*}

\noindent However, the first $O_p$ term converges to 0 by assumption, and since Assumption \ref{asm: reg conditions for CLT and var consistency}\ref{pi times one minus pi goes to infty} implies that $N_1\rightarrow\infty$, the second $O_p$ term converges to 0 as well.
\end{proof}

\paragraph{Proof of Proposition \ref{prop: local to zero coverage}}
\begin{proof}
From Proposition \ref{prop: clt for tauhat and var consistency}, we have that $\frac{\hat\tau - \expeR{\hat\tau} }{ \sqrt{ \vRapprox{\hat\tau} } } \xrightarrow{d} \normnot{0}{1}$.
Observe that we can write
$$ \frac{\hat\tau - \tau_{EATT} }{ \hat{s} } = \dfrac{ \sqrt{\expeRapprox{\hat{s}^2}}  }{ \hat{s} } \sqrt{ \dfrac{ \vRapprox{ \hat{\tau} }  }{\expeRapprox{\hat{s}^2}}  } \left(\frac{\hat\tau - \expeR{\hat\tau} }{ \sqrt{\vRapprox{\hat\tau} } } + \frac{b}{ \sqrt{\vRapprox{\hat\tau}} }  \right) , $$

\noindent where $\expeR{\hat\tau} = \tau_{EATT} + b$ by Proposition \ref{prop: expectation of sdim}. However, by Proposition \ref{prop: clt for tauhat and var consistency} and the continuous mapping theorem,

$$ \dfrac{ \sqrt{\expeRapprox{\hat{s}^2}}  }{ \hat{s} } \xrightarrow{p} 1. $$

\noindent It then follows from Slutky's lemma and the assumptions of the proposition that 

$$  \frac{\hat\tau - \tau_{EATT} }{ \hat{s} } \xrightarrow{d} r \cdot \left( \normnot{0}{1} + b^* \right) = \normnot{b^* \cdot r}{ r^2 }.$$ 
\end{proof}

\paragraph{Proof of Proposition \ref{prop: interpretation of ols w covariates}}
\begin{proof}
Let $E^*_{R}[\cdot \mid \cdot]$ denote the best linear projection under the randomization distribution with covariates.
That is, for unit-level variables $A_i \in \mathbb{R}$, $B_i \in \mathbb{R}^{p}$, $E^*_{R}[A_i \mid B_i] = \beta_{B}^\prime B_i$ for
$$
\beta_{B} := \arg \min_{\beta} \expeR{\frac{1}{N} \sum_{i=1}^{N} (A_i - \beta^\prime B_i)^2}.
$$
Define $\beta = \left( \beta_0, \beta_D, \beta_{W}^\prime \right)^\prime$ as the coefficients in the best linear projection of $Y_i$ on $(1, D_i, W_i^\prime)^\prime$
\begin{equation}\label{eqn: estimand, ols w covariates}
\beta := \arg \min_{\beta \in \mathbb{R}^{k+2}} \expeR{\frac{1}{N} \sum_{i=1}^{N} (Y_i - \beta'(1, D_i, W_i^\prime) )^2}.
\end{equation}

To prove the first claim, observe that 
$$
E^*_{R}[W_i | 1, D_i] = D_i \expesub{\pi}{W_i} + (1-D_i) \expesub{1-\pi}{W_i}.
$$
By the Frisch-Waugh-Lovell Theorem, 
$$
\beta_{W} = \expeR{ \frac{1}{N} \sum_i (W_i- E^*[W_i | 1,D_i ]) (W_i- E^*[W_i | 1,D_i ])^\prime }^{-1} \expeR{\frac{1}{N} \sum_i (W_i - E^*[W_i | 1,D_i ]) Y_i } = 
$$
$$
\expeR{\frac{1}{N}\sum_i D_i (W_i - \expesub{\pi}{W_i}) (W_i - \expesub{\pi}{W_i})^\prime + \frac{1}{N} \sum_i (1-D_i) (W_i - \expesub{1-\pi}{W_i}) (W_i - \expesub{1-\pi}{W_i})^\prime }^{-1} \times 
$$
$$
\hspace{1cm} \expeR{\frac{1}{N}\sum_i D_i (W_i - \expesub{\pi}{W_i}) Y_i(1) + \frac{1}{N} \sum_i (1-D_i) (W_i - \expesub{1-\pi}{W_i}) Y_i(0) } = 
$$
$$
\left( \frac{N_1}{N} \varsub{\pi}{W_i} + \frac{N_0}{N} \varsub{1-\pi}{W_i} \right)^{-1} \left( \frac{N_1}{N} \expesub{\pi}{(W_i - \expesub{\pi}{W_i}) Y_i(1)} + \frac{N_0}{N} \expesub{1-\pi}{ (W_i - \expesub{1-\pi}{W_i}) Y_i(0) } \right) . 
$$

\noindent Letting $\gamma(1) = \varsub{\pi}{W_i}^{-1} \covsub{\pi}{W_i}{Y_i(1)}$ be the $\pi$-weighted projection of $Y_i(1)$ on $W_i$, and likewise $\gamma(0) = \varsub{1-\pi}{W_i}^{-1} \covsub{1-\pi}{W_i}{Y_i(0)}$, the previous display implies that 
$$
\beta_W = \theta \gamma(1) + (I_k - \theta) \gamma(0) =: \gamma,
$$
\noindent for $\theta := \left( \frac{N_1}{N} \varsub{\pi}{W_i} + \frac{N_0}{N} \varsub{1-\pi}{W_i} \right)^{-1} \frac{N_1}{N} \varsub{\pi}{W_i} $.

Note, however, that $E^*_{R}[Y_i \mid 1, D_i, W_i] = E^*_{R}[Y_i - \beta_W'W_i \mid 1,D_i]$. It follows that
\begin{align*}
\beta_{D} &= \expeR{\frac{1}{N_1} \sum_i D_i (Y_i - \gamma' W_i) - \frac{1}{N_0} \sum_i (1-D_i) (Y_i - \gamma'W_i)}\\
&= \tau_{EATT} + \frac{N_1}{N} \frac{N_0}{N} \covsub{1}{\pi_i}{Y_i(0) - \gamma' W_i},
\end{align*}
\noindent where the last equality is obtained from applying Proposition \ref{prop: expectation of sdim} to the transformed outcome $Y_i - \gamma'W_i$.

To prove the second claim, by the Frisch-Waugh-Lovell Theorem,
$$
E^*_{R}[Y_i | D_i - \hat{\pi}_i ] = \beta_D (D_i - \hat{\pi}_i),
$$
and so
$$
\beta_D = \expeR{ \frac{1}{N} \sum_i (D_i - \hat{\pi}_i)^2 }^{-1} \expeR{ \frac{1}{N} \sum_i (D_i - \hat{\pi}_i) Y_i }.
$$
Writing $(D_i - \hat{\pi}_i)^2 = D_i - 2 D_i \hat{\pi}_i + \hat{\pi}_i^2$ and $Y_i = Y_i(0) + D_i \tau_i$ and evaluating the expectation over the randomization distribution yields
\begin{align*}
\beta_D &= \expesub{1}{\pi_i - 2 \pi_i \hat{\pi}_i + \hat{\pi}_i^2}^{-1} \expeR{  \frac{1}{N} \sum_i (D_i - \hat{\pi}_i) Y_i(0)} +  \\
& \hspace{1cm} \expesub{1}{\pi_i - 2 \pi_i \hat{\pi}_i + \hat{\pi}_i^2}^{-1} \expeR{  \frac{1}{N} \sum_i D_i (1 - \hat{\pi}_i) \tau_i} \\
&= \expesub{1}{\pi_i - 2 \pi_i \hat{\pi}_i + \hat{\pi}_i^2}^{-1} \expesub{1}{  (\pi_i - \hat{\pi}_i) Y_i(0)} +  \\
& \hspace{1cm} \expesub{1}{\pi_i - 2 \pi_i \hat{\pi}_i + \hat{\pi}_i^2}^{-1} \expesub{1}{ \pi_i (1 - \hat{\pi}_i) \tau_i}. \numberthis \label{eqn: beta D in proof}
\end{align*}
\noindent Note, however, that $\expesub{1}{\pi_i - \hat\pi_i} =0$, since a constant is included in $W_i$ and thus the regression residuals average to 0, and hence $\expesub{1}{  (\pi_i - \hat{\pi}_i) Y_i(0)} = \covsub{1}{\pi_i - \hat{\pi}_i}{Y_i(0)}$. Additionally, $$\expesub{1}{\pi_i - 2 \pi_i \hat{\pi}_i + \hat{\pi}_i^2} = \expesub{1}{\pi_i(1-\hat{\pi}_i)} + \expesub{1}{\hat\pi_i(\hat{\pi}_i - \pi_i)} = \expesub{1}{\pi_i(1-\hat{\pi}_i)},$$
\noindent where $\expesub{1}{\hat\pi_i(\hat{\pi}_i - \pi_i)} = 0$ since by construction regression residuals are orthogonal to the regressors. Substituting these expressions into (\ref{eqn: beta D in proof}) yields the desired result.
\end{proof}

\section{Relationship to Law of Total Variance}\label{section: variance conservativeness, conditional vs. unconditional frameworks}

In this section, we discuss how the conservativeness of the usual variance estimator $\hat{s}^2$ established in Proposition \ref{prop: shat conservative} is related to, but distinct from, the well-known fact that a conditional variance must on average be less than an unconditional one by the law of total variance. In order to do so, we nest our design-based framework within a super-population framework. 

Consider a super-population in which individuals are characterized by $(Y_i(1),Y_i(0),p_i, D_i) \sim P$, where $p_i$ is the (unconditional) individual-level probability of treatment and treatment is generated according to $D_i \mid p_i, Y_i(0), Y_i(1) \sim Bernoulli(p_i)$, and suppose we sample $N$ individuals i.i.d. from this super-population. 
The observed data is then $(Y_i,D_i) = (Y_i(1)D_i + Y_i(0) (1-D_i),D_i)$ for $i=1,...,N$.
The finite-population data-generating process we consider is equivalent to analyzing this sampling process conditional on $\mathcal{F}_N = \{ Y_1(\cdot),...,Y_N(\cdot), \sum_i D_i \}$.

We could of course analyze this sampling process without conditioning on $\mathcal{F}_N$ (i.e., unconditionally). In this case, the observable data satisfy $(Y_i,D_i) \overset{iid}{\sim} P^*$, where $P^*$ is the distribution of $(Y_i,D_i)$ induced by first sampling $(Y_i(1),Y_i(0),p_i, D_i) \sim P$ and then calculating $(Y_i,D_i) = (Y_i(1)D_i + Y_i(0) (1-D_i),D_i)$. As in the main text, let $\hat{s}^2 = \frac{1}{N_1}\hat{s}_1^2 + \frac{1}{N_0} \hat{s}_0^2$ be the standard variance estimator for the difference-in-means estimator $\hat\tau$, where $\hat{s}_d^2$ is the sample variance of $Y_i \mid D_i =d$.
Standard arguments for i.i.d. sampling imply that $(1+o(1))E_{P^*}[\hat{s}^2] = Var_{P^*}(\hat\tau)$, where the $o(1)$ term arises because for simplicity in the main text, we define $\hat{s}^2_d$ to be the sample variance without degrees of freedom adjustment (e.g. we use $N_1$ rather than $N_1-1$ in the denominator of $\hat{s}_1^2$). Observe that the law of total variance implies that $Var_{P^*}(\hat\tau) = E_{P^*}[ Var(\hat\tau \mid \mathcal{F}_N) ] + Var_{P^*}( E[\hat\tau \mid \mathcal{F}_N] )$.
Consequently, under $P^*$, the conditional variance of $\hat\tau$ must \emph{on average} be less than or equal to $(1+o(1))E_{P^*}[\hat{s}^2]$:
\begin{equation}
E_{P^*}[ Var(\hat\tau \mid \mathcal{F}_N) ] \leq (1+o(1)) E_{P^*}[\hat{s}^2]. \label{eqn: condl variance conservative on average}
\end{equation}
Notice, however, that \eqref{eqn: condl variance conservative on average} does not necessarily imply that $Var(\hat\tau \mid \mathcal{F}_N) \leq (1+o(1))E_{P^*}[\hat{s}]$ \emph{for all} $\mathcal{F}_N$, and furthermore the upper bound in \eqref{eqn: condl variance conservative on average} involves the \emph{unconditional} mean $E_{P^*}[\hat{s}^2]$. By contrast, our results in Section \ref{sec: distribution of DIM} establish that, for all $\mathcal{F}_N$,
\begin{equation}
Var(\hat\tau \mid \mathcal{F}_N) \leq  (1 +o(1)) E[\hat{s}^2 \mid \mathcal{F}_N] . \label{eqn: condl variance ineq}
\end{equation}
That is, while \eqref{eqn: condl variance conservative on average} bounds the average conditional variance of the difference-in-means estimator over realizations of $\mathcal{F}_N$, \eqref{eqn: condl variance ineq} holds for \textit{all realizations} $\mathcal{F}_N$. 
Moreover, the upper bound involves the conditional expectation of the variance estimator $E[\hat{s}^2 \mid \mathcal{F}_N]$ rather than the unconditional expectation $E_{P^*}[\hat{s}^2]$. 

\section{Berry-Esseen Type Bound on Quality of Normal Approximation}\label{section: berry-esseen}

In addition to the asymptotic results shown in Section \ref{sec: distribution of DIM} for the DIM estimator, we can also obtain Berry-Esseen type bounds on the quality of the normal approximation (using the approximate variance $\vRapprox{\tauhat})$ for a fixed finite population.
This result is attractive in the sense that it shows that the distribution of $\hat\tau$ will be approximately normally distributed in finite populations that are sufficiently large (relative to the fourth moment of the potential outcomes), without appealing to arguments involving a sequence of finite populations of increasing size.

\begin{prop} \label{prop: berryesseen}
Suppose Assumption \ref{asm: rejective assignment probability} holds.
Let $b_1,b_2$ be positive constants, and define $t = (\tauhat - \expeR{\tauhat} ) / \sqrt{\vRapprox{\tauhat}}.$ 
Then there exist constants $k$ and $\bar{N}$ such that $$\sup_{y}| \probsub{R}{ t \leq y  } - \Phi(y)  | \leq \frac{k}{\sqrt{N}}$$

\noindent for any finite population of size $N \geq \bar{N}$ such that $ \vRapprox{\hat\tau} = N b_1$ and $\expesub{1}{\left(\frac{1}{N_1} Y_i(1) + \frac{1}{N_0} Y_i(0) \right)^4 } < b_2.$

\begin{proof}
Viewing $\hat\tau$ as a Horvitz-Thompson estimator under rejective sampling once again, the result follows immediately from Theorem 3 in \citet{berger_rate_1998}. 
\end{proof}
\end{prop}

\section{Results for \citet{ImbensManski(04)} Intervals} \label{section: imbens-manski}
We provide more details on the \citet{ImbensManski(04)} robust confidence intervals described in Section \ref{sec: DIM sensitivity analyis} and implemented in our application in Section \ref{subsec: medicaid application}. We show that the Imbens-Manski intervals have correct but potentially conservative coverage under the imposed bound on $\covsub{1}{\pi_i}{Y_i(0)}$. We further show that breakdown values based on the Imbens-Manski intervals are likewise conservative. 

Recall from Section \ref{sec: DIM sensitivity analyis} that the \citet{ImbensManski(04)} CI for the parameter $\tau_{EATT}$ takes the from $\mathcal{C}(\hat\tau, \hat{s}) = [\hat\tau^{lb}_{EATT}- C \hat{s}, \hat\tau^{ub}_{EATT} + C \hat{s} ]$ for the constant $C$ that solves
\begin{equation}
\Phi\left( \frac{\Delta}{\hat{s}} + C  \right) - \Phi(-C) = 1-\alpha .     \label{eqn: eqn defining C}
\end{equation}

We first observe that the interval $\mathcal{C}$ becomes larger for larger values of $\hat{s}$, as formalized in the following lemma.

\begin{lem} \label{lem: imbens manski interval increasing}
For any $\hat\tau$ and $\Delta \geq 0$, if $\hat{s}_2 > \hat{s}_1 > 0$, then $\mathcal{C}(\hat\tau, \hat{s}_1) \subseteq \mathcal{C}(\hat\tau, \hat{s}_2)$. The inclusion is strict if $\Delta > 0$.
\end{lem}
\begin{proof}
From the definition of $\mathcal{C}(\hat\tau, \hat{s})$, it clearly suffices to show that the constant $C$ defined by \eqref{eqn: eqn defining C} is increasing in $\hat{s}$ (and strictly so if $\Delta>0$). Observe that \eqref{eqn: eqn defining C} defines $C$ by the equation $g(\hat{s},C) = 0$ for $g(\hat{s},C) = \Phi( \Delta/ \hat{s} +C) - \Phi(-C) - (1-\alpha) $. However, by the implicit function theorem, we then have that 
$$ \frac{d C}{d \hat{s}} = - \dfrac{ \frac{\partial g}{ \partial \hat{s} }  }{ \frac{\partial g}{ \partial C }  } = \dfrac{ \frac{\Delta}{ \hat{s}^2 } \cdot \phi\left( \frac{\Delta}{ \hat{s} } + C \right) }{ \phi\left( \frac{\Delta}{ \hat{s} } + C \right) + \phi(-C)  }  \geq 0, $$
\noindent where we use the fact that normal densities and $\Delta$ are weakly positive. The derivative is strictly positive if $\Delta > 0$. 
    
\end{proof}

With this result in hand, we can show that the Imbens-Manski intervals have correct but potentially conservative coverage under the imposed assumptions. Let $\hat{s}_*^2 = \vRapprox{\hat\tau} / \expeRapprox{\hat{s}^2}  \cdot  \hat{s}^2$ be the infeasible variance estimator that adjusts for the bias in $\hat{s}^2$. Our results imply that $\hat{s}_*^2 \leq \hat{s}^2$ and that $\hat{s}_*^2 / \vR{\hat\tau} \convp 1$. Hence, if $\hat\tau^{lb}_{EATT}$ satisfies a central limit theorem, the results in Imbens and Manski that assume a consistent variance estimator imply that 
$$\liminf_{N \to \infty} P_R( \tau_{EATT} \in \mathcal{C}(\hat\tau, \hat{s}_* )) \geq 1-\alpha. $$
\noindent However, since $\mathcal{C}(\hat\tau, \hat{s}_*) \subseteq \mathcal{C}(\hat\tau, \hat{s})$, it follows that
$$\liminf_{N \to \infty} P_R( \tau_{EATT} \in \mathcal{C}(\hat\tau, \hat{s} )) \geq 1-\alpha. $$

\noindent The regularity conditions to make this argument precise are formalized in the following lemma.
\begin{lem}[Coverage of Imbens-Manski Intervals]
Suppose Assumptions \ref{asm: rejective assignment probability} and \ref{asm: reg conditions for CLT and var consistency} hold, and that $N \vRapprox{\hat\tau} \to s_*^2 \in (0,\infty)$. If $\covsub{1}{\pi_i}{Y_i(0)} \in [\underline{b}, \overline{b}]$, then
$$\liminf_{N \to \infty} P_R( \tau_{EATT} \in \mathcal{C}(\hat\tau, \hat{s} )) \geq 1-\alpha. $$
\end{lem}
\begin{proof}
From Proposition \ref{prop: clt for tauhat and var consistency} part 1 along with the assumption that $N \vRapprox{\hat\tau} \to s_*^2$, we have that $\sqrt{N}(\hat\tau - \expeR{\hat\tau}) \convd \normnot{0}{s_*^2}$. Since $\hat\tau^{lb}_{EATT}$ simply shifts $\hat\tau$ by a deterministic constant, it follows that $\sqrt{N}(\hat\tau^{lb}_{EATT} - \expeR{\hat\tau^{lb}_{EATT}}) \convd \normnot{0}{s_*^2}$. Additionally, from Proposition \ref{prop: clt for tauhat and var consistency} part 2 along with the assumption that $N \vRapprox{\hat\tau} \to s_*^2$, we have that $$N \hat{s}_*^2 = N \vRapprox{\hat\tau} \cdot \frac{\hat{s}^2}{ \expeRapprox{\hat{s}^2} }  \convp s_*^2. $$ The interval $\mathcal{C}(\hat\tau, \hat{s}_*)$ thus corresponds to the interval proposed by \citet{ImbensManski(04)} for a setting with a consistently estimable variance. It follows from Lemma 4 in \citet{ImbensManski(04)} that
\begin{equation}
 \liminf_{N \to \infty} \inf_{\tau \in [\tau^{lb}_{EATT},\tau^{ub}_{EATT}]} P_R( \tau \in \mathcal{C}(\hat\tau, \hat{s}_* )) \geq 1-\alpha  \label{eqn: imbens manski coverage pts in id set}   
\end{equation}
\noindent and hence
$$\liminf_{N \to \infty} P_R( \tau_{EATT} \in \mathcal{C}(\hat\tau, \hat{s}_* )) \geq 1-\alpha $$
\noindent since $\tau_{EATT} \in [\tau_{EATT}^{lb},\tau_{EATT}^{ub}]$ when $\covsub{1}{\pi_i}{Y_i(0)} \in [\underline{b}, \overline{b}]$. However, by Proposition \ref{prop: shat conservative}, $\hat{s}_* \leq \hat{s}$, and thus $\mathcal{C}(\hat\tau, \hat{s}_*) \subseteq \mathcal{C}(\hat\tau, \hat{s})$ by Lemma \ref{lem: imbens manski interval increasing}. It follows that
$$\liminf_{N \to \infty} P_R( \tau_{EATT} \in \mathcal{C}(\hat\tau, \hat{s} )) \geq \liminf_{N \to \infty} P_R( \tau_{EATT} \in \mathcal{C}(\hat\tau, \hat{s}_* )) \geq 1-\alpha,$$
as we wished to show.
\end{proof}

We next study the properties of the ``breakdown'' values implied by sensitivity analyses using Imbens-Manski intervals. Let $\mathcal{I}(\tilde{b}) = [\tau_{EATT}^{lb}(\tilde{b}), \tau_{EATT}^{ub}(\tilde{b})]$ be the identified set for $\tau_{EATT}$ under the assumption that $\covsub{1}{\pi_i}{Y_i(0)} \in [-\tilde{b},\tilde{b}]$ (where we now write the bounds explicitly as a function of $\tilde{b}$). 
We define the ``breakdown'' value for a null effect to be the minimal value of $\tilde{b}$ such that that identified contains zero, i.e. $b^* = \inf \, \{ \tilde{b} \,:\, 0 \in \mathcal{I}(\tilde{b}) \}$. (One could analogously obtain breakdown values for the null hypothesis that $\tau_{EATT} = \tau^*$ for some other value $\tau^*$.) Let $\hat{b}^*$ be analogously defined as the minimum value of $\tilde{b}$ such that 0 is contained within the Imbens-Manski CI, $\hat{b}^* = \inf \, \{ \tilde{b} \,:\, 0 \in \mathcal{C}(\hat\tau, \hat{s}; \tilde{b}) \}$, where we now make the dependence of $\mathcal{C}$ on $\tilde{b}$ explicit. The following result shows that $\hat{b}^*$ is a valid, but potentially conservative, $1-\alpha$ level lower bound on $b^*$. 

\begin{lem}[Conservativeness of breakdown values]
 Suppose Assumptions \ref{asm: rejective assignment probability} and \ref{asm: reg conditions for CLT and var consistency} hold, and that $N \vRapprox{\hat\tau} \to s_*^2 \in (0,\infty)$. Then
 $$\liminf_{N \to \infty} P_R( \hat{b}^* \leq b^* ) \geq 1- \alpha.$$
\end{lem}

\begin{proof}
Note that by construction, $\hat{b}^* \leq b^*$ whenever $0 \in \mathcal{C}(\hat\tau,\hat{s}; b^*)$. Hence 
 $$\liminf_{N \to \infty} P_R( \hat{b}^* \leq b^* ) \geq \liminf_{N \to \infty} P_R( 0 \in \mathcal{C}(\hat\tau,\hat{s}; b^*) ) .$$
 \noindent However, the definition of $b^*$ combined with the continuity of the identified set bounds in $\tilde{b}$ implies that $0 \in [\tau_{EATT}^{lb}(b^*), \tau_{EATT}^{ub}(b^*)]$. It follows from \eqref{eqn: imbens manski coverage pts in id set} that 
 $$ \liminf_{N \to \infty} P_R( 0 \in \mathcal{C}(\hat\tau,\hat{s}; b^*) ) \geq 1-\alpha .$$
 \noindent Combining the previous two displays yields the desired result. 
\end{proof}

\section{Difference-in-Differences with Multiple Time Periods}\label{section: multi-period DiD}

We consider non-staggered, DID estimators with more than two time periods (e.g., Chapter 5 of \cite{AngristPischke(09)}), extending the simple two-period DID model discussed in Section \ref{sec: application to two period DiD}.
Suppose we observe panel data for a finite-population of $N$ units for periods $t = -\ubar{T}, \hdots, \bar{T}$. 
Units with $D_i = 1$ receive a treatment beginning at period $t=1$.
The observed outcome for unit $i$ at period $t$ is $Y_{it} = Y_{it}(D_i)$, and the treatment is assumed to have no effect prior to implementation so that $Y_{it}(1) = Y_{it}(0)$ for all $t < 1$ (``no-anticipation'').

Researchers commonly estimate the dynamic two-way fixed effects (TWFE) regression specification (sometimes called an ``event-study'')
\begin{equation}
Y_{it} = \alpha_i + \phi_t + \sum_{s \neq 0} D_i \times 1[s =t] \times \beta_s + \epsilon_{it}, \label{eqn: twfe specification}
\end{equation}
by OLS and causally interpret the regression coefficients $\{\hat\beta_t \colon t = 1, \hdots, \bar{T}\}$. 
The regression coefficients are numerically equivalent to the DID estimators $\betahat_t = \tauhat_t - \tauhat_0$ for $\tauhat_t = \frac{1}{N_1} \sum_i D_i Y_{it} - \frac{1}{N_0} \sum_i (1-D_i) Y_{it}$.

Under Assumption \ref{asm: rejective assignment probability}, Proposition \ref{prop: expectation of sdim} therefore implies, for all $t = -\underline{T}, \hdots, \bar{T}$,
$$
\expeR{\betahat_t} = \tau_{EATT,t} + \underbrace{ \frac{N}{N_0} \frac{N}{N_1} \covsub{1}{\pi_i}{ Y_{it}(0) - Y_{i0}(0) } }_{=:\delta_t},
$$
where $\tau_{EATT,t} = \frac{1}{N_1} \sum_i \pi_i (Y_{it}(1) - Y_{it}(0))$ is the EATT in period $t$ (which is equal to zero for $t<1$ by the no anticipation assumption). It follows that $\betahat_t$ is unbiased for $\tau_{EATT, t}$ under the design-based analog to the parallel trends assumption that $\delta_t = 0$. Furthermore, under additional regularity conditions (see Appendix \ref{section: appendix extension to multiple outcomes}), a multivariate, finite-population central limit theorem implies $\sqrt{N}(\boldsymbol{\betahat} - (\boldsymbol{\tau}_{EATT} + \boldsymbol{\delta})) \rightarrow_d \normnot{0}{\Sigma}$, where $\boldsymbol{\betahat}$, $\boldsymbol{\delta}$, and $\boldsymbol{\tau_{EATT}}$ respectively stack the $\betahat_t$, $\delta_t$, and $\tau_{EATT,t}$, and $\Sigma = \lim_{N\rightarrow \infty} N \vR{\betahat_t}$. 
Further, the cluster-robust variance estimator that clusters at the unit level \citep[][]{bertrand_how_2004} is consistent for an upper bound on the variance of $\boldsymbol{\betahat}$. Consequently, confidence intervals based on cluster-robust standard errors will have asymptotically correct but conservative coverage for the EATT when the design-based parallel trends assumption is satisfied.

\paragraph{Sensitivity analyses for the dynamic two-way fixed effects regression:}
Our results imply that existing sensitivity analyses for difference-in-differences settings developed from the super-population perspective can also be used in the design-based setting. \citet{rambachan_more_2023} introduce a sensitivity analysis framework for bounding causal estimands when the parallel trends assumption fails. In particular, they consider settings where the researcher has access to estimates $\boldsymbol{\betahat}$ such that $\sqrt{N}(\boldsymbol{\betahat}- (\boldsymbol{\delta} + \boldsymbol{\tau})) \xrightarrow{d} \normnot{0}{\Sigma}$, where $\boldsymbol{\tau}$ is a vector of causal effects of interest and $\boldsymbol{\delta}$ is a vector of biases. They then derive the identified set for parameters of the form $l' \boldsymbol{\tau}$, and show how inference on such parameters can be conducted using methods from the moment inequality literature when the variance is consistently estimable. Although their analysis is motivated by super-population sampling, our results illustrate that the same asymptotic approximation arises in the design-based setting. 
A subtlety in our design-based setting is that $\Sigma$ can only be conservatively estimated. Our results thus imply that sensitivity analyses based on \citet{rambachan_more_2023} will also be valid but potentially conservative from the design-based perspective provided the moment inequality method used remains valid given a conservative estimate of the variance. This property holds, for example, for tests based on the ``least-favorable'' critical values in \citet{ARP(23)}.

\section{Extension to General OLS estimators with Clustered Assignments}\label{section: appendix extension to OLS}

This section extends our analysis under the rejective assignment mechanism in two ways. 
First, we consider general regression estimators beyond the simple DIM. Second, we allow for clustered treatment assignment. This nests our results in the main text on the DIM under individual-level treatment assignment as a special case where (i) the regression estimator is the DIM, and (ii) each cluster corresponds with exactly 1 unit. 

As in Section \ref{sec: clustering}, suppose each unit $i = 1, \hdots, N$ belongs to one of $c = 1, \hdots, C$ clusters, where $c(i)$ denotes the cluster membership of unit $i$.
The treatment is assigned at the cluster level, where the cluster level treatment assignments $D := \left(D_1, \hdots, D_C \right)'$ follow a rejective assignment mechanism (Assumption \ref{asm: clustered rejective assignment}).  We denote by $N_c$ the number of units in cluster $c$, and let $C_0,C_1$ denote the number of untreated and treated clusters, respectively. Suppose that the researcher estimates the ordinary least squares (OLS) coefficients $\hat\beta$ from the regression $Y_{i} = X_{i}'\beta + \epsilon_i$, where $X_i = D_i X_{i}(1) + (1-D_i) X_{i}(0)$ is a vector of covariates potentially depending on $D_i$. Note that if $X_i(d) = (1,d)'$, then the second element of $\hat\beta$ corresponds with the DIM.

We analyze the properties of the OLS estimator along a sequence of finite-populations along which the number of clusters $C$ grows large, similar to the asymptotics in Section \ref{sec: distribution of DIM}. 
We provide the proofs of all results in Appendix \ref{sec: OLS w/ clustering proofs}.

Before stating our results, we introduce some notation. Let $\clustersumXX(d) = \sum_{i: c(i) =c} X_i(d) X_i(d)'$ and $\clustersumXY(d) = \sum_{i:c(i)=c} X_i(d) Y_i(d)$. 
For a cluster-level function of the potential outcome $A_c(d)$, we will write, $\expesub{w_c}{A_c(d)}$ to denote the sum $\frac{1}{\sum_c w_c} \sum_c A_c(d)$. Using this notation, $\betahat$ can be written as 
\begin{align*}
\hat\beta =& \left(\sum_i X_i X_i' \right)^{-1} \left(\sum_i X_i Y_i \right) \\
=& \left(\frac{C_1}{C} \frac{1}{C_1}\sum_c D_c \clustersumXX(1)  + \frac{C_0}{C} \frac{1}{C_0} \sum_c (1-D_c) \clustersumXX(0) \right)^{-1} \times \\& \left(\frac{C_1}{C} \frac{1}{C_1} \sum_c D_c \clustersumXY(1)  + \frac{C_0}{C} \frac{1}{C_0}\sum_c (1-D_c)  \clustersumXY(0) \right)
\end{align*}

Our first result shows $\betahat$ is consistent for 
\begin{align*}
\beta_{cluster} := &\left( \frac{C_1}{C} \expesub{\pi_c}{\clustersumXX(1)} + \frac{C_0}{C} \expesub{1-\pi_c}{\clustersumXX(0)} \right)^{-1} \left( \frac{C_1}{C} \expesub{\pi_c}{ \clustersumXY(1) } + \frac{C_0}{C} \expesub{1-\pi_c}{\clustersumXY(0)} \right),
\end{align*}
and asymptotically normally distributed under the clustered randomization distribution.

\begin{asm} \label{asm: conditions for ols consistency clustered} 
\hfill
\begin{enumerate}[label=(\alph*)]
    \item $\expesub{\pi_c}{\clustersumXY(1)}$, $\expesub{1-\pi_c}{\clustersumXY(0)}$, $\expesub{\pi_c}{\clustersumXX(1)}$, $\expesub{1-\pi_c}{\clustersumXX(0)}$, and $\frac{C_1}{C}$ have finite limits, with $\lim \frac{C_1}{C} \in (0,1)$. 
    
    \item $\frac{C_1}{C} \expesub{\pi}{ \clustersumXX(1) } + \frac{C_0}{C} \expesub{1-\pi}{ \clustersumXX(0) }$ has a full-rank limit.
    
    \item There exists $M < \infty$ such that $\varsub{\tilde\pi_c}{ (\clustersumXX(d))_{jk} } < M$ and $\varsub{\tilde\pi_c}{ (\clustersumXY(d))_{j} } < M$ for $d=0,1$ and $j,k=1,...,dim(X_{i})$.
    
    \item Assumption \ref{asm: multivariate lindeberg type condition} is satisfied for $\mathbf{Y}_i = \widetilde{X\epsilon}_c(1) - \widetilde{X\epsilon}_c(0) - \expesub{\pi_c}{\widetilde{X\epsilon}_c(1) - \widetilde{X\epsilon}_c(0)}$, where $\epsilon_i(d) = Y_i(d) - X_i(d)'\beta_{cluster}$ and $\widetilde{X\epsilon}_c(d) = \sum_{i:c(i) = c} X_i(d) \epsilon_i(d)$.
\end{enumerate}
\end{asm}

\begin{prop}[Consistency and asymptotic normality]\label{prop: clustered OLS with covar, consistency and CLT}
Suppose Assumption \ref{asm: clustered rejective assignment} holds, and assume $\sum_c \pi_c (1-\pi_c) \rightarrow \infty$. 
\begin{enumerate}
\item[(1)] If Assumption \ref{asm: conditions for ols consistency clustered} parts (i)-(iii) hold, $\betahat - \beta_{cluster} \xrightarrow{p} 0$.

\item[(2)] Define $V_{cluster} := C^{-1} \left( \sum_c \tilde{\pi}_c \right) \varsub{\tilde \pi_c}{ \sum_{i \colon c(i) = c} X_i(1) \epsilon_i(1) - X_i(0) \epsilon_i(0) }$. If Assumption \ref{asm: conditions for ols consistency clustered} holds, 
$$
\Omega_{cluster}^{-1/2} \sqrt{C} \left( \betahat - \beta_{cluster} \right) \xrightarrow{d} \normnot{0}{I},
$$
where $\Omega_{cluster} := \expesub{R}{\frac{1}{C}\sum_i X_i X_i^\prime}^{-1} V_{cluster} \expesub{R}{\frac{1}{C}\sum_i X_i X_i^\prime}^{-1}$.
\end{enumerate}

\end{prop}

We next analyze the cluster-robust variance estimator \citep{LiangZeger(86)}, 
\begin{equation}
\hat{\Omega}_{cluster} := \left( \frac{1}{C} \sum_i X_i X_i^\prime\right)^{-1} \hat{V}_{cluster} \left( \frac{1}{C} \sum_i X_i X_i^\prime \right)^{-1}, 
\end{equation}
where
\begin{equation}
\hat{V}_{cluster} := \frac{1}{C} \sum_{c} \clustersumXepshat  \clustersumXepshat^{\prime}
\end{equation}
for $\hat{\epsilon}_i = Y_i - X_i^\prime \betahat$ and $\clustersumXepshat = \sum_{i \colon c(i) = c} X_i \hat{\epsilon}_i$.
In the case with an individual-level treatment assignment (i.e., $C = N$), the cluster-robust variance estimator is equivalent to the Eicker-Huber-White heteroskedasticity-robust variance estimator.
Our next result establishes that $\hat{V}_{cluster}$ is consistent for an upper bound of $V_{cluster}$ defined in Proposition \ref{prop: clustered OLS with covar, consistency and CLT} in finite populations with a large number of clusters.

\begin{asm} \label{asm: conditions for cluster variance consistency}
\hfill
\begin{enumerate}[label=(\alph*)]
    \item
    $\expesub{\pi_c}{ \clustersumXeps(1) \clustersumXeps(1)'  }$ and $\expesub{1-\pi_c}{ \clustersumXeps(0) \clustersumXeps(0)'  }$ have limits.
    
    \item
    There exists $\tilde{M}_1 >0$ such that $\| \varsub{\tilde\pi_c}{ \clustersumXeps(d) \clustersumXeps(d)'} \| < \tilde{M}_1$ for $d=0,1$, where $\|A\|$ denotes the Frobenius norm of a matrix $A$.
    
    \item
    There exists $\tilde{M}_2 >0$ such that $\expesub{1}{ \| \clustersumXeps(d) \|^2 } < \tilde{M}_2$ and $\expesub{1}{ \|\clustersumXX(d)\|^2 } < \tilde{M}_2$ for $d=0,1$.
\end{enumerate}
\end{asm}

\begin{prop}[Variance consistency]\label{prop: clustered OLS with covar, var consistency and conservativeness}
If Assumptions \ref{asm: clustered rejective assignment}, \ref{asm: conditions for ols consistency clustered}(i)-(iii), and \ref{asm: conditions for cluster variance consistency} hold, and $\sum_c \pi_c (1-\pi_c) \rightarrow \infty$, then $\hat{V}_{cluster} - V_{cluster}^{est} \xrightarrow{p} 0$ for \begin{align*}
V_{cluster}^{est} := &\frac{C_1}{C} \expesub{\pi_c}{\widetilde{X\epsilon}_c(1) \widetilde{X\epsilon}_c(1)'} + \frac{C_0}{C} \expesub{1-\pi_c}{\widetilde{X\epsilon}_c(0) \widetilde{X\epsilon}_c(0)'} 
\end{align*}
Furthermore, $V_{cluster}^{est} \geq V_{cluster}$ (i.e., $V_{cluster}^{est} - V_{cluster}$ is positive semi-definite).
\end{prop}

\begin{cor}\label{cor: consistency of omega cluster}
Define $\Omega_{cluster}^{est} := \expeR{\sum_i X_i X_i}^{-1} V_{cluster}^{est} \expeR{\sum_i X_i X_i}^{-1}$. Under the same conditions as Proposition \ref{prop: clustered OLS with covar, var consistency and conservativeness}, $\hat{\Omega}_{cluster}-\Omega_{cluster}^{est} \xrightarrow{p} 0,$ and $\Omega_{cluster}^{est} \geq \Omega_{cluster}$.
\end{cor}

Recall that the DIM estimator $\hat\tau$ corresponds to $\hat\beta$ in the special case where $X_i = (1,D_i)'$. The following result summarizes the implications of our results on $\hat\beta$ for this special case. 

\begin{prop}[DIM Estimator Under Clustered Assignment] \label{prop: sdim limiting dist, clustering}
Suppose Assumption \ref{asm: clustered rejective assignment} and Assumption \ref{asm: conditions for ols consistency clustered} hold for $X_i(d) = (1,d)'$, and assume that $\sum_{c} \pi_c (1 - \pi_c) \rightarrow \infty$. Then:
\begin{enumerate}[label={(\roman*)}]
    \item \label{main text, cluster consistency}
    $\tauhat - (\tau_{EATT}^{cluster} + \delta_{cluster}) \xrightarrow{p} 0$, where $\tau_{EATT}^{cluster} = \expesub{\pi_{c(i)}}{\tau_i}$ and 
    $\delta_{cluster} = \frac{N}{N - \sum_{i} \pi_{c(i)} } \frac{N}{\sum_i \pi_{c(i)} } \covsub{1}{\pi_{c(i)}}{Y_i(0)}.$
    
    \item \label{main text, cluster limiting distribution}
    $\frac{ \sqrt{C} (\tauhat - \tau^{cluster}_{EATT} - \delta_{cluster}) }{ \sqrt{ \Omega_{cluster}(2,2) } } \xrightarrow{d} \normnot{0}{1}$,
    for $\Omega_{cluster}(2,2)$ the $(2, 2)$-th element of the matrix $\Omega_{cluster}$ defined in Proposition \ref{prop: clustered OLS with covar, consistency and CLT} (setting $X_i(d) = (1,d)'$).

    \item \label{main text, cluster robust ses}
    Let $\hat{\Omega}_{cluster}$ be the cluster-robust variance estimator \citep{LiangZeger(86)}. 
    If further Assumption \ref{asm: conditions for cluster variance consistency} holds with $X_i(d) = (1,d)'$, then $\hat{\Omega}_{cluster} - \Omega_{cluster}^{est} \xrightarrow{p} 0$, for a matrix  $\Omega_{cluster}^{est}$ such that $\Omega_{cluster}^{est} - \Omega_{cluster}$ is positive semi-definite.
\end{enumerate}
\end{prop}

Finally, we show that the Eicker-Huber-White (EHW) covariance estimator $\hat{V}_{EHW} = \frac{1}{N} \sum_{i} X_i X_i^\prime \hat{\epsilon}_i^2$ need not be valid under the clustered treatment assignment mechanism considered here (Assumption \ref{asm: clustered rejective assignment}). Under clustered treatment assignment mechanism, the EHW variance can be written as 
$$
\hat{V}_{EHW} = \frac{C_1}{N} \frac{1}{C_1} \sum_c D_c \left( \clustersumXXepshatsq(1) \right) + \frac{C_0}{N} \frac{1}{C_0} \sum_c (1-D_c) \left( \clustersumXXepshatsq(0) \right),
$$
where $\clustersumXXepshatsq(d) = \sum_{i \colon c(i) = c} X_i(d) X_i(d)^\prime \hat{\epsilon}_i^2$. 
Define $\clustersumXXepssq(d) = \sum_{i \colon c(i) = c} X_i(d) X_i(d)^\prime \epsilon_i(d)^2$ analogously.
Our next result characterizes the probability limit of $\hat{V}_{EHW}$.

\begin{asm}\label{asm: assumptions for EHW convergence, clustered}
\hfill
\begin{enumerate}
\item[(i)] $\expesub{\pi_c}{ \clustersumXXepssq(1)}$, $\expesub{1 - \pi_c}{ \clustersumXXepssq(0)}$, $N/C$, $C_1/C$ have finite limits with $\lim C_1/C \in (0, 1)$ and $\lim N/C < \infty$.

\item[(ii)] There exists $\tilde{M}_3$ such that $\|\varsub{\tilde \pi_c}{\clustersumXXepssq(d)} \| \leq \tilde{M}_3$ for $d = 0, 1$.

\item[(iii)] There exists $\tilde{M}_4$ such that $\expesub{1}{ \widetilde{ W(d) }_c } < \tilde{M}_4$ and $\expesub{1}{\widetilde{ V(d) }_c} < \tilde{M}_4$ for $d = 0, 1$, where $\widetilde{ W(d) }_c = \sum_{i \colon c(i) = c} \|X_i(1) \epsilon_i(d)\|^2$ and $\widetilde{ V(d) }_c = \sum_{i \colon c(i) = c} \| X_i(d) X_i(d)^\prime \|^2$.
\end{enumerate}
\end{asm}

\begin{prop}\label{prop: EHW invalid for clustered, regression}
If Assumptions \ref{asm: clustered rejective assignment}, \ref{asm: conditions for ols consistency clustered}, and \ref{asm: assumptions for EHW convergence, clustered}(i)-(iii) hold, and $\sum_c \pi_c (1-\pi_c) \rightarrow \infty$, then $\hat{V}_{EHW} - V^{EHW}_{cluster} \xrightarrow{p} 0$ for 
$$
V^{EHW}_{cluster} := \frac{C_1}{N} \expesub{\pi_c}{ \clustersumXXepssq(1)} + \frac{C_0}{N} \expesub{1 - \pi_c}{ \clustersumXXepssq(0)}.
$$
Furthermore, $V_{cluster} - \frac{N}{C} V_{cluster}^{EHW}$ equals 
$$
\frac{C_1}{C} \expesub{\pi_c}{\sum_{i \neq j \colon c(i), c(j) = c} \eta_i(1) \eta_j(1)^\prime} + \frac{C_0}{C} \expesub{1 - \pi_c}{\sum_{i \neq j \colon c(i), c(j) = c} \eta_i(0) \eta_j(0)^\prime} - 
$$
$$
\expesub{1}{( \pi_c \eta_c(1) + (1 - \pi_c) \eta_c(0) ) ( \pi_c \eta_c(1) + (1 - \pi_c) \eta_c(0) )^\prime } - \expesub{1}{\tilde \pi_c} \expesub{\tilde{\pi}_c}{ \eta_c(1) - \eta_c(0)} \expesub{\tilde{\pi}_c}{ \eta_c(1) - \eta_c(0)}^\prime
$$
where $\eta_i(d) = X_i(d) \epsilon_i(d)$ and $\eta_c(d) = \sum_{i \colon c(i) = c} \eta_i(d)$.
\end{prop}

Proposition \ref{prop: EHW invalid for clustered, regression} implies that the usual heteroskedasticity-robust variance estimator can be invalid in large populations if there is clustered treatment assignment (i.e., if $N \neq C$).
To see this, consider the DIM, which corresponds with $X_i = (1, D_i)^\prime$.
Suppose there is no within-cluster heterogeneity in potential outcomes (i.e., $Y_i(d) = Y_{c(i)}(d)$ for all $i$ and $d \in \{0, 1\}$) and all clusters are the same size (i.e., $N_c = N/C$). In this case, $V_{cluster}^{est} = \frac{N}{C} V_{cluster}^{EHW}$. 
If further there is no across-cluster treatment effect heterogeneity nor heterogeneity in cluster-specific treatment probabilities, $V_{cluster} = V_{cluster}^{est}$ by the same logic as Corollary \ref{cor: shat conservative under constant TEs} in the main text for the non-clustered case, and the heteroskedasticity-robust variance estimator is thus too small whenever $N/C > 1$. 
If there is either treatment effect heterogeneity or heterogeneity in cluster-specific treatment probabilities, then $V_{cluster} \leq V_{cluster}^{est}$ (generally with strict inequality), in which case the heteroskedasticity-robust variance estimator is valid whenever $C/N \geq V_{cluster}/V_{cluster}^{est}$.

\subsection{Proofs of Results for General OLS Estimators under Clustering}\label{sec: OLS w/ clustering proofs}

\paragraph{Proof of Proposition \ref{prop: clustered OLS with covar, consistency and CLT}}
\begin{proof}
To establish claim (1), let $p_c^*$ be the limit of $\frac{C_1}{C}$, let $\mu_{\pi_c}\left[\clustersumXX(1) \right]$ be the limit of $\expesub{\pi_c}{\clustersumXX(1)}$, and define $\mu_{\pi_c}\left[\cdot\right]$ and $\mu_{1-\pi_c}\left[\cdot\right]$ of other variables analogously. 
Let 
$$ \beta^*_{cluster} = \left( p_c^* \mu_{\pi_c}\left[\clustersumXX(1)^\prime\right] + (1-p_c^*) \mu_{1-\pi_c}\left[\clustersumXX(0)^\prime\right] \right)^{-1} \left( p_c^* \mu_{\pi_c}\left[\clustersumXY(1) \right] + (1-p_c^*) \mu_{1-\pi_c}\left[\clustersumXY(0)\right] \right).
$$
\noindent It is immediate from Assumption \ref{asm: conditions for ols consistency clustered}(i)-(ii) that $\beta_{cluster} \rightarrow \beta^*_{cluster}$, so it suffices to show that $\betahat \xrightarrow{p} \beta^*_{cluster}$. 
Note that we can write $\hat\beta$ as 
$$
\left( \frac{C_1}{C} \frac{1}{C_1} \sum_c D_c \widetilde{XX'}(1) + \frac{C_0}{C} \frac{1}{C_0} \sum_c (1-D_c) \clustersumXX(0) \right)^{-1} \left( \frac{C_1}{C} \frac{1}{C_1} \sum_c D_c \clustersumXY(1) + \frac{C_0}{C} \frac{1}{C_0} \sum_c (1-D_c) \clustersumXY(0) \right).
$$
Using Theorem 6.1 in \citet{hajek_asymptotic_1964} as in the proof to Lemma \ref{lem: variance of tauhat}, we have that 
\begin{align*}
\varsub{R}{ \frac{1}{C_1} \sum_c D_c (\clustersumXX(1))_{jk}} &= (1+o(1)) C_1^{-2} \left(\sum_c \tilde\pi_c \right) \varsub{\tilde\pi_c}{ (\clustersumXX(1))_{jk}} \\
&\leq (1+o(1))C_1^{-1} M \rightarrow 0, 
\end{align*}

\noindent where we obtain the inequality from Assumption \ref{asm: conditions for ols consistency clustered}(iii) combined with the fact that $\tilde\pi_c \leq \pi_c$ for all $c$ and thus $\sum_c \tilde\pi_c \leq \sum_c \pi_c = C_1$. Combining the previous display with Chebyshev's inequality, we obtain that $\frac{1}{C_1} \sum_c D_c \clustersumXX(1) - \expeR{\frac{1}{C_1} \sum_c D_c \clustersumXX(1)} \xrightarrow{p} 0$. 
But $\expeR{\frac{1}{C_1} \sum_c D_c \clustersumXX(1)} = \expesub{\pi_c}{\clustersumXX(1)} \rightarrow \mu_{\pi_c}\left[\clustersumXX(1)\right] $, and hence $\frac{1}{C_1} \sum_c D_c \clustersumXX(1)  \xrightarrow{p} \mu_{\pi_c}\left[\clustersumXX(1)\right] $. 
An analogous argument yields that $\frac{1}{C_0} \sum_c (1-D_c) \clustersumXX(0) \xrightarrow{p} \mu_{1-\pi_c}\left[\clustersumXX(0) \right]$, $\frac{1}{C_1} \sum_c D_c \clustersumXY(1) \xrightarrow{p} \mu_{\pi_c}\left[\clustersumXY(1)\right]$, and $\frac{1}{C_0} \sum_c (1-D_c) \clustersumXY(0) \xrightarrow{p} \mu_{1-\pi_c}\left[\clustersumXY(0)\right]$. 
These convergences together with the continuous mapping theorem yield that $\betahat \xrightarrow{p} \beta^*_{cluster}$, as we wished to show. 

To show the second claim, define $\epsilon_i = D_i \epsilon_i(1) + (1 - D_i) \epsilon_i(0)$ (and recall that  $\epsilon_i(d) = Y_i(d) - X_i(d)^\prime \beta_{cluster}$), so that 
$$
\betahat = \beta_{cluster} + \left( \frac{1}{C} \sum_i X_i X_i^\prime \right)^{-1} \left( \frac{1}{C} \sum_i X_i \epsilon_i \right).
$$
and 
$$
\sqrt{C} (\betahat - \beta_{cluster}) = \left( \frac{1}{C} \sum_i X_i X_i^\prime \right)^{-1} \left( \frac{1}{\sqrt{C}} \sum_i X_i \epsilon_i \right).
$$
In the proof of claim (1), we established that $\left( \frac{1}{C} \sum_i X_i X_i^\prime \right)^{-1}$ is consistent for $\expeR{\frac{1}{C} \sum_i X_i X_i^\prime}^{-1}$. 
We therefore focus on establishing the asymptotic normality of $\frac{1}{\sqrt{C}} \sum_i X_i \epsilon_i$. 
Towards this, notice that standard arguments for linear projections imply that 
\begin{equation}\label{eqn: cluster CLT eqn 1}
\expesub{R}{\frac{1}{C} \sum_i X_i \epsilon_i} = \frac{C_1}{C} \expesub{\pi_c}{ \clustersumXeps(1)} + \frac{C_0}{C} \expesub{1-\pi_c}{\clustersumXeps(0)} = 0,
\end{equation}
where $\clustersumXeps(d) = \sum_{i \colon c(i) = c} X_i(d) \epsilon_i(d)$ as before.
By adding/subtracting $C_1 \expesub{\pi_c}{\clustersumXeps(0)}$ from the previous display and applying the identity $C_1 \expesub{\pi_c}{v_c} + C_0 \expesub{1-\pi_c}{v_c} = C \expesub{1}{v_c}$ for any cluster-level attribute $v_c$, we obtain that
$$
C_1 \expesub{\pi_c}{ \clustersumXeps(1) - \clustersumXeps(0) } + \sum_c  \clustersumXeps(0) = 0.
$$
It therefore follows that 
\begin{align*}
\sum_i X_i \epsilon_i &= \sum_c D_c \clustersumXeps(1) + \sum_c (1-D_c) \clustersumXeps(0) \\
&= \sum_c D_c \left( \left( \clustersumXeps(1) - \clustersumXeps(0) \right) -  \expesub{\pi_c}{ \clustersumXeps(1) - \clustersumXeps(0) } \right)
\end{align*}
Therefore, $\sum_i X_i \epsilon_i$ can be represented as Horvitz-Thompson estimator under clustered rejective sampling. 
Applying the multivariate generalization of Theorem 1 in \citet{berger_rate_1998} as in the proof to Proposition \ref{subprop: multivariate clt}, we therefore conclude that 
$$
V_{cluster}^{-1/2} \frac{1}{\sqrt{C}} \sum_i X_i \epsilon_i \xrightarrow{d} \normnot{0}{I},
$$
where $V_{cluster}$ is defined in the statement of claim (2). Claim (2) follows by applying Slutsky's lemma.
\end{proof}

\paragraph{Proof of Proposition \ref{prop: clustered OLS with covar, var consistency and conservativeness}}
\begin{proof}
To show the first claim, observe that 
$$\hat{V}_{cluster} = \frac{C_1}{C} \frac{1}{C_1} \sum_c D_c \clustersumXepshat(1) \clustersumXepshat(1)^\prime + \frac{C_0}{C} \frac{1}{C_0} \sum_c (1-D_c) \clustersumXepshat(0) \clustersumXepshat(0)^\prime.$$
Furthermore, $\clustersumXepshat(d) = \clustersumXeps(d) - \clustersumXX(d) (\hat\beta- \beta_{cluster})$. 
It follows that
\begin{align*}
&\frac{1}{C_1} \sum_c D_c \clustersumXepshat(1) \clustersumXepshat(1)' =  \underbrace{\frac{1}{C_1} \sum_c D_c \clustersumXeps(1) \clustersumXeps(1)^\prime}_{=(A)}- \\ & \hspace{1cm} \underbrace{\frac{1}{C_1} \sum_c D_c \widetilde{X\epsilon}_c(1) (\hat\beta - \beta_{cluster})' \widetilde{XX'}_c(1)'}_{=(B)} - \underbrace{\frac{1}{C_1} \sum_c D_c \left(\widetilde{X\epsilon}_c(1) (\hat\beta - \beta_{cluster})' \widetilde{XX'}_c(1)'\right)'}_{=(B')}+\\& \hspace{1cm} 
\underbrace{\frac{1}{C_1} \sum_c D_c \clustersumXX(1)(\hat\beta - \beta_{cluster})(\hat\beta - \beta_{cluster})' \clustersumXX(1)'}_{=(C)} \numberthis \label{eqn: cluster var estimates terms}
\end{align*}

\noindent Consider the term labeled (A) in (\ref{eqn: cluster var estimates terms}) and observe that 
\begin{align*}
\left|\left| \vR{\frac{1}{C_1} \sum_c D_c \clustersumXeps(1) \clustersumXeps(1)^\prime} \right|\right| &= (1+o(1))C_1^{-2} (\sum_c \tilde\pi_c) \left|\left| \varsub{\tilde\pi_c}{ \clustersumXeps(1) \clustersumXeps(1)^\prime } \right|\right|    \\
&\leq (1+o(1)) C_1^{-1} \tilde{M}_1 \rightarrow 0,
\end{align*}
\noindent where we use Assumption \ref{asm: conditions for cluster variance consistency}(ii) to bound $||\varsub{\tilde\pi_c}{\clustersumXeps(1)  \clustersumXeps(1)^\prime }|| $. Hence, by Chebyshev's inequality, $\frac{1}{C_1} \sum_c D_c \clustersumXeps(1) \clustersumXeps(1)^\prime \xrightarrow{p} \mu_{\pi_c}\left[ \clustersumXeps(1) \clustersumXeps(1)^\prime \right]$, where we define $\mu_{\pi_c}[\cdot]$ as in the proof to Proposition \ref{prop: clustered OLS with covar, consistency and CLT}. 
Next, consider the term labeled $(C)$ in (\ref{eqn: cluster var estimates terms}). Recall that the Frobenius norm is sub-multiplicative, so that $\|QR\| \leq \|Q\| \|R\|$ for any matrices $Q,R$. Hence, we have that 
\begin{align*}
\|(C)\| &\leq \frac{1}{C_1} \sum_c D_c || \clustersumXX(1) (\hat\beta - \beta_{cluster})(\hat\beta - \beta_{cluster})' \clustersumXX(1)' ||\\
&\leq ||(\hat\beta - \beta_{cluster})(\hat\beta - \beta_{cluster})'|| \frac{1}{C_1} \sum_c D_c || \clustersumXX(1) ||^2 \\
&\leq ||(\hat\beta - \beta_{cluster})(\hat\beta - \beta_{cluster})'|| \frac{C}{C_1} \frac{1}{C}\sum_c || \clustersumXX(1) ||^2 \\
& \leq ||(\hat\beta - \beta_{cluster})(\hat\beta - \beta_{cluster})'|| \frac{C}{C_1} \tilde{M}_2 \xrightarrow{p} 0
\end{align*}
\noindent where the last inequality uses Assumption \ref{asm: conditions for cluster variance consistency}(iii), and we use the fact that $C/C_1$ has a finite limit by Assumption \ref{asm: conditions for ols consistency clustered}(i) and $\hat\beta - \beta_{cluster} \xrightarrow{p} 0$ by Proposition \ref{prop: clustered OLS with covar, consistency and CLT}. 
Finally, 
\begin{align*}
\|(B)\| & \leq  \frac{1}{C_1} \sum_c D_c || \widetilde{X\epsilon}_c(1) (\hat\beta - \beta_{cluster})' \widetilde{XX'}_c(1)' ||\\
&\leq \frac{1}{C_1} \sum_c D_c || \widetilde{X\epsilon}_c(1) || \cdot ||\clustersumXX(1)|| \cdot || (\hat\beta - \beta_{cluster})|| \\
&\leq \frac{C}{C_1}\frac{1}{C} \sum_c || \widetilde{X\epsilon}_c(1) || \cdot ||\clustersumXX(1)|| \cdot || (\hat\beta - \beta_{cluster})|| \\
&\leq \frac{C_1}{C} \sqrt{\frac{1}{C} \sum_c  || \widetilde{X\epsilon}_c(1) ||^2 }\cdot \sqrt{ \frac{1}{C} \sum_c  ||\clustersumXX(1)||^2 } \cdot || (\hat\beta - \beta_{cluster})|| \\
& \leq \frac{C_1}{C} \tilde{M}_2 ||\hat\beta - \beta_{cluster}|| \xrightarrow{p} 0,
\end{align*}
\noindent where the fourth inequality uses Cauchy-Schwarz, the fifth inequality uses Assumption \ref{asm: conditions for cluster variance consistency}(iii) and we use the fact that $\hat\beta - \beta_{cluster} \xrightarrow{p} 0$ as shown above. 
We have thus shown that $\frac{1}{C_1} \sum_c D_c \clustersumXepshat(1) \clustersumXepshat(1)^\prime \xrightarrow{p} \mu_{\pi_c}\left[ \clustersumXeps(1) \clustersumXeps(1)^\prime \right]$. 
By analogous argument, we can show that $\frac{1}{C_0} \sum_c (1-D_c) \clustersumXepshat(0) \clustersumXepshat(0)^\prime \xrightarrow{p} \mu_{1-\pi_c}\left[ \clustersumXeps(0) \clustersumXeps(0)^\prime \right]$. The first part of the result then follows from the continuous mapping theorem.

To show the second claim, let $\eta_c(d) = \sum_{i:c(i) = c} X_i(d) \epsilon_i(d)$, $\dot{\eta}_c(1) = \dot{\eta}_c(1) - \expesub{\pi_c}{\eta_c(1)}$, and $\dot{\eta}_c(0) = \dot{\eta}_c(0) - \expesub{1-\pi_c}{\eta_c(0)}$. Then,
\begin{align*}
V_{cluster} =& \frac{1}{C} \sum_c \pi_c(1-\pi_c) \left(\eta_c(1) - \eta_c(0) - \expesub{\tilde\pi_c}{\eta_c(1) - \eta_c(0)}\right)\left(\eta_c(1) - \eta_c(0) - \expesub{\tilde\pi_c}{\eta_c(1) - \eta_c(0)}\right)'\\
\leq& \frac{1}{C} \sum_c \pi_c(1-\pi_c) \left(\dot{\eta}_c(1) - \dot{\eta}_c(0)\right) \left(\dot{\eta}_c(1) - \dot{\eta}_c(0)\right)' \\
=&  \frac{1}{C} \left( \sum_c \pi_c \dot{\eta}_c(1) \dot{\eta}_c(1)' + \sum_c (1-\pi_c)\dot{\eta}_c(0)\dot{\eta}_c(0)' - \right. \\& \left. \hspace{1cm} \left(\sum_c \pi_c^2 \dot{\eta}_c(1)\dot{\eta}_c(1)' + \sum_c (1-\pi_c)^2 \dot{\eta}_c(0)\dot{\eta}_c(0)' + \sum_c \pi_c(1-\pi_c) (\dot{\eta}_c(1) \dot{\eta}_c(0)' + \dot{\eta}_c(0) \dot{\eta}_c(1)')  \right) \right) \\
=& \frac{C_1}{C} \varsub{\pi_c}{ \eta_c(1) } + \frac{C_0}{C} \varsub{1-\pi_c}{\eta_c(0)}  - \frac{1}{C}\sum_c (\pi_c \dot{\eta}_c(1) + (1-\pi_c) \dot{\eta}_c(0))(\pi_c \dot{\eta}_c(1) + (1-\pi_c)\dot{\eta}_c(0))'\\ 
\leq & \frac{C_1}{C} \expesub{\pi_c}{ \eta_c(1) \eta_c(1)' } + \frac{C_0}{C} \expesub{1-\pi_c}{\eta_c(0)\eta_c(0)'}= V_{cluster}^{est}.
\end{align*}
\end{proof}

\paragraph{Proof of Corollary \ref{cor: consistency of omega cluster}}
\begin{proof}
The proof is immediate from Proposition \ref{prop: clustered OLS with covar, var consistency and conservativeness} combined with the fact that $\frac{1}{C}\sum_i X_i X_i' - \expeR{\frac{1}{C}\sum_i X_i X_i'} \xrightarrow{p} 0$ as shown in the proof to Proposition \ref{prop: clustered OLS with covar, consistency and CLT}.
\end{proof}

\paragraph{Proof of Proposition \ref{prop: sdim limiting dist, clustering}}
\begin{proof}
To prove these results, we will show that the second-element of $\beta_{cluster}$ defined in Proposition \ref{prop: clustered OLS with covar, consistency and CLT} equals $\tau_{cluster}^{EATT} + \delta_{cluster}$ when $X_i(d) = (1, d)^\prime$ and $X_i = X_i(D_i) = (1, D_i)^\prime$. 
The stated claims then immediately follow by applying Proposition \ref{prop: clustered OLS with covar, consistency and CLT}.
Defining $N_1^C = \sum_c \pi_c N_c = \sum_i \pi_{c(i)}$, $N_0^C = N - N_1^C = \sum_i (1 - \pi_{c(i)})$,  observe that
\begin{align*}
& \left( \frac{C_1}{C} \expesub{\pi_c}{ \clustersumXX(1) } + \frac{C_0}{C} \expesub{1-\pi_c}{\clustersumXX(0)} \right)^{-1} = \frac{C}{N_0^C N_1^C} \begin{pmatrix} N_1^C & -N_1^C \\ -N_1^C & N \end{pmatrix} 
\end{align*}
and 
\begin{align*}
& \frac{C_1}{C} \expesub{\pi_c}{ \clustersumXY(1)  } + \frac{C_0}{C} \expesub{1-\pi_c}{\clustersumXY(0)} = C^{-1} \sum_i \begin{pmatrix} Y_i(0) + \pi_{c(i)} \tau_i \\ \pi_{c(i)} (Y_i(0) + \tau_i). \end{pmatrix}
\end{align*}
Multiplying out, we therefore arrive at
$$
\beta_{cluster} = \left( \frac{C_1}{C} \expesub{\pi_c}{\clustersumXX(1)} + \frac{C_0}{C} \expesub{1-\pi_c}{\clustersumXX(0)} \right)^{-1} \left( \frac{C_1}{C} \expesub{\pi_c}{ \clustersumXY(1) } + \frac{C_0}{C} \expesub{1-\pi_c}{\clustersumXY(0)} \right) = 
$$
$$
\frac{1}{N_0^C N_1^C} \begin{pmatrix} N_1^C & -N_1^C \\ -N_1^C & N \end{pmatrix} \sum_i \begin{pmatrix} Y_i(0) + \pi_{c(i)} \tau_i \\ \pi_{c(i)} (Y_i(0) + \tau_i) \end{pmatrix} = 
\begin{pmatrix} \frac{1}{N_0^C} \sum_i (1 - \pi_{c(i)}) Y_i(0) \\ \frac{1}{N_1^C} \sum_i \pi_{c(i)} \tau_i + \sum_i \left( \frac{\pi_{c(i)}}{N_1^C} - \frac{1 - \pi_{c(i)}}{N_0^C} \right) Y_i(0) \end{pmatrix}.
$$
Re-arranging the second element then yields
$$
\beta_{cluster, 2} = \expesub{\pi_{c(i)}}{\tau_i} + \frac{N}{\sum_i \pi_{c(i)}} \frac{N}{N - \sum_{i} \pi_{c(i)}} \covsub{1}{\pi_{c(i)}}{Y_i(0)},
$$
which gives the first claim in the Proposition. The second and third claims follow immediately from Proposition \ref{prop: clustered OLS with covar, var consistency and conservativeness} with $X_i(d) = (1, d)^\prime$ and $X_i = X_i(D_i) = (1, D_i)^\prime$.
\end{proof}

\paragraph{Proof of Proposition \ref{prop: EHW invalid for clustered, regression}}
\begin{proof}
To show the first claim, it is immediate from Assumption \ref{asm: assumptions for EHW convergence, clustered}(i) that $V^{EHW}_{cluster}$ converges to 
$$
(1/n_c^*) p_c^* \mu_{\pi_c}[\clustersumXXepssq(1)] + (1/n_c^*) (1 - p_c^*) \mu_{1-\pi_c}[\clustersumXXepssq(0)],
$$
where $n_c^* = \lim N/C$, $p_c^* = \lim C_1/C$, and $\mu_{\pi_c}[\cdot]$ is defined as in the proof to Proposition \ref{prop: clustered OLS with covar, consistency and CLT}.
It therefore suffices to show that $\hat{V}_{EHW}$ converges in probability to the same limit.
To show this, recall that $\hat\epsilon_i = D_i \hat{\epsilon}_i(1) + (1-D_i) \hat\epsilon_i(0)$ for $\hat{\epsilon}_i(d) = \epsilon_i(d) - X_{i}(d)' (\betahat - \beta_{cluster})$ and $X_i(d) \hat{\epsilon}_i(d) = X_i(d) \epsilon_i(d) - X_i(d) X_i(d)^\prime (\betahat - \beta_{cluster})$. 
Therefore, we can write $\frac{C_1}{N} \frac{1}{C_1} \sum_c D_c \left( \clustersumXXepshatsq(1) \right)$ as 
$$
\frac{C}{N} \frac{C_1}{C} \underbrace{\frac{1}{C_1} \sum_c D_c \clustersumXXepssq(1)}_{(A)} + \frac{C}{N} \underbrace{ \frac{1}{C} \sum_c D_c  \sum_{i \colon c(i) = c} X_i(1) \epsilon_i(1) (\betahat - \beta_{cluster})^\prime X_i(1) X_i(1)^\prime }_{(B)} +
$$
$$
\frac{C}{N}  \underbrace{\frac{1}{C} \sum_c D_c \sum_{i \colon c(i) = c} X_i(1) X_i(1)^\prime (\betahat - \beta_{cluster}) X_i^\prime(1) \epsilon_i(1) }_{(B\prime)} + 
$$
$$
\frac{C}{N} \underbrace{\frac{1}{C} \sum_c D_c \left( \sum_{i \colon c(i) = c} X_i(1) X_i^\prime(1) (\betahat - \beta_{cluster}) (\betahat - \beta_{cluster})^\prime X_i(1) X_i^\prime(1) \right)}_{(C)}.
$$
First, consider the term (A), and observe that 
\begin{align*}
\left|\left| \vR{\frac{1}{C_1} \sum_c D_c \clustersumXXepssq(1)}\right|\right| &= (1 + o(1)) C_1^{-2} \left( \sum_c \tilde{\pi}_c \right) \left|\left| \varsub{\tilde \pi_c}{\clustersumXXepssq(1)} \right|\right| \\
&\leq (1 + o(1)) C_1^{-1} \tilde{M}_3 \rightarrow 0,
\end{align*}
where we use Assumption \ref{asm: assumptions for EHW convergence, clustered}(ii) to bound $\left|\left| \varsub{\tilde \pi_c}{\clustersumXXepssq(1)}\right|\right|$. 
Hence, $\frac{1}{C_1} \sum_c D_c \clustersumXXepssq(1) \xrightarrow{p} \mu_{\pi_c}\left[\clustersumXXepssq\right]$ by Chebyshev's Inequality. 
Next, consider term (B) and observe that
\begin{align*}
\|(B)\| &\leq  \frac{1}{C} \sum_c D_c \sum_{i \colon c(i) = c} \|  X_i(1) \epsilon_i(1) (\betahat - \beta_{cluster})^\prime X_i(1) X_i(1)^\prime \|  \\
&\leq \| \betahat - \beta_{cluster} \| \left( \frac{1}{C} \sum_c D_c \sum_{i \colon c(i) = c} \|  X_i(1) \epsilon_i(1) \| \| X_i(1) X_i(1)^\prime \| \right) \\
&\leq \| \betahat - \beta_{cluster} \| \left( C^{-1} \sum_c \widetilde{ W(1) }_c \widetilde{V(1)}_c \right) \\
&\leq \| \betahat - \beta_{cluster} \| \sqrt{ C^{-1} \sum_c \widetilde{ W(1) }_c } \sqrt{C^{-1} \sum_c \widetilde{V(1)}_c } \\
& \leq \| \betahat - \beta_{cluster} \|  \tilde{M}_4
\end{align*}
where the first inequality applies the triangle inequality, the second inequality applies the submultiplicative property of the Frobenius norm, the third inequality uses the positivity of the norm, and the fourth inequality uses the Cauchy-Schwarz inequality. Since $\betahat - \beta_{cluster} \xrightarrow{0}$, it follows that $\|(B)\| \xrightarrow{p} 0$  by Assumption \ref{asm: assumptions for EHW convergence, clustered}(iii).
The analogous argument gives that (B') converges in probability to zero. 
Finally, consider term (C) and observe that 
\begin{align*}
\| (C) \| &\leq \frac{1}{C_1} \sum_c D_c \sum_{i \colon c(i) = c} \| X_i(1) X_i^\prime(1) (\betahat - \beta_{cluster}) (\betahat - \beta_{cluster})^\prime X_i(1) X_i^\prime(1) \| \\
&\leq \|(\betahat - \beta_{cluster}) (\betahat - \beta_{cluster})^\prime\| \left( \frac{1}{C_1} \sum_c D_c \sum_{i \colon c(i) = c} \| X_i(1) X_i^\prime(1)\|^2 \right) \\
&= \| (\betahat - \beta_{cluster}) (\betahat - \beta_{cluster})^\prime \| \left( \frac{1}{C_1} \sum_c D_c \widetilde{ V(d) }_c \right) \\
&\leq \| (\betahat - \beta_{cluster}) (\betahat - \beta_{cluster})^\prime \| \frac{C}{C_1} \left( \frac{1}{C} \sum_c \widetilde{ V(d)}_c \right) \\
& \leq \| (\betahat - \beta_{cluster}) (\betahat - \beta_{cluster})^\prime \| \frac{C}{C_1} \tilde{M}_4, 
\end{align*}
which converges in probability to zero since $\betahat - \beta_{cluster} \xrightarrow{p} 0$ and $\frac{C_1}{C}$ has a finite limit.
Putting this together, it follows that 
$\frac{C}{N} \frac{C_1}{C} \frac{1}{C_1} \sum_c D_c \left( \clustersumXXepshatsq(1) \right) \xrightarrow{p} (1/n_c^*) p_c^* \mu_{\pi_c}[\clustersumXXepssq(1)]$ by the continuous mapping theorem. By the same argument, we can show
$ \frac{C}{N} \frac{C_0}{C} \frac{1}{C_0} \sum_c (1-D_c) \left( \clustersumXXepshatsq(0) \right) \xrightarrow{p} (1/n_c^*) (1 - p_c^*) \mu_{1-\pi_c}[\clustersumXXepssq(0)].$
The first claim then follows by another application of the continuous mapping theorem.

To show the second claim, we first observe that $V_{cluster}$ can be expanded into
$$
C^{-1} \sum_{c} \pi_c (1 - \pi_c) \left( \eta_c(1) - \eta_c(0) - \expesub{\tilde{\pi}_c}{ \eta_c(1) - \eta_c(0)} \right) \left( \eta_c(1) - \eta_c(0) - \expesub{\tilde{\pi}_c}{ \eta_c(1) - \eta_c(0)} \right)^\prime = 
$$
$$
\underbrace{C^{-1} \sum_c \pi_c (1 - \pi_c) (\eta_c(1) - \eta_c(0)) (\eta_c(1) - \eta_c(0))^\prime}_{(a)} - \left( C^{-1} \sum_c \tilde \pi_c \right) \expesub{\tilde{\pi}_c}{ \eta_c(1) - \eta_c(0)} \expesub{\tilde{\pi}_c}{ \eta_c(1) - \eta_c(0)}^\prime.
$$
Further expanding out, notice that (a) equals
$$
C^{-1} \sum_c \pi_c (1 - \pi_c) \left( \eta_c(1) \eta_c(1)^\prime + \eta_c(0) \eta_c(0)^\prime - \eta_c(1) \eta_c(0)^\prime - \eta_c(0) \eta_c(1)^\prime \right) = 
$$
$$
C^{-1} \sum_{c} \pi_c \eta_c(1) \eta_c(1)^\prime + C^{-1} \sum_c (1 - \pi_c) \eta_c(0) \eta_c(0)^\prime - 
$$
$$
C^{-1} \sum_c \left( \pi_c^2 \eta_c(1) \eta_c(1)^\prime + (1 - \pi_c)^2 \eta_c(0) \eta_c(0)^\prime + \pi_c (1 - \pi_c) (\eta_c(1) \eta_c(0)^\prime + \eta_c(0) \eta_c(1)^\prime) \right) = 
$$
$$
\underbrace{C^{-1} \sum_{c} \pi_c \eta_c(1) \eta_c(1)^\prime + C^{-1} \sum_c (1 - \pi_c) \eta_c(0) \eta_c(0)^\prime}_{(b)} - C^{-1} \sum_c ( \pi_c \eta_c(1) + (1 - \pi_c) \eta_c(0) ) ( \pi_c \eta_c(1) + (1 - \pi_c) \eta_c(0) )^\prime.
$$
Then, using the identity $\eta_c(d) \eta_c(d)^\prime = \sum_{i \colon c(i) = c} \sum_{j \colon c(j) = c} \eta_i(d) \eta_j(d)^\prime = \sum_{i \colon c(i) = c} \eta_i(d) \eta_i(d)^\prime  + \sum_{i \neq j: c(i), c(j) = c} \eta_i(d) \eta_j(d)^\prime$, we further expand out (b) as
$$
C^{-1} \sum_c \pi_c \eta_c(1) \eta_c(1)^\prime + C^{-1} \sum_c (1 - \pi_c) \eta_c(0) \eta_c(0)^\prime = 
$$
$$
C^{-1} \sum_c \pi_c \sum_{i \colon c(i) = c} \eta_i(1) \eta_i(1)^\prime + C^{-1} \sum_c (1 - \pi_c) \sum_{i \colon c(i) = c} \eta_i(0) \eta_i(0)^\prime + 
$$
$$
C^{-1} \sum_{c} \pi_c \sum_{i\neq j \colon c(i), c(j) = c} \eta_i(1) \eta_j(1)^\prime + C^{-1} \sum_{c} (1-\pi_c) \sum_{i \neq j \colon c(i), c(j) = c} \eta_i(0) \eta_j(0)^\prime = 
$$
$$
\frac{N}{C} V_{cluster}^{EHW} + \frac{C_1}{C} \expesub{\pi_c}{\sum_{i \neq j \colon c(i), c(j) = c} \eta_i(1) \eta_j(1)^\prime} + \frac{C_0}{C} \expesub{1 - \pi_c}{\sum_{i \neq j \colon c(i), c(j) = c} \eta_i(0) \eta_j(0)^\prime}.
$$
Putting this altogether, we therefore have shown that $V_{cluster}$ equals 
$$
\frac{N}{C} V_{cluster}^{EHW} + \frac{C_1}{C} \expesub{\pi_c}{\sum_{i \neq j \colon c(i), c(j) = c} \eta_i(1) \eta_j(1)^\prime} + \frac{C_0}{C} \expesub{1 - \pi_c}{\sum_{i \neq j \colon c(i), c(j) = c} \eta_i(0) \eta_j(0)^\prime} - 
$$
$$
\expesub{1}{( \pi_c \eta_c(1) + (1 - \pi_c) \eta_c(0) ) ( \pi_c \eta_c(1) + (1 - \pi_c) \eta_c(0) )^\prime } - \expesub{1}{\tilde \pi_c} \expesub{\tilde{\pi}_c}{ \eta_c(1) - \eta_c(0)} \expesub{\tilde{\pi}_c}{ \eta_c(1) - \eta_c(0)}^\prime.
$$
\end{proof}

\section{Extension to Vector-Valued Outcomes}\label{section: appendix extension to multiple outcomes}

In this appendix, we generalize our results for the DIM estimator in Sections \ref{sec: expectation of sdim}-\ref{sec: distribution of DIM} to the vector-valued outcomes case. We apply these results to analyze IV estimators from a design-based perspective in Section \ref{sec: iv} of the main text, and non-staggered DID estimators with multiple time periods in \ref{section: multi-period DiD}.

We extend our notation from the main text, so that $\mathbf{Y}_i \in \reals^K$ is the vector-valued outcome. For a fixed vector-valued characteristic $\mathbf{X}_i$, $\expesub{w}{\boldsymbol X_i} := \frac{1}{\sum_i w_i} \sum_i w_i \boldsymbol X_i$ and $\varsub{w}{\boldsymbol X_i} = \frac{1}{\sum_i w_i} \sum_i \left( \boldsymbol X_i - \expesub{w}{\boldsymbol X_i} \right)\left( \boldsymbol X_i - \expesub{w}{\boldsymbol X_i} \right)'$.
Further, as shorthand, define $S_{1,w} := \varsub{w}{\boldsymbol Y_i(1)}$, $S_{0,w} := \varsub{w}{\boldsymbol Y_i(0)}$, $S_{10,w} := \expesub{w}{ (\mathbf{Y}_i(1) - \expesub{w}{ \mathbf{Y}_i(1) })(\mathbf{Y}_i(0) - \expesub{w}{ \mathbf{Y}_i(0) })' }$ to be the weighted finite-population variances and covariance of $\boldsymbol Y_i(1)$ and $\boldsymbol Y_i(0)$. Finally, the vector-valued ATE is $\boldsymbol{\tau}_{ATE} := \frac{1}{N} \sum_i (\mathbf{Y}_i(1) - \mathbf{Y}_i(0))$, and the vector-valued EATT is $\boldsymbol{\tau}_{EATT} := \frac{1}{N_1} \sum_i \pi_i (\mathbf{Y}_i(1) - \mathbf{Y}_i(0))$. 

We analyze the behavior over the randomization distribution (Assumption \ref{asm: rejective assignment probability}) of the vector-valued DIM estimator $\boldtauhat  = \frac{1}{N_1} \sum_i D_i \boldsymbol Y_i - \frac{1}{N_0} \sum_i (1-D_i) \boldsymbol Y_i$ and associated variance estimators
\begin{align*}
& \mathbf{\hat{s}} := \frac{1}{N_1} \mathbf{\hat{s}}_1 + \frac{1}{N_0} \mathbf{\hat{s}}_0, \\
& \mathbf{\hat{s}}_1 := \frac{1}{N_1} \sum_i D_i (\boldsymbol Y_i - \mathbf{ \bar{Y} }_1) (\boldsymbol Y_i - \mathbf{ \bar{Y} }_1)' , \hspace{0.5cm} \mathbf{\hat{s}}_0 := \frac{1}{N_0} \sum_i (1-D_i) (\boldsymbol Y_i - \mathbf{ \bar{Y} }_0) (\boldsymbol Y_i - \mathbf{ \bar{Y} }_0)', 
\end{align*}
where $\mathbf{ \bar{Y} }_1 := \frac{1}{N_1}\sum_i D_i \boldsymbol Y_i$ and $ \mathbf{ \bar{Y} }_0 := \frac{1}{N_0}\sum_i (1-D_i) \boldsymbol Y_i$.

We introduce the following regularity conditions on the sequence of finite populations.

\begin{asm} \label{asm: variance matrices converge}
Suppose $N_1 / N \rightarrow p_1 \in (0,1)$, and $S_{1,w}, S_{0,w}, S_{10,w}$ have finite limits for $w \in \{\pi, 1-\pi, \tilde\pi \}$.
\end{asm}
\begin{asm}\label{asm: asm for consistent variance estimation - vector version}
$\max_{1 \leq i \leq N} || \mathbf Y_i(1) - \expesub{\pi}{\mathbf Y_i(1)} ||^2/N \rightarrow 0$ and $\max_{1 \leq i \leq N} || \mathbf Y_i(0) - \expesub{1-\pi}{\mathbf Y_i(0)} ||^2/N \rightarrow 0$, where $|| \cdot ||$ is the Euclidean norm.
\end{asm} 

\begin{asm}\label{asm: multivariate lindeberg type condition}
Let $\boldsymbol \Ytilde_i = \frac{1}{N_1} \boldsymbol Y_i(1) + \frac{1}{N_0} \boldsymbol Y_i(0)$, and let $\lambda_{min}$ be the minimal eigenvalue of $\Sigma_{\tilde{\pi}} = \varsub{\tilde{\pi}}{ \boldsymbol \Ytilde_i }$. Assume $\lambda_{min} >0$ and for all $\epsilon > 0$,
$$\frac{1}{ \lambda_{min} } \expesub{\tilde{\pi}}{ \left|\left| \boldsymbol \Ytilde_i - \expesub{\tilde{\pi}}{ \boldsymbol \Ytilde_i }\right|\right|^2 \cdot 1\left[ \left|\left| \boldsymbol \Ytilde_i - \expesub{\tilde{\pi}}{\boldsymbol \Ytilde_i}\right|\right| > \sqrt{ \sum_i \pi_i (1-\pi_i)  \cdot \lambda_{min}} \cdot \epsilon \right]  } \rightarrow 0.$$
\end{asm}

\noindent Assumption \ref{asm: variance matrices converge} requires that the fraction of treated units and the (weighted) variance and covariances of the potential outcomes have finite limits along the sequence of finite populations. 
Assumption \ref{asm: asm for consistent variance estimation - vector version} is a multivariate analog of Assumption \ref{asm: reg conditions for CLT and var consistency}\ref{asm for consistent variance estimation} in that it requires that no single observation dominate the $\pi$ or $(1-\pi)$-weighted variance of the potential outcomes. Assumption \ref{asm: multivariate lindeberg type condition} is a multivariate generalization of the Lindeberg-type condition in Assumption \ref{asm: reg conditions for CLT and var consistency}\ref{lindeberg type condition}.

\begin{prop}[Results for vector-valued outcomes] \label{prop: results for multivariate outcomes}
\hfill
\begin{enumerate}
\item Under Assumption \ref{asm: rejective assignment probability},
\begin{align*}
\expeR{\boldtauhat } &= \boldsymbol{\tau}_{ATE} + \frac{N}{N_0} \left( \frac{1}{N}\sum_i \left( \pi_i - \frac{N_1}{N}  \right) \mathbf{Y}_i(0) \right) + \frac{N}{N_1} \left( \frac{1}{N}\sum_i \left( \pi_i - \frac{N_1}{N}  \right) \mathbf{Y}_i(1) \right), \\
&= \boldsymbol \tau_{EATT} + \frac{N}{N_0} \frac{N}{N_1} \left( \frac{1}{N}\sum_i \left( \pi_i - \frac{N_1}{N}  \right) \mathbf{Y}_i(0) \right).
\end{align*}
    
\item  \label{subprop: multivariate variance expression and inequality}
Under Assumptions \ref{asm: rejective assignment probability}, \ref{asm: reg conditions for CLT and var consistency}\ref{pi times one minus pi goes to infty} and \ref{asm: variance matrices converge},
\begin{align*}
    \vR{ \boldtauhat } + o(N^{-1}) &=  \dfrac{ \frac{1}{N} \sum_{k=1}^{N} \pi_k (1-\pi_k) }{ \frac{N_0}{N} \frac{N_1}{N}  } \left[ \frac{1}{N_1} \varsub{\tilde{\pi}}{\boldsymbol Y_i(1)} + \frac{1}{N_0}\varsub{\tilde{\pi}}{\boldsymbol Y_i(0)} - \frac{1}{N} \varsub{\tilde{\pi}}{\boldsymbol \tau_i} \right] \\
    & \leq \frac{1}{N_1} \varsub{\pi}{\boldsymbol Y_i(1)} + \frac{1}{N_0}\varsub{1-\pi}{\boldsymbol Y_i(0)},
\end{align*}
\noindent where $A \leq B$ if $B - A$ is positive semi-definite.

\item \label{subprop: multivariate variance consistency}
Under Assumptions \ref{asm: rejective assignment probability}, \ref{asm: reg conditions for CLT and var consistency}\ref{pi times one minus pi goes to infty}, \ref{asm: variance matrices converge}, and \ref{asm: asm for consistent variance estimation - vector version}, $$\mathbf{\hat{s}}_1 - \varsub{\pi}{\mathbf{Y}_i(1)} \xrightarrow{p} 0, \hspace{1cm} \mathbf{\hat{s}}_0 - \varsub{1-\pi}{\mathbf{Y}_i(0)} \xrightarrow{p} 0.$$ 

\item \label{subprop: multivariate clt}
Under Assumptions \ref{asm: rejective assignment probability}, \ref{asm: reg conditions for CLT and var consistency}\ref{pi times one minus pi goes to infty},  \ref{asm: variance matrices converge}, and \ref{asm: multivariate lindeberg type condition},
$$\vR{\boldtauhat}^{-\frac{1}{2}} (\boldtauhat - \boldsymbol{\tau} ) \xrightarrow{d} \normnot{0}{I}.$$
\noindent Assumption \ref{asm: variance matrices converge} implies $\Sigma_\tau = \lim_{N\rightarrow\infty} N \vR{\boldtauhat}$ exists, so the previous display can alternatively be written as
$$\sqrt{N} (\boldtauhat - \boldsymbol{\tau} ) \xrightarrow{d} \normnot{0}{\Sigma_{\boldsymbol{\tau}}}.$$
\end{enumerate}

\begin{proof}
The proof of claim (1) is analogous to the proof of Proposition \ref{prop: expectation of sdim} in the scalar case.

We next prove claim (2). For simplicity, let $A_n = \vR{\tauhat}$, let $B_n$ be the right-hand-side of the first equality in claim (2), and let $C_n$ be the right-hand side of the inequality in claim (2). We first prove the inequality. Note that by the definition of a semi-definite matrix, it suffices to show that $l'B_n l \leq l'C_n l$ for all $l \in \reals^K$. However, letting $Y_i(d) = l'\boldsymbol Y_i(d)$, the desired inequality follows from Proposition \ref{prop: shat conservative}. Next, observe that $A_n - B_n = o(N^{-1})$ if and only if $D_n := NA_n - NB_n = o(1)$, which holds if and only if $l'D_n l = o(1) $ for all $l \in L := \{e_j \,|\, 1\leq j \leq K \} \cup \{e_j - e_{j'} \,|\, 1 \leq j, j' \leq K\}$, where $e_j$ is the $j$th basis vector in $\reals^K$. To obtain the last equivalence, note that $e_j' D_n e_j = [D_n]_{jj}$ (the $(j,j)$ element of $D_n$), whereas exploiting the fact that $D_n$ is symmetric, $(e_j - e_{j'})' D_n (e_j - e_{j'}) = [D_n]_{jj} + [D_n]_{j'j'} - 2 [D_n]_{jj'}$, and so convergence of $l'D_n l$ to zero for all $l\in L$ is equivalent to convergence of each of the elements of $D_n$. Next, note that if $Y_i(d) = l' \mathbf{Y}_i(d)$, then $\tauhat$ as defined in (\ref{eqn: defn of tauhat}) is equal to $l'\boldtauhat$ and $\varsub{\tilde{\pi}}{Y_i(d)} = l'\varsub{\tilde{\pi}}{\boldsymbol Y_i(d)} l$. It follows from Proposition \ref{prop: formula for asymptotic variance} that
\begin{equation} 
N \cdot l'\vR{ \boldtauhat }l [1 + o(1)] =   \dfrac{ \frac{1}{N} \sum_{k=1}^{N} \pi_k (1-\pi_k) }{ \frac{N_0}{N} \frac{N_1}{N}  } l' \left[ \frac{N}{N_1} \varsub{\tilde{\pi}}{ \mathbf{Y}_i(1) } + \frac{N}{N_0}\varsub{\tilde{\pi}}{ \boldsymbol Y_i(0)} -  \varsub{\tilde{\pi}}{\boldsymbol \tau_i } \right] l, 
\end{equation}
\noindent which implies that $ l'D_n l =  l'(N A_n) l \cdot o(1) $. However, Assumption \ref{asm: variance matrices converge}, together with the inequality in claim (2), implies that the right-hand side of the previous display is $O(1)$, and thus $l'(N A_n) l = O(1)$, from which the desired result follows. 

The proof of claim (3) is similar to the proof of Lemma A3 in \citet{li_general_2017}, which gives a similar result in the case of completely randomized experiments. We provide a proof for the convergence of $\boldsymbol{\hat{s}}_1$; the convergence of $\boldsymbol{\hat{s}}_0$ is similar. As in the proof to claim (2), it suffices to show that $l'\boldsymbol{\hat{s}}_1 l - l'\varsub{\pi}{\boldsymbol Y_i(1)} l \rightarrow_p 0$ for all $l \in L$. Let $Y_i(d) = l'\boldsymbol Y_i(1)$. Then 
\begin{align*}
l'\boldsymbol{\hat{s}}_1 l &= \frac{1}{N_1} \sum_i D_i (l'\mathbf{Y}_i(1) - \frac{1}{N_1} \sum_j D_j l' \boldsymbol Y_j(1))^2 \\
&=\left( \frac{1}{N_1} \sum_i D_i (l'\mathbf{Y}_i(1) - l'\expesub{\pi}{\boldsymbol Y_i(1)})^2 \right)+ \left( \frac{1}{N_1} \sum_i D_i l' \boldsymbol Y_i(1) - \expesub{\pi}{l'\boldsymbol Y_i(1)}\right)^2, \numberthis \label{eqn: l shat l decom - vector variance consistency}
\end{align*}

\noindent where the second line uses the bias variance decomposition. The first term can be viewed as a Horvitz-Thompson estimator of $\frac{1}{N_1} \sum_i \pi_i (l' \boldsymbol Y_i(1) - \expesub{\pi}{l'\boldsymbol Y_i(1)})^2 = \varsub{\pi}{l' \boldsymbol Y_i(1)}$ under rejective sampling, and thus has variance equal to $$(1+o(1)) \frac{1}{N_1^2} \left( \sum_{i} \pi_i (1-\pi_i)\right) \varsub{\tilde{\pi}}{(l' \boldsymbol Y_i(1) - \expesub{\pi}{l'\boldsymbol Y_i(1)})^2}.$$ Further, observe that 
\begin{align*}
&\frac{1}{N_1^2} \left( \sum_{i} \pi_i (1-\pi_i)\right) \varsub{\tilde{\pi}}{(l' \boldsymbol Y_i(1) - \expesub{\pi}{l'\boldsymbol Y_i(1)})^2} \leq \\
&\frac{1}{N_1}  \expesub{\pi}{(l' \boldsymbol Y_i(1) - \expesub{\pi}{l'\boldsymbol Y_i(1)})^4} \leq \\
&\frac{1}{N_1} \max_i \left\{ (l' \boldsymbol Y_i(1) - \expesub{\pi}{l'\boldsymbol Y_i(1)})^2 \right\} \cdot \varsub{\pi}{l' \boldsymbol Y_i(1) } \leq \\
&\left[ ||l||^2  \frac{N}{N_1} \right] \left[ \max_i || \boldsymbol Y_i(1) - \expesub{\pi}{\boldsymbol Y_i(1)}||^2/N \right] \cdot \left[ l'\varsub{\pi}{\boldsymbol Y_i(1)}l \right] = o(1)
\end{align*}
\noindent where the first inequality is obtained using the fact that $\varsub{\tilde{\pi}}{X} \leq \expesub{\tilde{\pi}}{X^2}$, expanding the definition of $\expesub{\tilde{\pi}}{\cdot}$, and using the inequality $\pi_i (1-\pi_i) \leq \pi_i$, analogous to the argument in the proof to Proposition \ref{prop: clt for tauhat and var consistency} in the scalar case; the final inequality uses the Cauchy-Schwarz inequality and factors out $l$; and we obtain that the final term is $o(1)$ by noting that the first and final bracketed terms are $O(1)$ by Assumption \ref{asm: variance matrices converge} and the middle term is $o(1)$ by Assumption \ref{asm: asm for consistent variance estimation - vector version}. Applying Chebyshev's inequality, it follows that the first term in (\ref{eqn: l shat l decom - vector variance consistency}) is equal to $\varsub{\pi}{l' \mathbf{Y}_i(1)} + o(1)$. 

To complete the proof of the claim, we show that the second term in (\ref{eqn: l shat l decom - vector variance consistency}) is $o(1)$. Note that we can view $\frac{1}{N_1} \sum_i D_i l'\boldsymbol Y_i(1)$ as a Horvitz-Thompson estimator of $\expesub{\pi}{l'\mathbf{Y}_i}$. Following similar arguments to that in the proceeding paragraph, we have that its variance is bounded above by $\frac{1}{N_1} l'\varsub{\pi}{\boldsymbol Y_i(1)}l$, which is $o(1)$ by Assumption \ref{asm: variance matrices converge} combined with the fact that Assumption \ref{asm: reg conditions for CLT and var consistency}\ref{pi times one minus pi goes to infty} implies $N_1 \rightarrow \infty$. Applying Chebyshev's inequality again, we obtain that the second term in (\ref{eqn: l shat l decom - vector variance consistency}) is $o(1)$, as needed.

To prove claim (4), appealing to the Cramer-Wold device, it suffices to show that for any $l \in \reals^K \setminus \{0\}$, $Y_i = l'\mathbf{Y}_i$, and $\tauhat$ as defined in (\ref{eqn: defn of tauhat}), $\vR{\tauhat}^{-\frac{1}{2}} (\tauhat - \tau) \rightarrow_d \normnot{0}{1}$. 
This follows from Proposition \ref{prop: clt for tauhat and var consistency}, provided that we can show that Assumption \ref{asm: reg conditions for CLT and var consistency}\ref{asm: multivariate lindeberg type condition} implies that Assumption \ref{lindeberg type condition} holds when $Y_i = l'\boldsymbol Y_i$ for any conformable vector $l$. Indeed, recall that $\sigma_{\tilde{\pi}}^2 = l' \Sigma_{\tilde{\pi}} l \geq \lambda_{min} ||l||^2$, and hence $\frac{1}{\lambda_{min}} \geq \frac{1}{||l||^2} \frac{1}{\sigma_{\tilde{\pi}}^2}$. From the Cauchy-Schwarz inequality $$ \left|\left| \boldsymbol \Ytilde_i - \expesub{\tilde{\pi}}{ \boldsymbol \Ytilde_i }\right|\right|^2 \cdot ||l||^2 \geq (\Ytilde_i - \expesub{\tilde{\pi}}{\Ytilde_i})^2 .$$
Together with the previous inequality, this implies that

\begin{align*}
& \frac{1}{ \lambda_{min} } \expesub{\tilde{\pi}}{ \left|\left| \boldsymbol \Ytilde_i - \expesub{\tilde{\pi}}{ \boldsymbol \Ytilde_i }\right|\right|^2 \cdot 1\left[ \left|\left| \boldsymbol \Ytilde_i - \expesub{\tilde{\pi}}{\boldsymbol \Ytilde_i}\right|\right| \geq \sqrt{ \sum_i \pi_i (1-\pi_i)  \cdot \lambda_{min}} \cdot \epsilon \right]  }    \geq
\\&\frac{1}{ \sigma_{\tilde{\pi}}^2 }\expesub{\tilde{\pi}}{ (\Ytilde_i - \expesub{\tilde{\pi}}{\Ytilde_i})^2 \cdot 1\left[ \left| (\Ytilde_i - \expesub{\tilde{\pi}}{\Ytilde_i}) \right| \geq \sqrt{ \sum_i \pi_i (1-\pi_i)  } \cdot \sigma_{\tilde{\pi}} \epsilon \right]  }, 
\end{align*}
\noindent from which the result follows.
\end{proof}
\end{prop}

\paragraph{Implications for instrumental variables:} 
Consider the IV setting in Section \ref{sec: iv}. We can view the realizations $(D(Z_i),Y(Z_i))$ as the realizations of a vector of potential outcomes as a function of the ``treatment'' $Z_i$ (note that Assumption \ref{asm: rejective assignment for instrument} is analogous to Assumption \ref{asm: rejective assignment probability}, just relabeling the treatment $D_i$ as the instrument $Z_i$.) In particular, if we let $\mathbf{Y}_i(\cdot) = (Y_i(\cdot),D_i(\cdot))$, then $\boldtauhat = (\tauhat_{RF},\tauhat_{FS})'$. Proposition \ref{prop: results for multivariate outcomes} then provides regularity conditions on $\mathbf{Y}_i(\cdot)$ under which $\sqrt{N} \left( \tauhat_{RF} - \expeR{\tauhat_{RF}}, \tauhat_{FS} - \expeR{\tauhat_{FS}} \right)^\prime \xrightarrow{d} \normnot{0}{\Sigma_\tau}$. Provided the sequence of finite-populations further satisfies $(\expeR{\tauhat_{RF}},\expeR{\tauhat_{FS}}) \rightarrow (\tau_{RF}^*,\tau_{FS}^*)$ with $\tau_{FS}^* > 0$, then the uniform delta method (e.g., Theorem 3.8 in \citet{vaart_asymptotic_2000}) implies $\sqrt{N} (\betahat_{2SLS} - \beta_{2SLS}) \rightarrow_d N(0,g' \Sigma_\tau g),$ where $g$ is the gradient of $h(x,y) = x/y$ evaluated at $(\tau_{RF}^*,\tau_{FS}^*)$. Likewise, under these conditions, Proposition \ref{prop: results for multivariate outcomes} implies that the delta-method standard errors $\hat{g}' \mathbf{\hat{s}} \hat{g}$, for $\hat{g} = \nabla h(\boldtauhat)$, are consistent for an upper bound on the variance of $\hat\beta_{2SLS}$. Typical delta-method standard errors for IV will therefore be correct for $\beta_{2SLS}$ but potentially conservative in large finite-populations with a strong first-stage. We note that if one is concerned about a weak first-stage, one could construct \citet{anderson_estimation_1949}-style confidence sets by inverting tests of the form $H_0: \expeR{\tauhat_{RF}} - \beta_{2SLS} \expeR{\tauhat_{FS}} = 0$, in which case the strong first-stage assumption is not needed. 

\section{Additional Monte Carlo Simulations}\label{sec: appendix simulations}

This appendix provides additional results and extensions to the simulations in Section \ref{subsec: qwi monte carlo sims}. Figure \ref{fig: state level did distribution, qwi} plots the distribution of the DID estimator over the randomization distribution in our main specification. The remainder of the section presents extensions where (i) the number of treated units varies, (ii) there is treatment effect heterogeneity, and (iii) the size of the finite population varies.

\begin{figure}[htbp!]
    \centering
    \subfloat[Log employment]{\includegraphics[width = 3in]{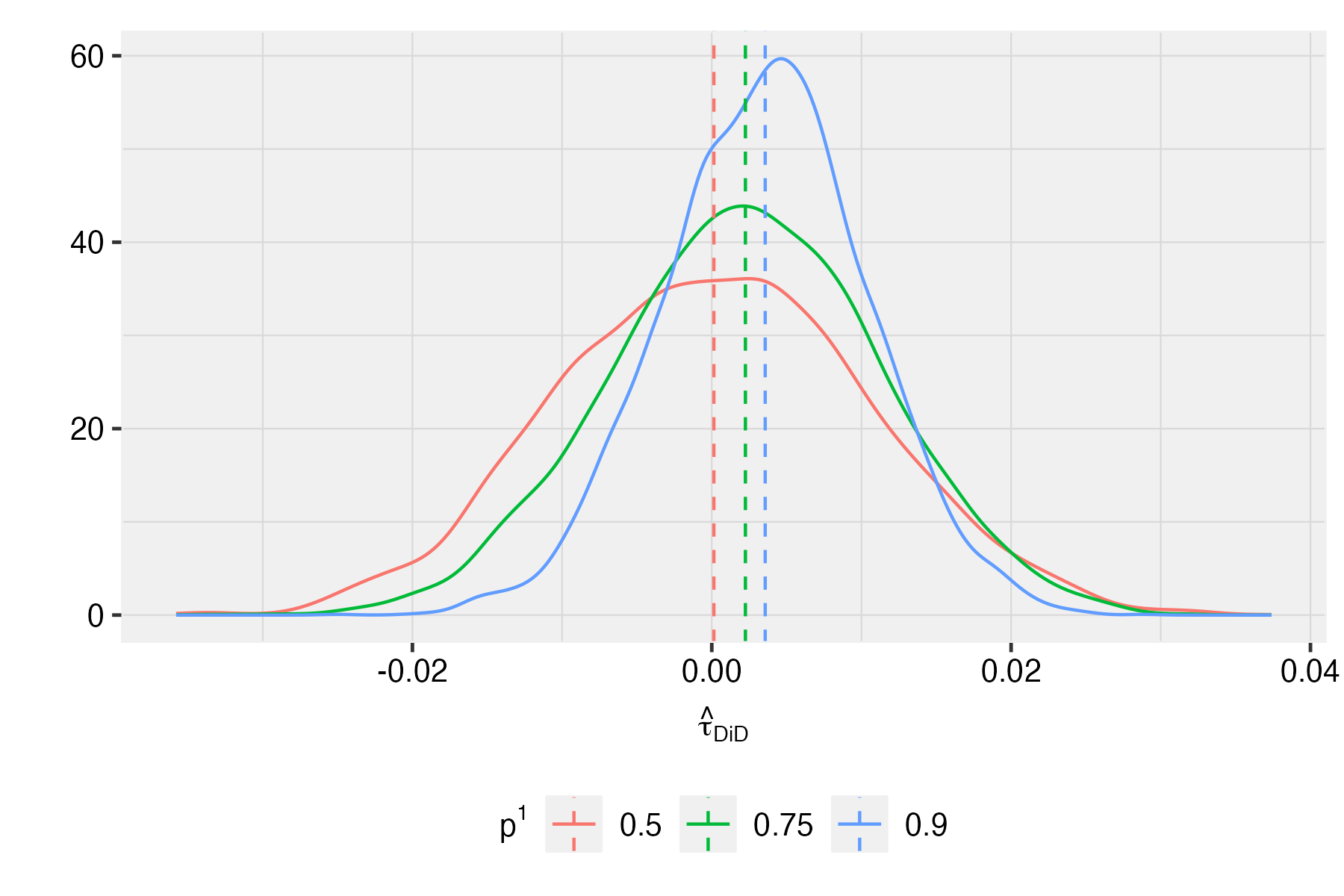}} \hspace{5mm}
    \subfloat[Log earnings]{\includegraphics[width = 3in]{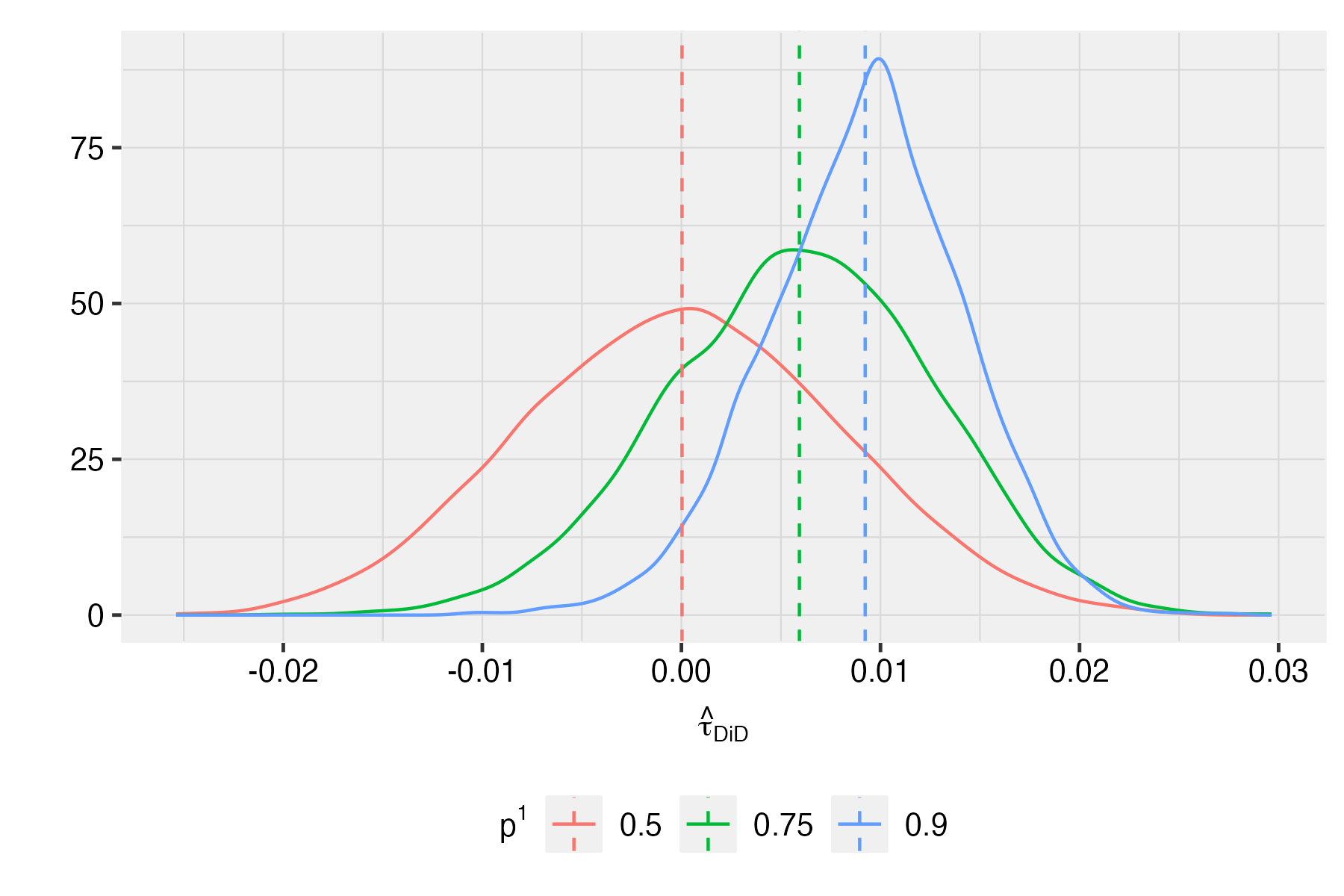}}
    \caption{Behavior of DID estimator $\tauhat_{DID}$ over the randomization distribution.}
    \floatfoot{\textit{Notes}: 
    This figure plots the behavior of the DID estimator $\tauhat_{DID}$ over the randomization distribution.
    The treatment probability for Democratic states, $p^1$, varies over $\{0.5, 0.75, 0.9\}$ (colors), holding fixed the number of treated units $N_1 = 25$.
    The results are computed over 5,000 simulations. The vertical dashed lines show the mean of the estimator for the relevant parameter values.}
    \label{fig: state level did distribution, qwi}
\end{figure} 

\subsection{Varying the Number of Treated Units}

In Section \ref{subsec: qwi monte carlo sims} of the main text, we report Monte Carlo simulations that documented the behavior of two-period DID estimates for the effect of a placebo law on state-level log average employment and state-level log average monthly earnings from the QWI when the number of treated and untreated units was approximately equal ($\frac{N_1}{N} = \frac{25}{51}$). 
We report the same results for the fraction of treated units varying over $N_1 \in \{ \lfloor 0.4 \, N \rfloor, \lfloor 0.6 \, N \rfloor \}$ in Table \ref{tab: state level did, appendix table treated frac varies}, where $\lfloor \cdot \rfloor$ is the floor function. 
The results are qualitatively similar as the case with $N_1 = \lfloor 0.5 \, N \rfloor$ in the main text.

\begin{table}[!ht]
    \centering
    \subfloat[Log employment with $N_1 = \lfloor 0.4 \, N \rfloor$]{\includegraphics[width=0.45\textwidth]{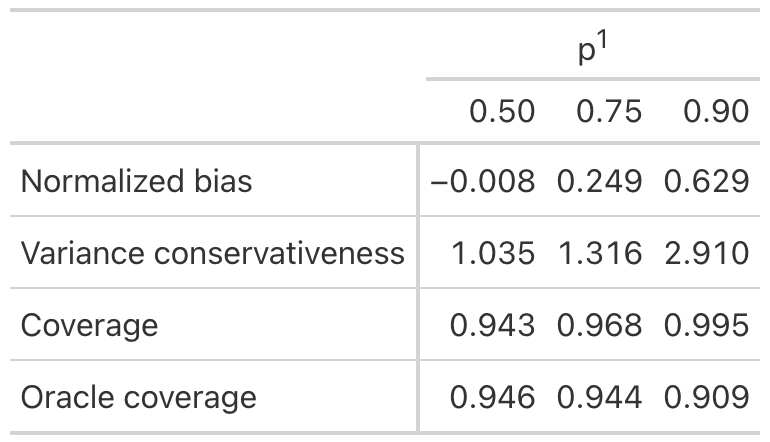}} \hspace{5mm}
    \subfloat[Log earnings with $N_1 = \lfloor 0.4 \, N \rfloor$]{\includegraphics[width=0.45\textwidth]{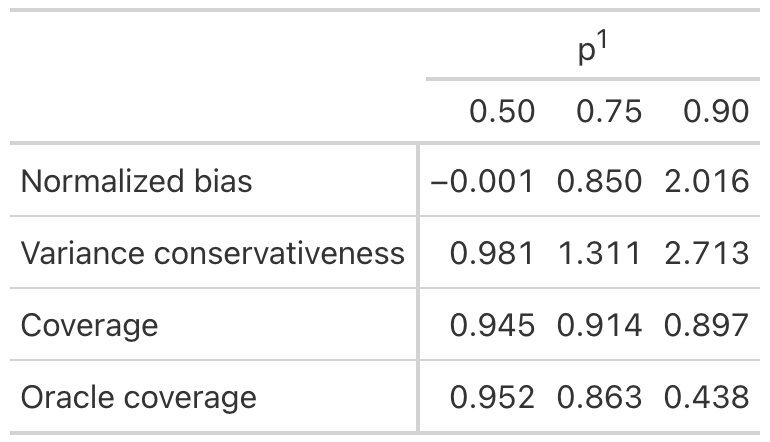}} \\
    \subfloat[Log employment with $N_1 = \lfloor 0.6 \, N \rfloor$]{\includegraphics[width=0.45\textwidth]{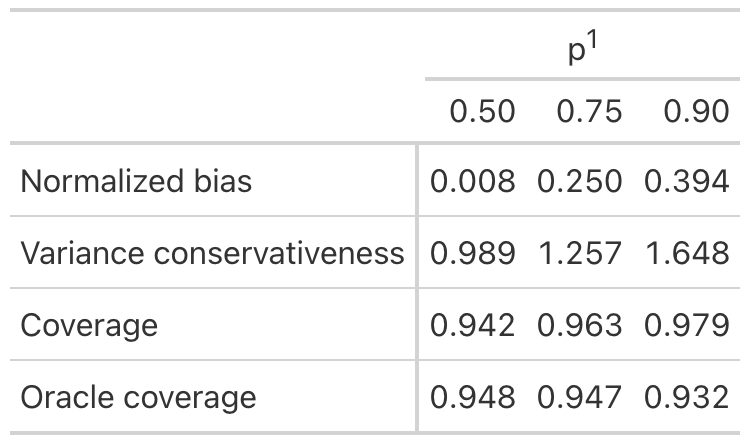}} \hspace{5mm}
    \subfloat[Log earnings with $N_1 = \lfloor 0.6 \, N \rfloor$]{\includegraphics[width=0.45\textwidth]{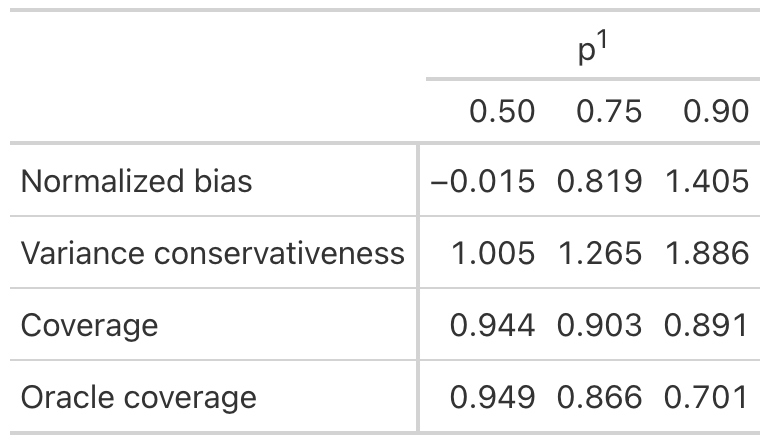}}
    \caption{Normalized bias, variance conservativeness, and coverage in Monte Carlo simulations with 
    $N_1 \in \{\lfloor 0.4 \, N \rfloor, \lfloor 0.6 \, N \rfloor$\}.}
    \floatfoot{\textit{Notes}: 
    Row 1 reports the normalized bias of the DID estimator ($\expesub{R}{\tauhat_{DID}}/\sqrt{\varsub{R}{\tauhat_{DID}}}$) for the EATT over the randomization distribution.
    Row 2 reports the estimated ratio $\frac{\expesub{R}{\hat{s}^2}}{\varsub{R}{\tauhat_{DID}}}$ across simulations, which measures the conservativeness of the heteroskedasticity-robust variance estimator. 
    Row 3 reports the estimated coverage rate of a 95\% confidence interval for the EATT based on the limiting normal approximation of the randomization distribution of the DID estimator and the heteroskedasticity-robust variance estimator $\hat{s}^2$. 
    Row 4 reports the coverage rate of an ``oracle'' 95\% confidence interval of the form $\hat\tau_{DID} \pm z_{0.975} \, \sqrt{\vR{\tauhat_{DID}}}$.
    The columns report results as the treatment probability $p^1$ for Democratic states varies over $\{0.5, 0.75, 0.9\}$.
    The results are computed over 5,000 simulations with $N = 51$.
    }
    \label{tab: state level did, appendix table treated frac varies}
\end{table}

\begin{table}[htbp!]
    \centering
    \subfloat[Log employment with $N_1 = \lfloor 0.4 \, N \rfloor$]{\includegraphics[width=0.45\textwidth]{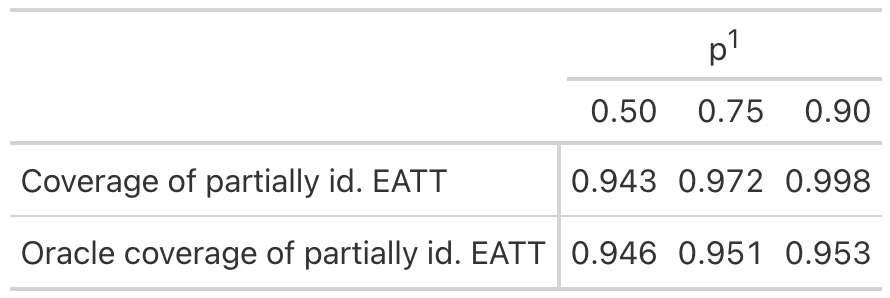}} \hspace{5mm}
    \subfloat[Log earnings with $N_1 = \lfloor 0.4 \, N \rfloor$]{\includegraphics[width=0.45\textwidth]{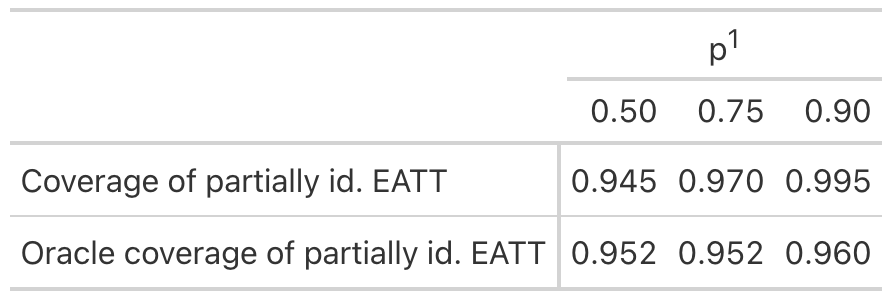}} \\
    \subfloat[Log employment with $N_1 = \lfloor 0.6 \, N \rfloor$]{\includegraphics[width=0.45\textwidth]{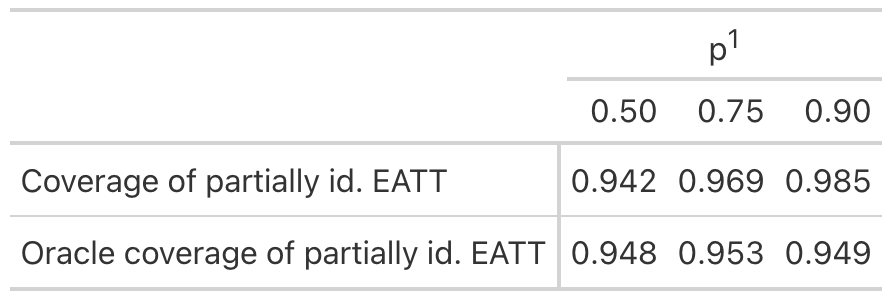}} \hspace{5mm} 
    \subfloat[Log earnings with $N_1 = \lfloor 0.6 \, N \rfloor $]{\includegraphics[width=0.45\textwidth]{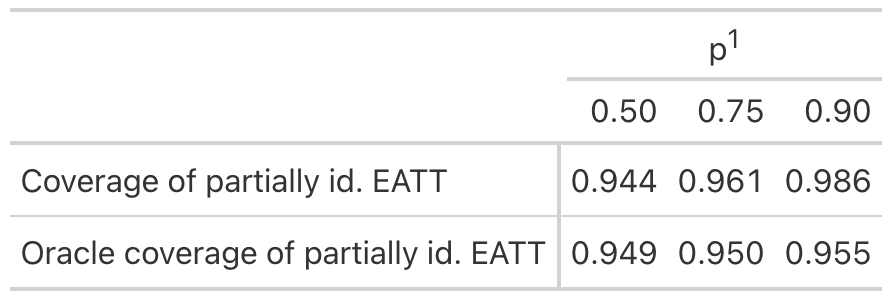}} 
    \caption{Coverage for the partially identified causal estimand in Monte Carlo simulations with $N_1 \in \{\lfloor 0.4 \, N \rfloor, \lfloor 0.6 \, N \rfloor\}$.}
    \floatfoot{\textit{Notes}: 
    Row 1 reports the coverage rate of a 95\% confidence interval for the partially identified EATT based on the construction in \cite{ImbensManski(04)} (see Section \ref{sec: DIM sensitivity analyis}). 
    Row 2 reports the coverage rate of an ``oracle'' 95\% confidence interval that uses the true variance rather than an estimated one.
    The bounds are chosen such that $\frac{N}{N_1} \frac{N}{N_0} \overline{b} = |\expesub{R}{\tauhat_{DID}}|$ and $\frac{N}{N_1} \frac{N}{N_0} \underline{b} = -|\expesub{R}{\tauhat_{DID}}|$.
    The columns report results as the treatment probability $p^1$ for Democratic states varies over $\{0.5, 0.75, 0.9\}$.
    When $p^1 = 0.5$, the upper bound $\tilde{b}$ equals zero, and the  \citet[][]{ImbensManski(04)} confidence interval is equivalent to a standard, nominal 95\% confidence interval.
    The results are computed over 5,000 simulations with $N = 51$.
    }
    \label{tab: state level did, appendix table treated frac varies, coverage of id set}
\end{table}

\subsection{Treatment Effect Heterogeneity}
In Section \ref{subsec: qwi monte carlo sims} of the main text, we report Monte Carlo simulations that documented the behavior of two-period DID estimators for the effect of a placebo law on state-level average employment and state-level log average monthly earnings from the QWI. 
These simulations were conducted without treatment effect heterogeneity, setting $Y_{it}(1) = Y_{it}(0)$ both to equal the observed state-level outcomes $Y_{it}$.

We report results from Monte Carlo simulations that incorporate treatment effect heterogeneity. 
As in the main text, we use aggregate data on the 50 U.S. states and Washington D.C. from the QWI (indexed by $i = 1, \hdots, N$) for the years 2012 and 2016 (indexed by $t = 1, 2$).
For each state and year, we set the untreated potential outcome $Y_{it}(0)$ equal to the state's observed outcome in the QWI. 
We impose ``no-anticipation'' by setting $Y_{i1}(1) = Y_{i1}(0)$.
We draw the treated potential outcome at $t = 2$ as $Y_{i2}(1) = Y_{i1}(0) + \lambda \sqrt{\varsub{1}{Y_{i2}(0) - Y_{i1}(0)}} Z_{i}$, where $Z_{i}$ is drawn from a standard normal distribution and $\lambda \in \{0.5, 1\}$. We draw the $Z_i$ once and hold them fixed throughout the simulations. To ease interpretation, we recenter the draws of the unit-specific treatment effects $\lambda \sqrt{\varsub{1}{Y_{i2}(0) - Y_{i1}(0)}} Z_{i}$ so that the EATT $\tau_{EATT,2}$ equals zero.

We simulate $D$ from the rejective assignment mechanism using the state-level results in the 2016 presidential election as in the main text, and we fix the number of treated states at $N_1 = \lfloor 0.5 \, N \rfloor$.
We again report results for two choices of the outcome $Y_{it}$: the log employment level for state $i$ in period $t$, and the log of state-level average quarterly earnings for state $i$ in year $t$.

\paragraph{Simulation results:}
Table \ref{tab: state level did, appendix table} summarizes the normalized bias, variance conservativeness, and coverage in the Monte Carlo simulations. 
The first row illustrates results in Table \ref{tab: state level did, main text table} without treatment effect heterogeneity (i.e., $\lambda = 0$). 
This table differs from Table \ref{tab: state level did, main text table} in the main text since these results are associated with a different simulation seed, although we see the same qualitative results.
For a particular choice of the treatment probabilities $p^1$, the bias of the two-period DID estimator for the EATT is fixed as the standard deviation of unit-specific treatment effects varies in these simulations.
But, as the standard deviation of unit-specific treatment effects increases, the standard errors become noticeably more conservative. For example, for the log earnings outcome and $p^1 = 0.75$, the variance estimator is approximately 1.4 times too large when $\lambda =0$, approximately 1.5 times too large when $\lambda = 0.5$, and approximately 2 times too large when $\lambda = 1$. As a result of this conservativeness, coverage rates increase for both outcomes as $\lambda$ increases: e.g., for log-earnings with $p^1 = 0.75$, coverage is 91.7\% with $\lambda =0$, 93.5\% with $\lambda = 0.5$, and $97.4\%$ with $\lambda = 1$. 

\begin{table}[!ht]
    \centering
    \subfloat[Log employment with $\lambda = 0$]{\includegraphics[width=0.45\textwidth]{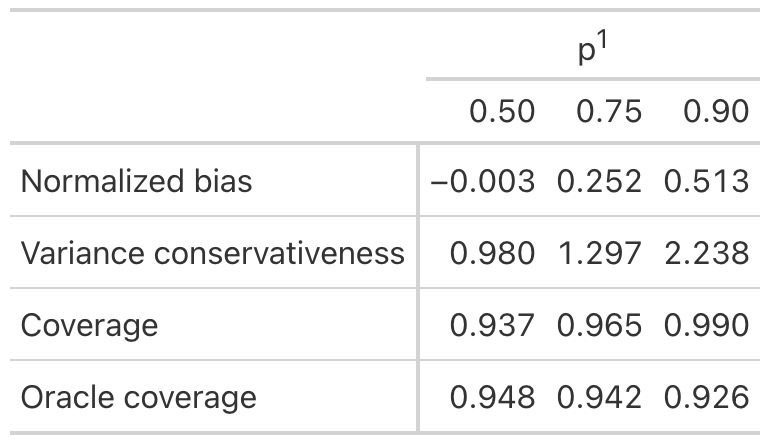}} \hspace{5mm}
    \subfloat[Log earnings with $\lambda = 0$]{\includegraphics[width=0.45\textwidth]{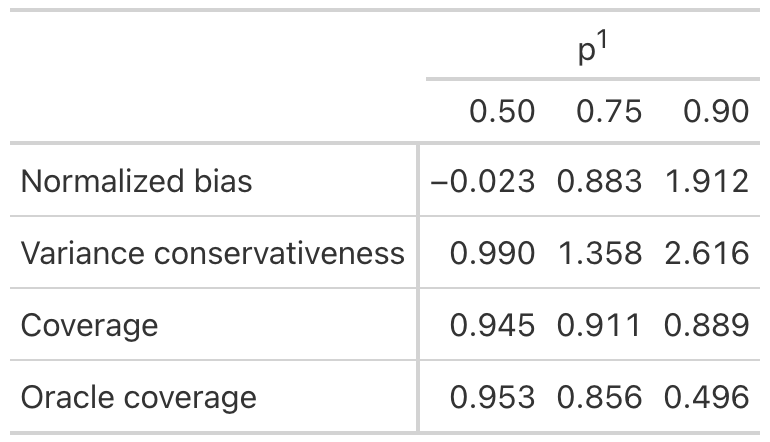}} \\
    \subfloat[Log employment with $\lambda = 0.5$]{\includegraphics[width=0.45\textwidth]{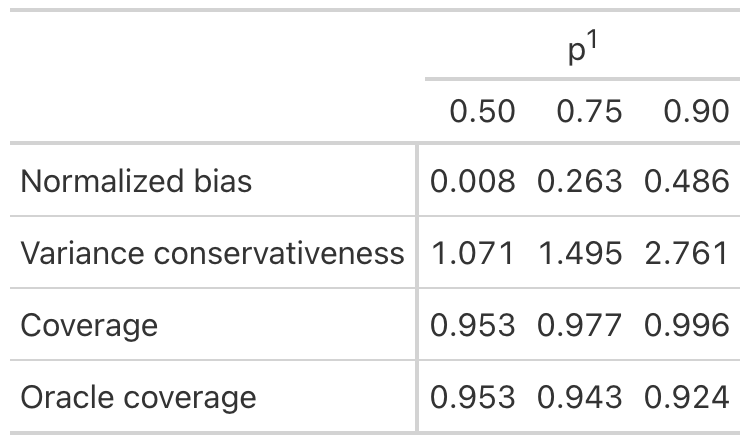}} \hspace{5mm}
    \subfloat[Log earnings with $\lambda = 0.5$]{\includegraphics[width=0.45\textwidth]{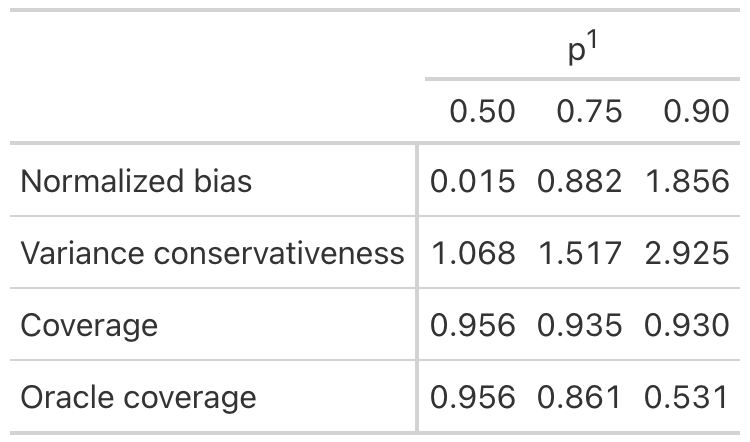}} \\
    \subfloat[Log employment with $\lambda = 1$]{\includegraphics[width=0.45\textwidth]{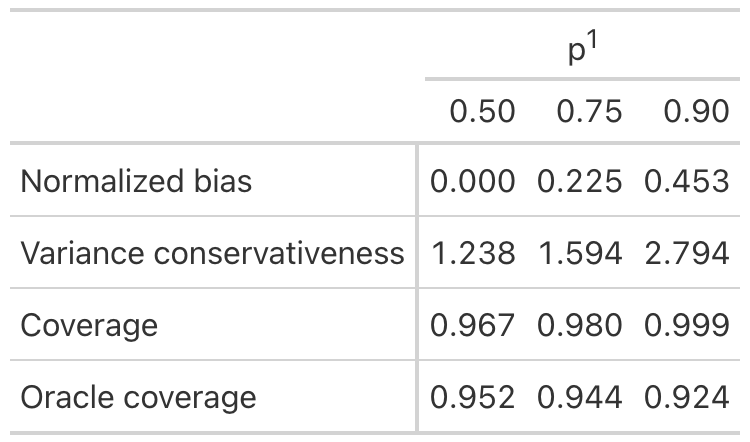}} \hspace{5mm}
    \subfloat[Log earnings with $\lambda = 1$]{\includegraphics[width=0.45\textwidth]{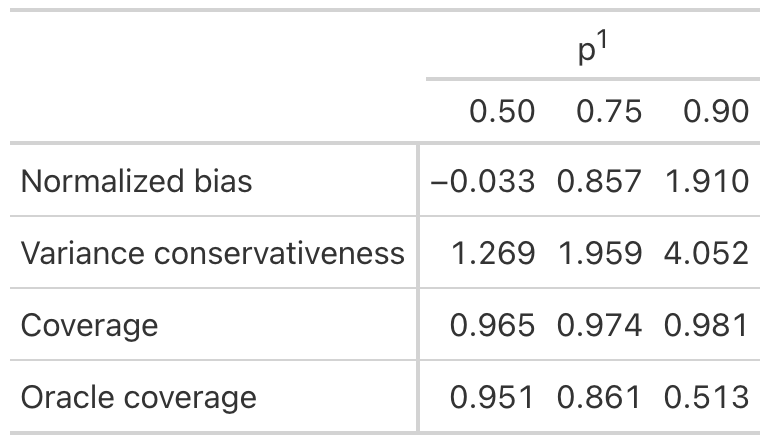}}
    \caption{Normalized bias, variance conservativeness, and coverage in Monte Carlo simulations with treatment effect heterogeneity.}
    \floatfoot{\textit{Notes}: 
    Within a particular table, Row 1 reports the normalized bias of the DID estimator ($\expesub{R}{\tauhat_{DID}}/\sqrt{\varsub{R}{\tauhat_{DID}}}$) for the EATT over the randomization distribution; Row 2 reports the estimated ratio $\frac{\expesub{R}{\hat{s}^2}}{\varsub{R}{\tauhat_{DID}}}$ across simulations, which measures the conservativeness of the heteroskedasticity-robust variance estimator; Row 3 reports the coverage rate of a nominal 95\% confidence interval of the form $\hat\tau_{DID} \pm z_{0.975} \, \hat{s}$; and Row 4 reports coverage of an oracle confidence interval that uses the true variance rather than an estimated one. The columns report results as the treatment probability $p^1$ for Democratic states varies over $\{0.5, 0.75, 0.9\}$.
    The results are computed over 5,000 simulations with $N_1 = \lfloor 0.5 \, N \rfloor$ and $N = 51$. Panels (a)-(f) vary the outcome and the degree of treatment effect heterogeneity ($\lambda$).
    }
    \label{tab: state level did, appendix table}
\end{table}

\begin{table}[htbp!]
\centering
\subfloat[Log employment with $\lambda = 0$]{\includegraphics[width=0.45\textwidth]{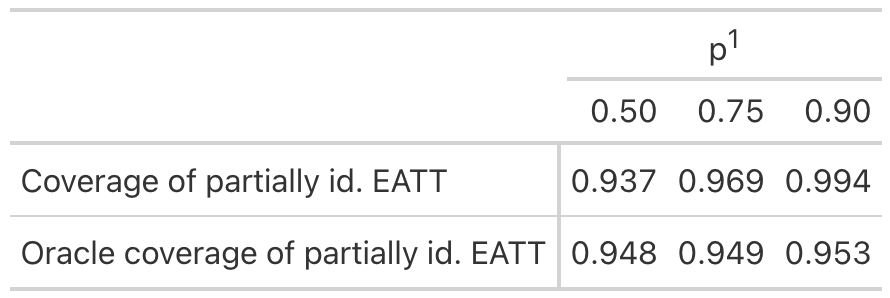}} \hspace{5mm}
\subfloat[Log earnings with $\lambda = 0$]{\includegraphics[width=0.45\textwidth]{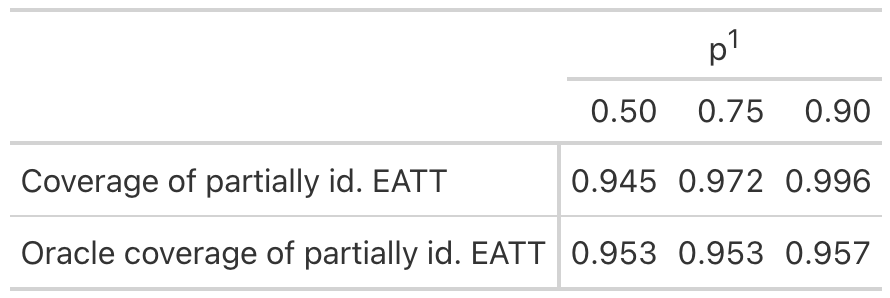}} \\
\subfloat[Log employment with $\lambda = 0.5$]{\includegraphics[width=0.45\textwidth]{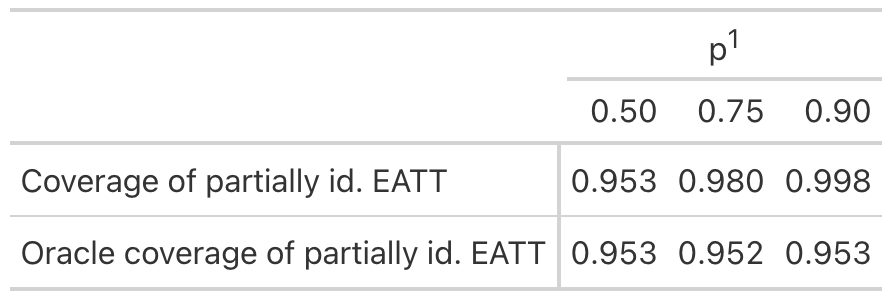}} \hspace{5mm}
\subfloat[Log earnings with $\lambda = 0.5$]{\includegraphics[width=0.45\textwidth]{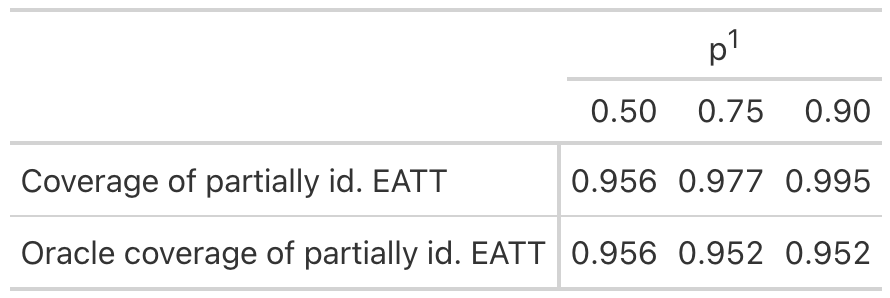}} \\
\subfloat[Log employment with $\lambda = 1$]{\includegraphics[width=0.45\textwidth]{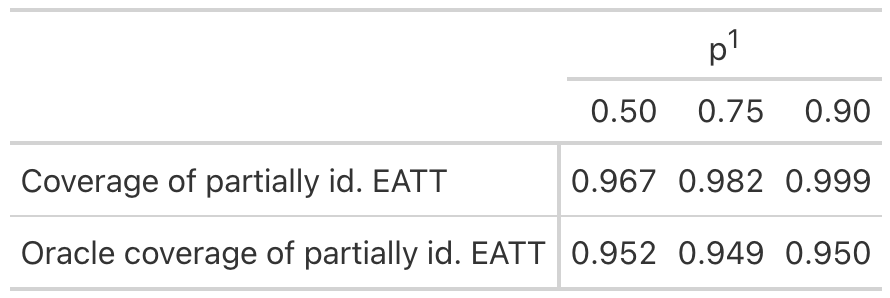}} \hspace{5mm} 
\subfloat[Log earnings with $\lambda = 1$]{\includegraphics[width=0.45\textwidth]{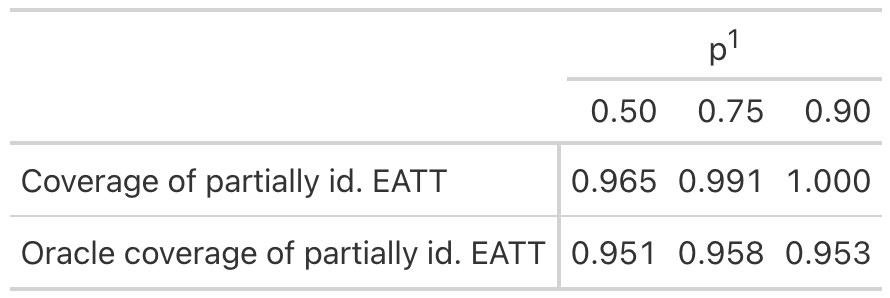}} 
\caption{Coverage for the partially identified causal estimand in Monte Carlo simulations with treatment effect heterogeneity.}
\floatfoot{\textit{Notes}: Row 1 reports the coverage rate of a 95\% confidence interval for the partially identified EATT based on the construction in \cite{ImbensManski(04)} (see Section \ref{sec: DIM sensitivity analyis} for details). 
Row 2 reports the coverage rate of an ``oracle'' 95\% confidence interval that uses the true variance rather than an estimated one.
The columns report results as the treatment probability $p^1$ for Democratic states varies over $\{0.5, 0.75, 0.9\}$.
When $p^1 = 0.5$, the upper bound $\tilde{b}$ equals zero, and the  \citet[][]{ImbensManski(04)} confidence interval is equivalent to a standard, nominal 95\% confidence interval.
The results are computed over 5,000 simulations with $N_1 = \lfloor 0.5 \, N \rfloor$ and $N = 51$. 
Panels (a)-(f) vary the outcome and the degree of treatment heterogeneity ($\lambda$).
}
\label{tab: state level did, appendix table, coverage of id set}
\end{table}

In Figure \ref{fig: state level did distribution, QWI, appendix, vary p1 and tau}, we plot how the randomization distribution of the DID estimator varies as we vary both the individual treatment probabilities and the standard deviation of unit-specific treatment effects.

\begin{figure}[htbp!]
    \centering
    \subfloat[Log employment]{\includegraphics[width = 3in]
    {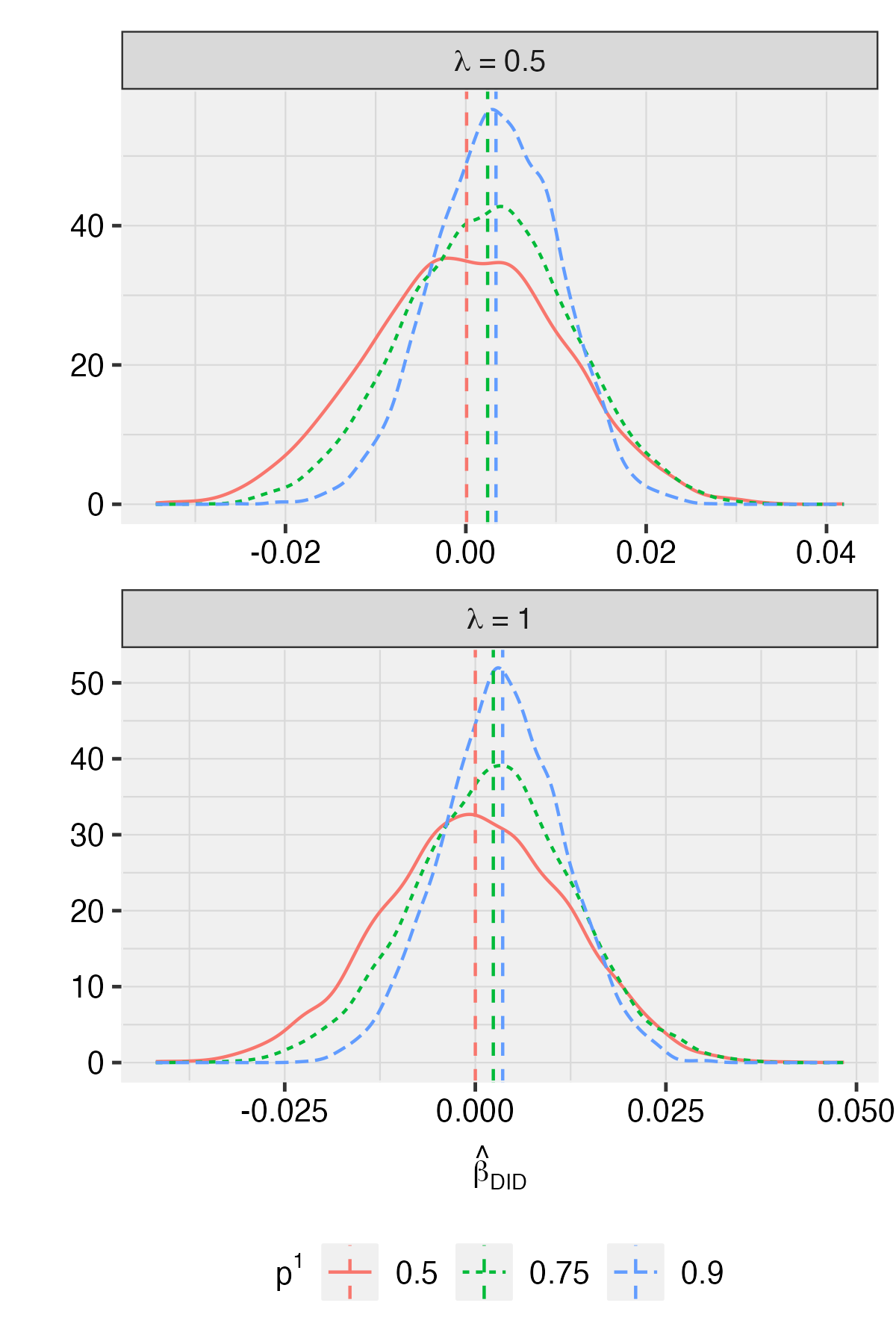}} \hspace{5mm}
    \subfloat[Log earnings]{\includegraphics[width = 3in]{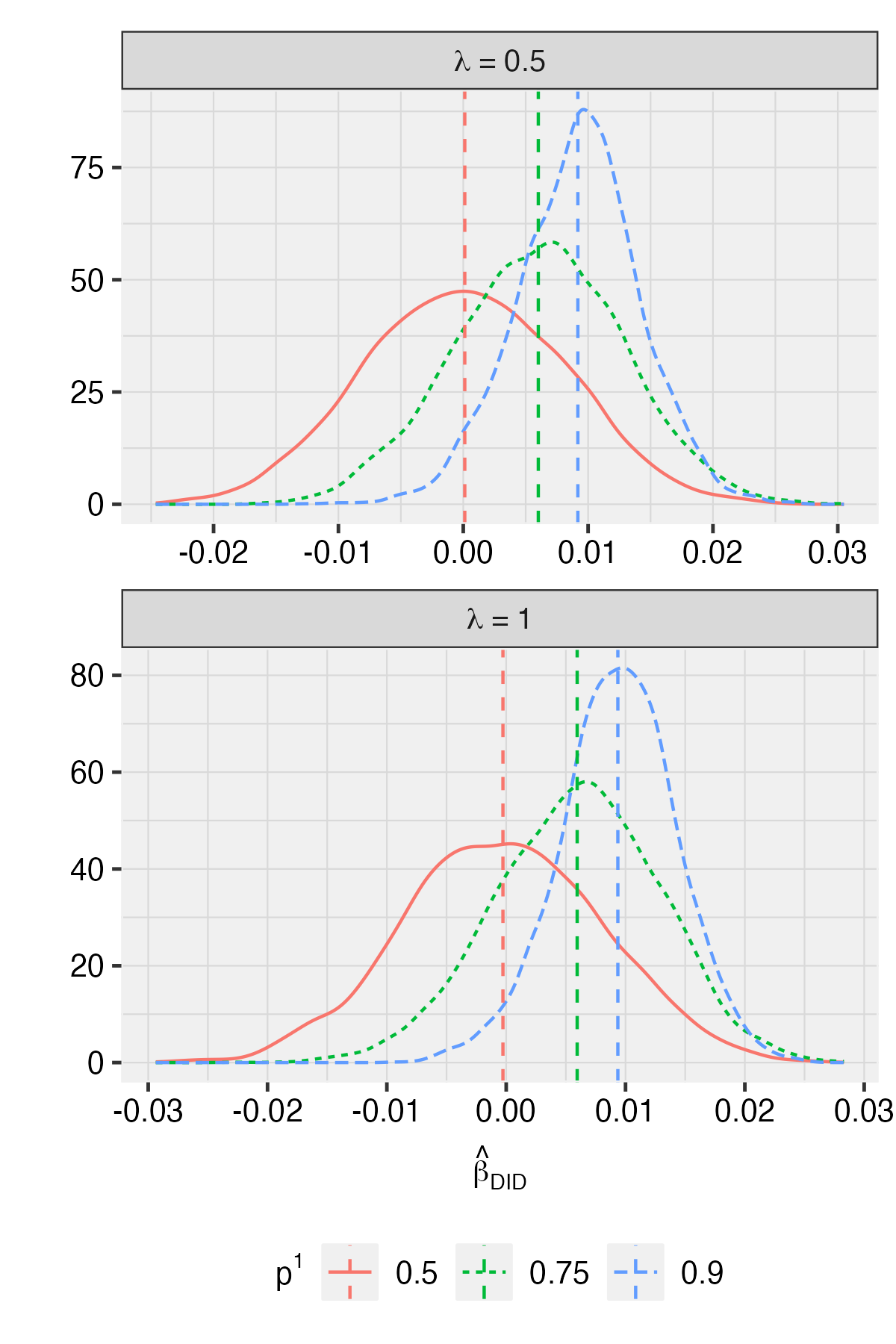}} 
    \caption{Behavior of DID estimator $\tauhat_{DID}$ over the randomization distribution with treatment effect heterogeneity.}
    \floatfoot{\textit{Notes}: This figure plots the behavior of the DID estimator $\hat{\tau}_{DID}$ over the randomization distribution. The individual treatment probabilities $p^1$ varies over $\{0.5, 0.75, 0.9\}$ (colors), and the standard deviation of unit-specific treatment effects $\lambda$ varies over $\{0.5, 1\}$ (columns). The results are computed over $5,000$ simulations with $N_1 = \lfloor 0.5 \, N \rfloor$ and $N = 51$.}
    \label{fig: state level did distribution, QWI, appendix, vary p1 and tau}
\end{figure}

\subsection{Varying Population Sizes}
In Section \ref{subsec: qwi monte carlo sims}, we reported results where the finite population was the 50 U.S. states and Washington D.C. We report simulations where the size of the finite population varies. Specifically, we consider simulations designs with $N \in \{10,26,51\}$, where the smaller populations are obtained by choosing a subset of the 51 units in ascending order of their associated FIPS codes.

In Figure \ref{fig: state level did distribution, QWI, appendix, vary p1 and N}, we fix the standard deviation of unit-specific treatment effects to be $\lambda = 0$, and plot how the randomization distribution of the two-period DID estimator varies as we vary both the individual treatment probabilities $p^1$ and the total number of states $N$.
For $N = 10$, the distributions appear to be symmetric, but have oscillations that are not characteristic of a normal distribution (particularly for $p^1 = 0.9$). 
But, as $N$ is increased to 26 (or 51), the distributions appear to be approximately normally distributed, illustrating the finite-population central limit theorem in Proposition \ref{prop: clt for tauhat and var consistency}.
Table \ref{tab: coverage, QWI, appendix, vary p and N} summarizes how the coverage rate of a nominal 95\% confidence interval of the form $\hat\tau_{DID} \pm z_{0.975} \, \hat{s}$ varies. Interestingly, for $N_c = 10$, despite the non-normal distribution we find that the coverage rate never drops below 91.9\% for the log employment outcome and 92.3\% for the log earnings outcome.

\begin{figure}[!ht]
    \centering
    \subfloat[Log employment]{\includegraphics[width = 3in]{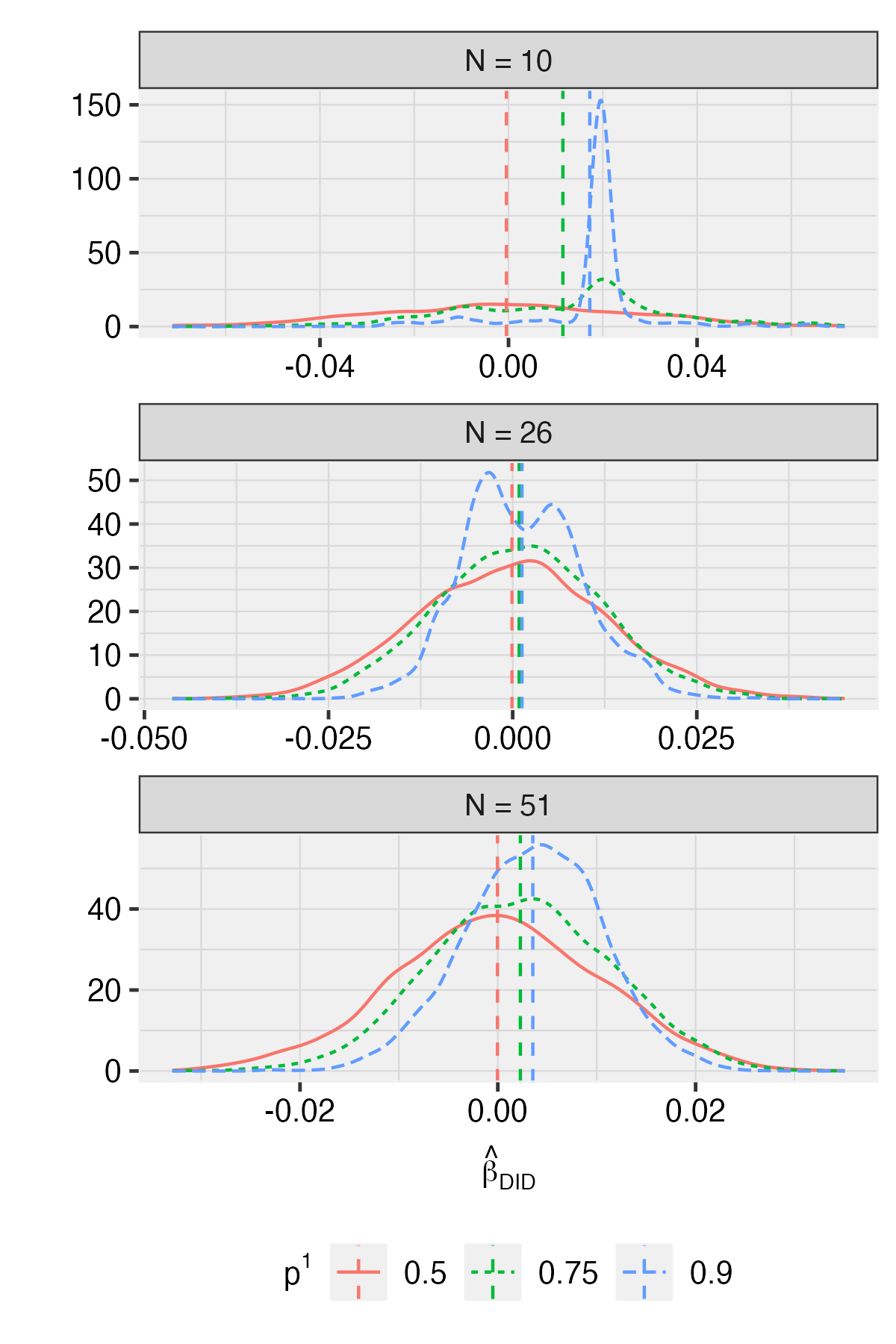}} \hspace{5mm}
    \subfloat[Log earnings]{\includegraphics[width = 3in]{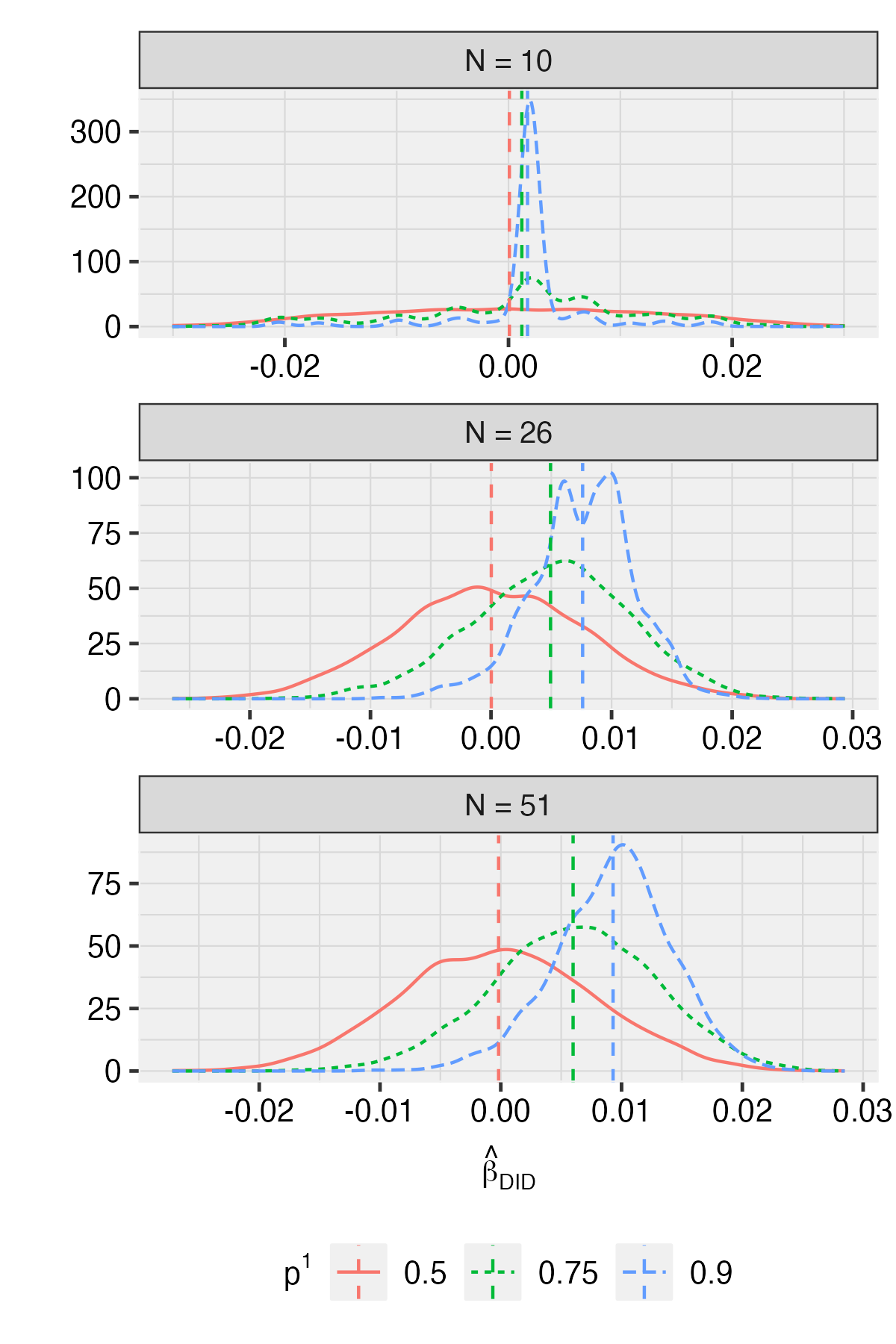}} 
    \caption{Behavior of DID estimator $\tauhat_{DID}$ over the randomization distribution varying the size of the finite population.}
    \floatfoot{\textit{Notes}: This figure plots the behavior of the DID estimator $\hat{\tau}_{DID}$ over the randomization distribution. The individual treatment probabilities $p^1$ varies over $\{0.5, 0.75, 0.9\}$ (colors), and the total number of units $N$ varies over $\{10, 26, 51\}$ (columns). The results are computed over $5,000$ simulations with $N_1 = \lfloor 0.5\, N \rfloor$ and $\lambda = 0$.}
    \label{fig: state level did distribution, QWI, appendix, vary p1 and N}
\end{figure}

\begin{table}[htbp!]
    \centering
    \subfloat[Log employment with $\lambda = 0$]{\includegraphics[width=0.45\textwidth]{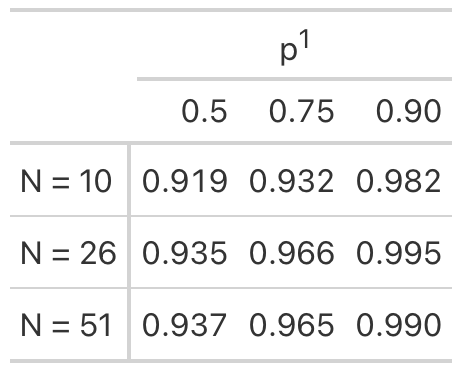}} \hspace{5mm}
    \subfloat[Log earnings with $\lambda = 0$]{\includegraphics[width=0.45\textwidth]{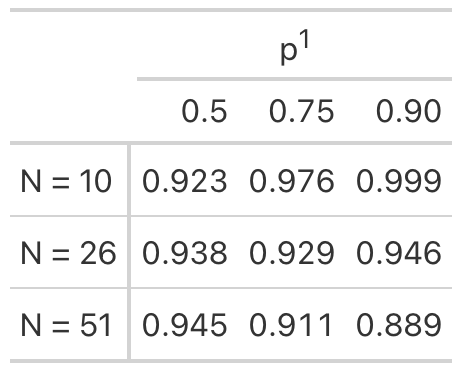}}
    \caption{Coverage in Monte Carlo simulations varying the size of the finite population.}
    \floatfoot{\textit{Notes}:
    This table reports the coverage rate of a nominal 95\% confidence interval of the form $\hat\tau_{DID} \pm z_{0.975} \, \hat{s}$ as the size of the finite population $N$ varies over $\{10, 26, 51\}$ (rows) and the treatment probability $p^1$ for Democratic states varies over $\{0.5, 0.75, 0.9\}$ (columns).
    The results are computed over 5,000 simulations with  with $N_1 = \lfloor 0.5\, N \rfloor$ and $\lambda = 0$.
    }
    \label{tab: coverage, QWI, appendix, vary p and N}
\end{table}

\end{document}